\documentclass[useAMS]{emulateapj}
\usepackage{latexsym,graphicx,natbib}
\usepackage{amsmath}

\newcommand{\gap}{\;\rlap{\lower 2.5pt \hbox{$\sim$}}\raise 1.5pt\hbox{$>$}\;}
\newcommand{\lap}{\;\rlap{\lower 2.5pt \hbox{$\sim$}}\raise 1.5pt\hbox{$<$}\;}

\def\beq{\begin{equation}}
\def\eeq{\end{equation}}

\newcommand\msun{\, \rm M_\odot} 
\newcommand\mpc{{\, \rm mpc}}
\newcommand\pc{{\, \rm pc}}
\newcommand\yr{{\, \rm yr}}

\newcommand\msbh{M_\bullet}

\newcommand\mg{M_{\rm gal}}
\newcommand\rh{r_{\rm h}}
\newcommand\ri{r_{\rm m}}
\newcommand\tnb{T_{\rm NB}}
\newcommand\thd{T_{\rm HD}}
\newcommand\tr{T_{\rm r}}

\def\aj{AJ}%
\def\apj{ApJ}%
\def\apjl{ApJL}%
\def\apjs{ApJS}%
\def\aap{A\&A}%
\def\mnras{MNRAS}%
\def\nat{Nature}%
\def\na{New A}%
\def\physrep{Phys.~Rep.}%
\shortauthors{Gualandris and Merritt}
\shorttitle{Mergers of galaxies with collisionally relaxed nuclei}

\begin{document}

\title{Long-term evolution of massive black hole binaries.
IV. Mergers of galaxies with collisionally relaxed nuclei}

\author{Alessia Gualandris\altaffilmark{1} and David Merritt\altaffilmark{2}}

\altaffiltext{1}{Max-Planck Institut f\"{u}r Astrophysik,
  Karl-Schwarzschild-Str. 1, D-85748 Garching, Germany}
\altaffiltext{2}{Department of Physics and Center for Computational Relativity
  and Gravitation, Rochester Institute of Technology, 54 Lomb Memorial
  Drive, Rochester, NY}

\email{alessiag@mpa-garching.mpg.de}

\begin{abstract}
We simulate mergers between galaxies containing collisionally-relaxed
nuclei around massive black holes (MBHs). Our galaxies contain four
mass groups, representative of old stellar populations; a primary goal
is to understand the distribution of stellar-mass black holes (BHs)
after the merger. Mergers are followed using direct-summation $N$-body
simulations, assuming a mass ratio of 1:3 and two different
orbits. Evolution of the binary MBH is followed until its separation
has shrunk by a factor of 20 below the hard-binary separation. During
the galaxy merger, large cores are carved out in the stellar
distribution, with radii several times the influence radius of the
massive binary. Much of the pre-existing mass segregation is erased
during this phase. We follow the evolution of the merged galaxies for
approximately three, central relaxation times after coalescence of the
massive binary; both standard, and top-heavy, mass functions are
considered. The cores that were formed in the stellar distribution
persist, and the distribution of the stellar-mass black holes evolves
against this essentially fixed background. Even after one central
relaxation time, these models look very different from the relaxed,
multi-mass models that are often assumed to describe the distribution
of stars and stellar remnants near a massive BH.  While the stellar
BHs do form a cusp on roughly a relaxation time-scale, the BH density
can be much smaller than in those models. We discuss the implications
of our results for the EMRI problem and for the existence of
Bahcall-Wolf cusps.
\end{abstract}


\section{Introduction}
\label{sec:intro}

The massive black holes (MBHs) that reside at the centers of some
nearby galaxies are believed to grow together with their hosts through
mergers: MBHs grow partly as a result of gas accretion, and partly by
coalescence with other MBHs that are brought into the nucleus during
the merger process \citep{BBR80}.  The detailed assembly history of
MBHs is poorly understood; major uncertainties include the ``seed''
mass distribution of MBHs at high redshift, the typical gas accretion
efficiency, and the frequency with which MBHs are ejected due to
gravitational-wave recoil \citep{KH00,VHM03}.  But a robust prediction
of the hierarchical models is that galaxies hosting MBHs in the nearby
universe were formed from less massive systems, at least some of which
already contained MBHs.

Binary MBHs created during galaxy mergers leave imprints on the
stellar distribution: for instance, they create low-density cores, by
exchanging energy with passing stars \citep{BBR80,mm01}.  Such cores
are observed to be ubiquitous in stellar spheroids brighter than $\sim
10^{10}L_\odot$ \citep{ferr1994,Lauer95} and their sizes -- of order
the influence radius of the (presumably single) MBH -- are consistent
with the predictions of merger models \citep{graham2004,merritt06}.
Here we define the influence radius as
\begin{equation}
\label{eq:rh}
\rh \equiv \frac{G\,M_\bullet}{\sigma^2}
\end{equation}
where $\sigma$ is the rms velocity of stars in any direction at $r\gap
\rh$.  Cores of radius $\sim \rh$ become difficult to resolve in
galaxies beyond the Local Group if the MBH mass is below $\sim
10^8\msun$.  Even in the nearest nucleus, that of the Milky Way, the
presence of a parsec-scale core around Sgr A$^*$ was only clearly
established in the last few years \citep{buchholz09}.

In the absence of MBHs, mergers tend to preserve the form of the
stellar distribution near the centers of galaxies \citep{Dehnen2005}.
Binary MBHs, however, are efficient at erasing the structure that was
present on scales $\lap\rh$ \citep{mc2001}, and this fact precludes
drawing definite conclusions about the nuclear properties of the
galaxies that preceded the hosts of observed MBHs.  On the other hand,
it is well established that low-luminosity galaxies have higher
central densities than high-luminosity galaxies \citep{Kormendy1985}.
This is true both in terms of the mean density within the half-mass
radius, and also in terms of the density on the smallest resolvable
scales: low-luminosity spheroids often contain dense, nuclear stars
clusters (NSCs), with sizes of order $10\pc$ and masses $\lap 1\%$ the
mass of the galaxy \citep{Boeker2010}.  NSC masses are therefore
comparable to, or somewhat greater than, MBH masses
\citep{Ferrarese2006}, although NSCs have been shown to coexist with
MBHs in only a handful of galaxies \citep{Seth2008,GS2009}.  The NSC
in the Milky Way is believed to be a representative example: its
half-light radius is $3-5\pc$, or $1-2\rh$, and its mass is a few
times $10^7\msun$, or several times $M_\bullet$
\citep{GS2009,Schoedel2011}.

In the NSC of the Milky Way, the two-body relaxation time is
comparable with the age of the universe, and this is consistent with
the persistence of a core (as opposed to a \citet{BW76} cusp) in the
late-type stars \citep{merritt2010}.  But in fainter systems, central
relaxation times are shorter.  For instance, in the Virgo cluster,
galaxies with NSCs have nuclear half-light relaxation times that scale
with host-galaxy luminosity as \citep{merritt2009} \beq
T_{r,\mathrm{NSC}}\approx 1.2\times 10^{10} \mathrm{yr}
\left(\frac{L_\mathrm{gal}}{10^{10}L_\odot}\right).  \eeq By
comparison, the mean time between ``major mergers'' (mergers with mass
ratios $3:1$ or less) of dark-matter haloes in the hierarchical models
varies from $\sim 0.2$ Gyr at redshift $z=10$ to $\sim 10^{10}$ yr at
$z=1$, with a weak dependence on halo mass \citep{Fakhouri2010}.  This
comparison suggests that the progenitors of many spheroids in the
current universe may have been galaxies containing nuclei that were
able to attain a collisionally-relaxed state before the merger that
formed them took place.

In the absence of a MBH, collisional relaxation implies mass
segregation and core collapse.  If a MBH is present, mass segregation
still occurs, but core collapse is inhibited by the fixed potential
due to the MBH.  A collisional steady state, which is reached by a
time $\sim T_r(\rh)$ at radii $r\lap \rh$, is characterized by a
Bahcall-Wolf, $n\sim r^{-7/4}$ cusp in the dominant component at
$r\lap 0.2\rh$.  If there is a mass spectrum, less-massive objects
follow a shallower profile, $n\sim r^{-3/2}$, while more-massive
objects follow a steeper profile, $n\sim r^{-2}$ \citep{BW77,HA06}.

As a first approximation, the mass spectrum of an evolved stellar
population can be represented in terms of just two components: objects
of roughly one solar mass or less (main-sequence stars, white dwarfs,
neutron stars); and remnant black holes (BHs) with masses
$10-20\msun$.  Standard initial mass functions predict that roughly
$1\%$ of the total mass will be in stellar BHs \citep{Alexander2005};
so-called ``top-heavy'' mass functions \citep[e.g.][]{Bartko10}
predict a larger fraction.  In a collisionally relaxed nucleus, the
density of stellar BHs will rise more steeply toward the center than
the density of the stars.  Mergers between galaxies with such nuclei
would be expected to modify these steady-state distributions
substantially, and also to affect (increase) the time scale over which
a collisionally relaxed cusp could be regenerated following the merger
\citep{MS06}.

These arguments motivated us to carry out merger simulations between
galaxies containing multi-component, mass-segregated nuclei around
MBHs.  As in previous papers from this series
\citep{mm01,ber05,mms07}, our merger simulations are purely
stellar-dynamical.  In some galaxies, torques from gas would assist in
the evolution of binary MBHs \citep{escala05,dotti07,cuadra09}.  Gas
also implies star formation, and there is evidence for complex star
formation histories in many NSCs \citep{Walcher2006}.  But $N$-body
simulations that allow for non-spherical geometries
\citep{ber06,khan2011} have shown that purely dissipationless energy
exchange with ambient stars can bring binary MBHs to milliparsec
separations on time scales much shorter than galaxy lifetimes.
%suggesting that gas dynamics may be of secondary importance even when
%gas is present.  This is particularly likely to be true when the
%initial stellar densities are high, as in our models.  
Unless the late evolution of the binary MBH is greatly accelerated by
torques from the gas, the influence of the binary on the distribution
of the {\it stellar} populations should be accurately reproduced by
our dissipationless models.  With respect to star formation,
population synthesis of NSC spectra suggest that most of the mass
typically resides in stars with ages of order 5-10 Gyr
\citep{figer2004,Boeker2010}, i.e. an old population.

Simulating the merger of galaxy-sized systems, while enforcing the
spatial and temporal resolution required to faithfully reproduce the
dynamics of stars on scales $\ll\rh$ around the central MBH, is
computationally demanding.  Our simulations used $\sim 10^6$ particles
per galaxy, and the models were advanced using a parallel,
direct-summation $N$-body code \citep{Harfst2007}.  The integrations
were accelerated using special-purpose hardware.  The galaxy models
contained four mass groups, representing an evolved stellar
population.  Initial conditions of the merging galaxies were
constructed in a two-stage process: models of mass-segregated nuclei
around a MBH were first constructed, then these collisionally-relaxed
models were imbedded into larger, spheroid-sized models.  Mergers were
then carried out, assuming a galaxy mass ratio of $1:3$.

A major motivation for our new simulations was the need to understand
the distribution of stellar remnants, particularly stellar-mass BHs,
near the centers of galaxies.  Knowledge of the BH density well inside
$\rh$ is crucial for predicting the rates of many astrophysically
interesting processes; in particular, the rate of capture of
stellar-mass BHs by MBHs, or EMRIs \citep{EMRIreview}.  Published EMRI
rate calculations almost always assume a state of mass segregation,
implying a high density of stellar remnants near the MBH
\citep{HA06,Hopman2009}. However, in a nucleus formed via a merger,
any pre-existing mass segregation would have been disrupted by the
binary MBH when it created a core; whether or not the massive remnants
would have had time to re-segregate following the merger is difficult
to assess without full $N$-body simulations.

Our simulations followed the evolution of the binary MBHs for a time
almost long enough that gravitational wave emission would dominate the
binaries' evolution.  We then combined the two MBH particles into one,
simulating gravitational wave coalescence, and continued the $N$-body
integrations for a time corresponding to several relaxation times at
the (new) influence radius.  In this post-merger evolutionary phase,
we also considered the consequences of varying the relative numbers of
the different mass components.  In this way, we were able, for the
first time, to observe how rapidly the stellar BHs would re-segregate
following a merger.  We found substantially longer time scales for
this evolution than in earlier simulations that started from
physically less motivated initial conditions.

The paper is organized as follows. In Section~\ref{sec:models} we
describe the procedure to generate equilibrium segregated models
starting from single-component models. Scaling to physical units is
discussed in Section~\ref{sec:scaling}. Evolution of the binary MBH
and its effects on the underlying stellar distribution are described
in Section~\ref{sec:premerger}. In Section~\ref{sec:postmerger} we
describe the evolution of the light and heavy objects after the
massive binary has undergone coalescence.  Section~\ref{sec:shape}
describes the shapes and kinematics of the merger
remnants. Section~\ref{sec:disc} discusses the implications of our
results for the formation and observability of Bahcall-Wolf cusps, and
for the distribution of stellar remnants.

\section{Initial models and numerical methods}
\label{sec:models}
Multi-mass Fokker-Planck models have been constructed for stars around
a MBH \citep[e.g.][]{HA06}.  Extending these models beyond $\rh$ --
where the stellar distributions are expected to be unrelaxed and where
the gravitational potential contains contributions from stars as well
as from the MBH -- is problematic.  Instead, we used $N$-body
integrations to create models of galaxies with collisionally-relaxed
nuclei.  The major difficulty was obtaining sufficient resolution on
the scale of the relaxed density cusp, without using a prohibitively
large number of particles overall.  In brief, we proceeded as follows.

1. A mass-segregated model of the inner parts of a galaxy containing a
MBH was created via $N$-body integrations, starting from a
configuration in which the different mass groups all had the same
phase-space distribution.  This model had a total mass of $50\msbh$;
scaled to a galaxy like the Milky Way, the outer radius of the model
would be $\sim 10\pc$.

2. Smooth representations of the density profiles were constructed for
each of the $N$-body species at the end of the integration.  These
functions were then ``spliced'' onto a larger, unrelaxed model, at a
radius where the effects of mass segregation were essentially zero.
This larger model had a mass of $200\msbh$, or roughly $1/5$ the mass
of an entire stellar spheroid.

3. The smooth functions representing the different mass components in
this larger model were used to generate Monte-Carlo positions and
velocities as initial conditions for the $N$-body integrations.

4. Two such $N$-body models, with different total masses and radii,
were placed in orbit around each other and integrated forward until
the two MBH particles had formed a tightly-bound pair at the center.
At this time, the merged galaxy contained a large, low-density core
created by the binary MBH.

5. The two MBH particles were combined into a single particle and the
merged galaxy was re-sampled using a smaller $N$.  This model was then
integrated forward for a few central relaxation times, allowing the
different mass groups to again evolve toward a collisional steady
state near the center.

In more detail, the mass-segregated models described in step 1 were
created as follows.

\begin{figure}
  \begin{center}
    \includegraphics[width=7cm]{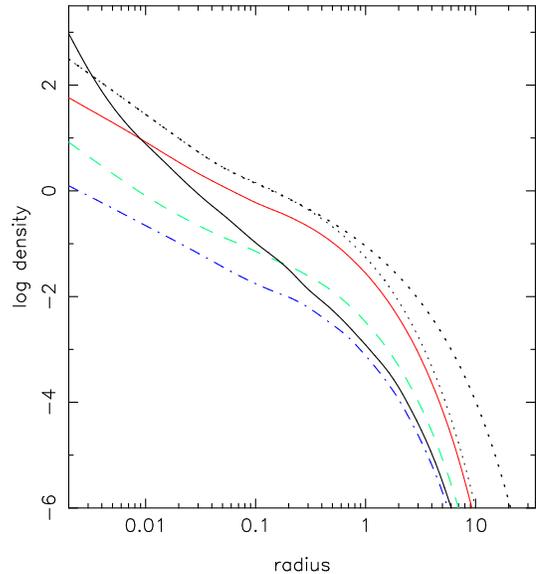}
  \end{center}
  \caption{Thin, dotted (black) curve is the density profile of the
    initial, non-mass-segregated model, equation\,(\ref{eq:CPS}).
    Other curves show mass density profiles of the four species after
    integration for $\sim 1.5$ central relaxation times: $1\msun$
    main-sequence stars (thin, red); $0.6\msun$ white dwarfs (dashed,
    green); $1.6\msun$ neutron stars (dash-dotted, blue) and $10\msun$
    stellar black holes (thick, black).  Each curve represents the
    combined density of eight independent integrations with 132k
    particles.  Scaled to the Milky Way, the unit of length is
    approximately $10\pc$ and the total mass is approximately $2\times
    10^8\msun$.  Thick dotted (black) curve shows the second, more
    extended analytical model into which the mass-segregated cusp was
    imbedded; this model has four times the mass of the first model.
  }
  \label{fig:rhoinit}
\end{figure}

Initial conditions were generated from a density law having roughly
the expected, steady-state distribution near the MBH.  We used the
modified Prugniel-Simien (1997) model described by \citet{tg05}, which
has a central density cusp of adjustable slope:
\begin{eqnarray}
\label{eq:CPS}
\rho(r)&=&\rho'\left[1+\left(\frac{r_s}{r}\right)^\alpha\right]^{\gamma/\alpha}
\left[\left(r^\alpha+r_s^\alpha\right)/r_\mathrm{PS}^\alpha\right]^{-p/\alpha}
\nonumber \\ &\times&
\exp\left\{-b\left[(r^\alpha+r_s^\alpha)/r_\mathrm{PS}^\alpha\right]^{1/{n\alpha}}
\right\}.
\end{eqnarray}
Setting $\gamma=3/2$ in the first term gives $\rho\sim r^{-3/2}$ near
the center, which is close to the collisionally-relaxed density
profile expected for the dominant population in a multi-mass cusp
\citep[e.g.][]{HA06}.  The parameter $r_s$ determines the extent of
the cusp; in relaxed, single-component models, $r_s\approx 0.2\rh$
\citep[e.g.][]{MS06}.  The parameter $p$ sets the power-law slope
beyond the central cusp; we set $p=0.5$, i.e. a relatively constant
density like that of the nuclear stellar disk of the Milky Way.  The
two final terms on the right hand side of equation\,(\ref{eq:CPS})
mimic a deprojected S\'{e}rsic-law galaxy, with $n$ the S\'{e}rsic index; the
parameter $b$ can be related to $n$ if $r_\mathrm{PS}$ is identified
with the effective radius \citep{tg05}.  In our case, the exponential
term is invoked only to provide a sharp truncation to the model
outside of $\sim$ a few $\rh$; we set $n=1.5$.  Given these
parameters, and setting $\alpha=4$, we could then solve for the cusp
radius $r_s$ in units of the model scale-length $r_\mathrm{PS}$:
$r_s\approx 0.05 r_\mathrm{PS}$.  The result is shown as the thin
dotted curve in Figure\,\ref{fig:rhoinit}.
%%% figure 1 %%%

This initial model was assigned a mass of $50\msbh$.  If we equate
$\msbh$ and $\rh$ with their values in the Milky Way (respectively
$\sim 4\times 10^6\msun$ and $\sim 2.5\pc$), the unit of length is
$\sim 10\pc$ and the total mass is $\sim2\times 10^8\msun$.
Henceforth we refer to this subsystem as the NSC.

The NSC model was assumed to be made up of four discrete mass groups.
The relative values of the particle masses were $1:0.6:1.4:10$.  These
represent, respectively, one-solar-mass main sequence stars (MS);
white dwarfs (WD); neutron stars (NS); and $10\msun$ black holes (BH).
The relative numbers of the four populations were set to
\begin{equation}
N_{MS}:N_{WD}:N_{NS}:N_{BH} = 1:0.2:0.02:0.005,
\label{eq:nfrac1}
\end{equation}
independent of radius.
In reality, these fractions would depend on 
the initial mass function and the star formation history.
Standard values are
\begin{equation}
N_{MS}:N_{WD}:N_{NS}:N_{BH} \approx 1:0.1:0.01:0.001
\label{eq:nfrac2}
\end{equation}
\citep[e.g.][]{Alexander2005}.  By comparison, our model contains
roughly twice the numbers of WDs and NSs and five times the number of
BHs.  This was done in order to improve the statistics for the remnant
populations, particularly the BHs.  Our choices could also be seen as
corresponding to a ``top-heavy'' initial mass function
\cite[e.g.][]{Maness2007}.  
In our post-merger simulations, we explored the consequences of varying these
ratios.

Initial positions and velocities of particles from the four mass
groups were generated in a standard way: (i) The gravitational
potential $\Phi(r)$ was computed from the known mass distribution
(equation\,\ref{eq:CPS}) plus the central MBH particle.  (ii) The function
$N(<v,r)$, the cumulative distribution of velocities at each radius,
was computed as in \citet{Szell2005} from $\rho(r)$ and $\Phi(r)$,
assuming an isotropic velocity distribution.  (iii) Monte-Carlo
positions and velocities were generated from $N(<v,r)$ for each of the
four mass groups, with the relative numbers given by
equation\,\ref{eq:nfrac1}.

We generated eight such models, each containing $131071$ particles,
using different initial seeds for the random number generator in each
case.  We then independently integrated these eight models forward for
a time corresponding to roughly $1.5$ central relaxation times as
defined by equation\,(\ref{eq:tr_spitz}) (using for $m$ the mass of a
MS particle).  The initial relaxation time (which is independent of
radius near the MBH in a $\rho\sim r^{-3/2}$ cusp) was roughly 130 in
model units $(G=M=r_{PS}=1)$.  We used the direct-summation $N$-body
code $\phi$GRAPE \citep{Harfst2007}, integrating each model on one
node of the RIT GRAPE cluster.
%%% figure 2 %%%
\begin{figure}
  \begin{center}
    \includegraphics[width=8cm]{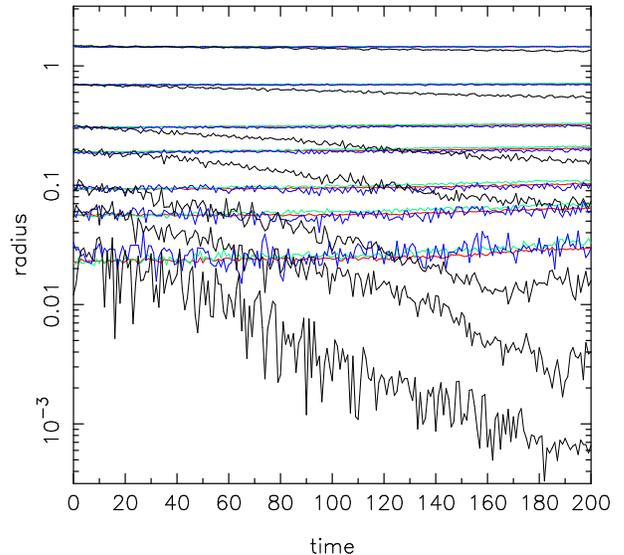}
  \end{center}
  \caption{Lagrange radii of the four stellar species during the
    $N$-body integrations of the initial model shown in
    Figure\,\ref{fig:rhoinit}.  This plot is based on the roughly one
    million particles in the 8, independent integrations of that model
    generated with different random number seeds.  {\it Red}: $1\msun$
    main-sequence stars; {\it green}: $0.6\msun$ white dwarfs; {\it
      blue}: $1.6\msun$ neutron stars; {\it black}: $10\msun$ stellar
    BHs.  The BHs form a dense cluster around the MBH in approximately
    one, central relaxation time as defined by the main-sequence
    stars.  The lighter populations are pushed slightly outward during
    this time.  Scaled to the Milky Way, the approximate unit of
    length is $10\pc$. }
  \label{fig:lagrange}
\end{figure}
Figure\,\ref{fig:lagrange} shows the evolution of the Lagrange radii
for the four mass groups.  This figure combines data from the eight
independent $N$-body integrations; the total number of particles
represented is roughly one million.  The stellar BHs (of total number
$8\times 635=5080$) accumulate toward the center; by $t=200$, their
distribution appears to have reached a steady state inside $\sim
0.1\approx 1\pc$.  The lighter populations evolve less, as expected,
since their distribution was close to the steady-state form at the
start (by design).  The final density profiles of the four populations
are shown in Figure\,\ref{fig:rhoinit}.  The BHs follow $\rho\sim
r^{-2}$ at $r\lap 0.1\approx 1\pc$, and they dominate the total
(mass) density inside $r\approx 0.007 \approx 0.07\pc$.  The density
cusp defined by the less massive populations is nearly unchanged aside
from a slight expansion due to heating by the BHs.

The mass-segregated model was then imbedded into a more extended,
unsegregated model with four times the total mass.  The template for
this larger model is shown as the thick dotted line in
Figure\,\ref{fig:rhoinit}.  It follows the same density law as in
equation\,(\ref{eq:CPS}), but with larger scale length $r_{PS}'$.  The
imbedding was achieved as follows: Smoothed density profiles were
constructed from each of the four mass groups in the evolved $N$-body
model (the same functions plotted in Figure\,\ref{fig:rhoinit}).  These
smooth profiles were then matched onto the density of the extended
model at a radius $\sim 0.3r_{PS}$; beyond this radius, essentially no
evolution occurred in the $N$-body integrations and so the $N$-body
density profiles (with appropriate vertical scalings) matched well
onto the analytic profile.  A small degree of smoothing was
nevertheless necessary near the matching radius to keep the first
derivatives of the density continuous.  Finally, small (a few percent)
adjustments were made in the vertical normalizations of the four
density profiles in order to recover precisely the original ratios
between the total numbers of the four species.

The total mass of this extended model was $200$ times the MBH mass.
Since observed bulge masses are $\sim 10^3\msbh$, such a model can be
interpreted as comprising the innermost $\sim 20\%$ of a real bulge.
In what follows, we refer to this model as ``the bulge.''

$N$-body realizations of the bulge models were then constructed in the
same way as described above.  Two such models were required for each
merger simulation.  We considered unequal-mass mergers with a mass
ratio of $3:1$. The radius of the smaller bulge was scaled as the
square-root of the bulge mass.  Particle masses (aside from the MBH
particle) were the same in the two bulges, i.e., particle number
scaled linearly with total mass.

Table\,\ref{tab:models1} lists the important parameters of the merging
systems: the mass ratio of the two MBHs (equal to the galaxy mass
ratio), the ratio of MBH mass to host bulge mass, the ratio of the MS
stars mass to MBH mass, the number of particles in each galaxy, and
the initial separation between the bulge centers, $\Delta r$,
expressed in units of the outer radius of the larger bulge, $R_1$.
The table also gives the influence radius associated with the MBH in
the larger galaxy, under two definitions.  In addition to the first
definition given in equation~\ref{eq:rh}, we also compute a second,
mass-based influence radius: the radius $\ri$ that contains a mass in
stars equal to twice the MBH mass:
\begin{equation}
\label{eq:DefrM} M_\star(r<\ri) = 2M_\bullet.  
\end{equation}
The mass-based definition is the most straightforward to apply in
$N$-body models like ours; in computing $\rh$, a choice must be made
about the radius at which to evaluate $\sigma$, and this, combined
with the need to bin particles, can lead to factor of $\sim$ two
uncertainties in $\rh$.  The values of $\rh$ and $\ri$ in
Table\,\ref{tab:models1} refer to the larger of the two galaxies.

The two bulge models were placed far enough apart initially that there
was no overlap, but not much farther, in order to minimize the
integration time.  Two relative orbits were considered: a circular
orbit (Model A), and an eccentric orbit in which the initial relative
velocity was set to $0.7$ times the circular value (Model B).

\begin{table}
\begin{center}
\caption{Galaxy merger parameters}
\label{tab:models1}
\begin{tabular}{cccccccc}
\hline 
$M_1 : M_2$ & $\msbh$ / $\mg$ & $M_{\rm MS} / \msbh$  & $N_1$ & $N_2$ & $\Delta
r/R_1$ & $\rh$ & $\ri$ \\
\hline 
3:1 & 0.005 & 0.0005 & 400k & 120k & 1.5 & 0.10 & 0.28 \\
\hline
\end{tabular}
\end{center}
\end{table}

The merger simulations were also carried out using $\phi$GRAPE, both
in combination with GRAPE hardware at the Rochester Institute of
Technology and with GPU hardware at the Max-Planck Institute for
Astrophysics in Garching by means of the $\tt Sapporo$ library
\citep{sapporo2009}. We conservatively set the accuracy parameter
\citep{Harfst2007} to 0.005 and the softening to $10^{-4}$ (in
$N$-body units). Such small values of the softening and accuracy
parameters were necessary to accurately follow the dynamics of the
massive binary.

\begin{table}
\begin{center}
\caption{Post-merger models}
\label{tab:models2}
\begin{tabular}{cccc}
\hline 
Model & $N$ & Fractions & $\msbh$/$\mg$\\
\hline 
A1  &  250k  & 1:0.2:0.02:0.005  & 0.0050 \\
A2  &  225k  & 1:0.1:0.01:0.001  & 0.0055 \\
B1  &  250k  & 1:0.2:0.02:0.005  & 0.0050 \\
B2  &  225k  & 1:0.1:0.01:0.001  & 0.0055 \\
\hline
\end{tabular}
\end{center}
\end{table}

The simulations were continued until the two MBH particles had formed
a tight binary and the binary separation had shrunk by an additional
factor of $\sim 20$.   At
this time, the two MBH particles were combined into a single particle,
which was placed at the center-of-mass position and velocity of the
binary.  A random subset of particles was then chosen from this model,
yielding a new model with the same phase-space distributions of the
four components but a smaller total $N$. This ``merged galaxy'' model
was then integrated forward, for a time corresponding to a few central
relaxation times. The subsampling was a necessary compromise to keep
the physical integration time from becoming prohibitively long while
still allowing the simulations to proceed for more than one relaxation
time.  For each of the models presented in Table\,\ref{tab:models1} we
generated two different submodels with different number fractions, as
illustrated in Table\,\ref{tab:models2}. In order to generate the
models with the standard fractions $N_{MS}:N_{WD}:N_{NS}:N_{BH}
\approx 1:0.1:0.01:0.001$, we deleted an appropriate number of
particles in each mass group from the models with
$N_{MS}:N_{WD}:N_{NS}:N_{BH} = 1:0.2:0.02:0.005$.

\section{Scaling}
\label{sec:scaling}

In any $N$-body simulation, an important consideration is how to
relate the computational units of mass, length and time to physical
units.  In our simulations, the mass scale is most naturally set by
the mass of the particle(s) representing the MBH(s), and the length
scale by the radius that encloses a mass in stars that is some
multiple of the MBH mass.  A natural choice for the latter is $\ri$,
as defined in equation\,(\ref{eq:DefrM}).  Identifying $\ri$ in the
simulations with $\ri$ in a real galaxy is only justified, of course,
to the extent that the radial dependence of the density near the MBH
is similar in both systems.

Scaling of the time is more subtle.  Two basic time scales are of
interest: the crossing time, which is determined by the total mass and
size of the model; and the relaxation time, which depends as well on
the masses of the particles, or equivalently on $N$:
\begin{equation}
\label{eq:tr_spitz}
\tr = \frac{0.33\,\sigma^3}{G^2\,n\,m^2\ln \Lambda}
\end{equation}
\citep{spitzer1987}.  
Here, $n$ is the stellar number density, $m$ is the mass of one star,
and $\ln\Lambda$ is the Coulomb logarithm.  The particle masses in our
simulations have the correct {\it ratios} with respect to one another,
but their masses in relation to the total galaxy mass is much larger
than in a real galaxy -- due of course to the fact that our $N$ is
much smaller than $10^{11}$.  While there is a well-defined relaxation
time in the models, that time is much shorter relative to the crossing
time than it would be in real galaxies.

These statements apply to many $N$-body simulations that extend over
relaxation time scales.  A novel feature in our simulations is the
treatment of the galaxy merger.  Such a merger requires of order a few
crossing times in order to reach a (collisionless) steady state.
Since crossing times are much shorter than relaxation times, both in
reality and in our models, a galaxy would not undergo a significant
amount of collisional relaxation during the merger.  During this phase
of the simulations, therefore, the appropriate unit of time is the
crossing time.  In the phases preceding and following the merger, the
appropriate unit of time is the relaxation time.

Subtleties arise when one considers the massive binary.  After the
galaxy merger is essentially complete, the binary MBH continues to
evolve.  In a spherical nucleus, the binary quickly ejects stars on
intersecting orbits, and continued hardening takes place on a
relaxation time scale, as stars are scattered by other stars onto
previously-depleted orbits that intersect the binary
\citep{makinofunato2004,mms07}.  In such models, there is effectively
just one time scale -- the relaxation time -- that determines both the
rate of collisional evolution of the galaxy and of the central binary
following the merger.

Another mode of evolution is possible, if the galaxy potential is
significantly nonspherical \citep{merrittpoon2004,ber06} In this case,
orbital angular momenta of stars near the massive binary evolve due
both to encounters, and to torques from the overall stellar potential.
Typically the latter dominates, and the supply of stars to the massive
binary remains high in spite of ongoing, slingshot ejections.  The
binary evolves at a rate that is fixed essentially by stellar orbital
periods, i.e. by the crossing time, and not by the relaxation time.
Such evolution appears to be the norm when the galaxy hosting the
massive binary was formed in a realistic merger simulation
\citep{khan2011,preto2011}, and we find the same result in our new
simulations.  This result simplifies the time scaling of our models,
since it means that a single physical time -- the crossing time --
sets the rate of evolution, both during the galaxy merger, and
immediately afterwards, as the massive binary hardens.

One potential difficulty does arise, however.  If hardening of the
massive binary is simulated for a time that is comparable with the
$N$-body central relaxation time, some collisional evolution in the
stellar distribution will occur.  This may or may not be realistic,
since in a real galaxy, the ratio of relaxation time to crossing time
is much larger than in the simulations.  For this reason, the models
that we adopt at the start of the final phase of our simulations --
single galaxies containing merged MBHs -- may exhibit overly-segregated
nuclei, causing the subsequent, collisional relaxation to occur more
quickly than it would in a real galaxy.

%%% figure 3 %%%
\begin{figure*}
  \begin{center}
    \includegraphics[width=5cm]{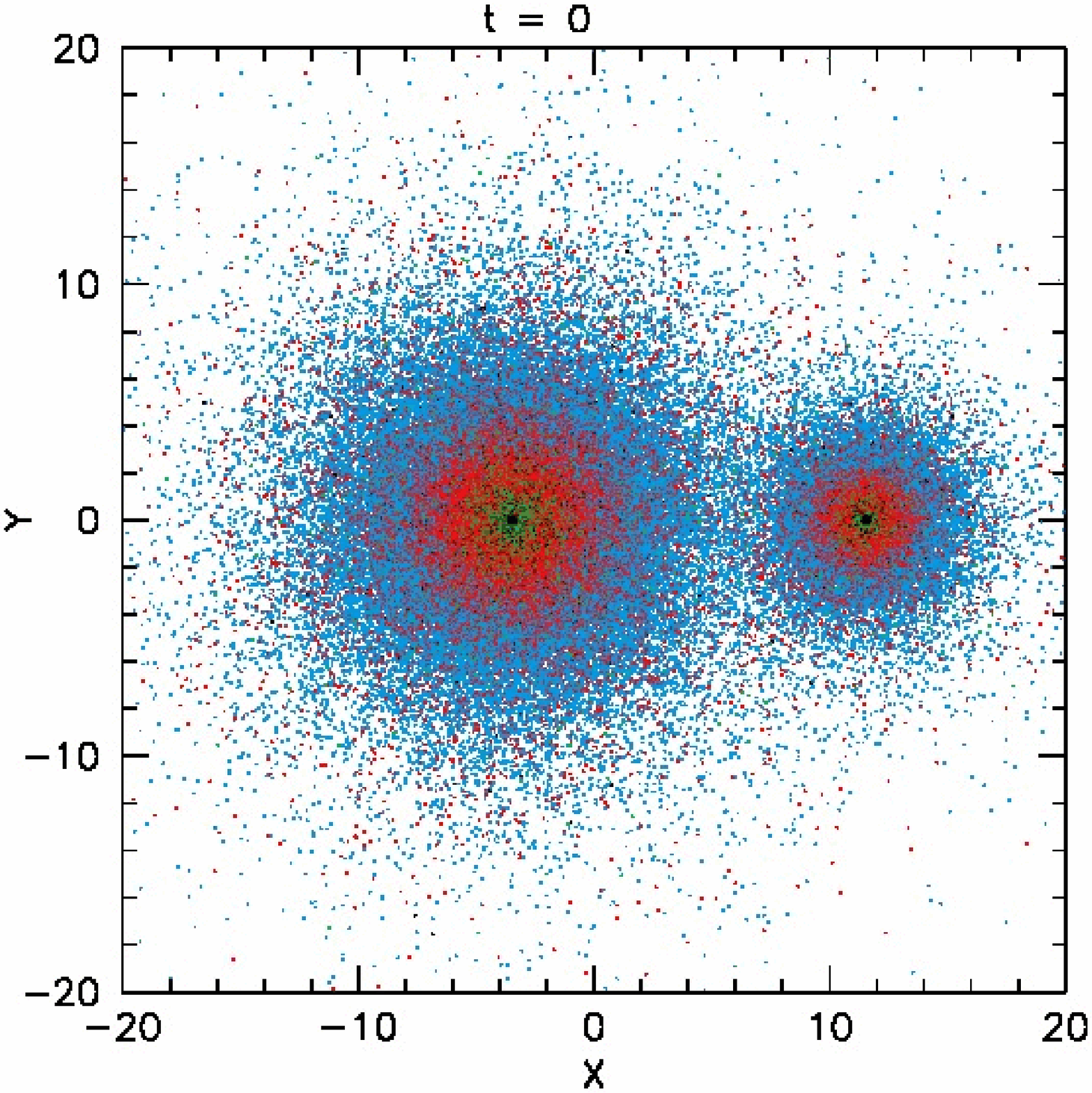}
    \includegraphics[width=5cm]{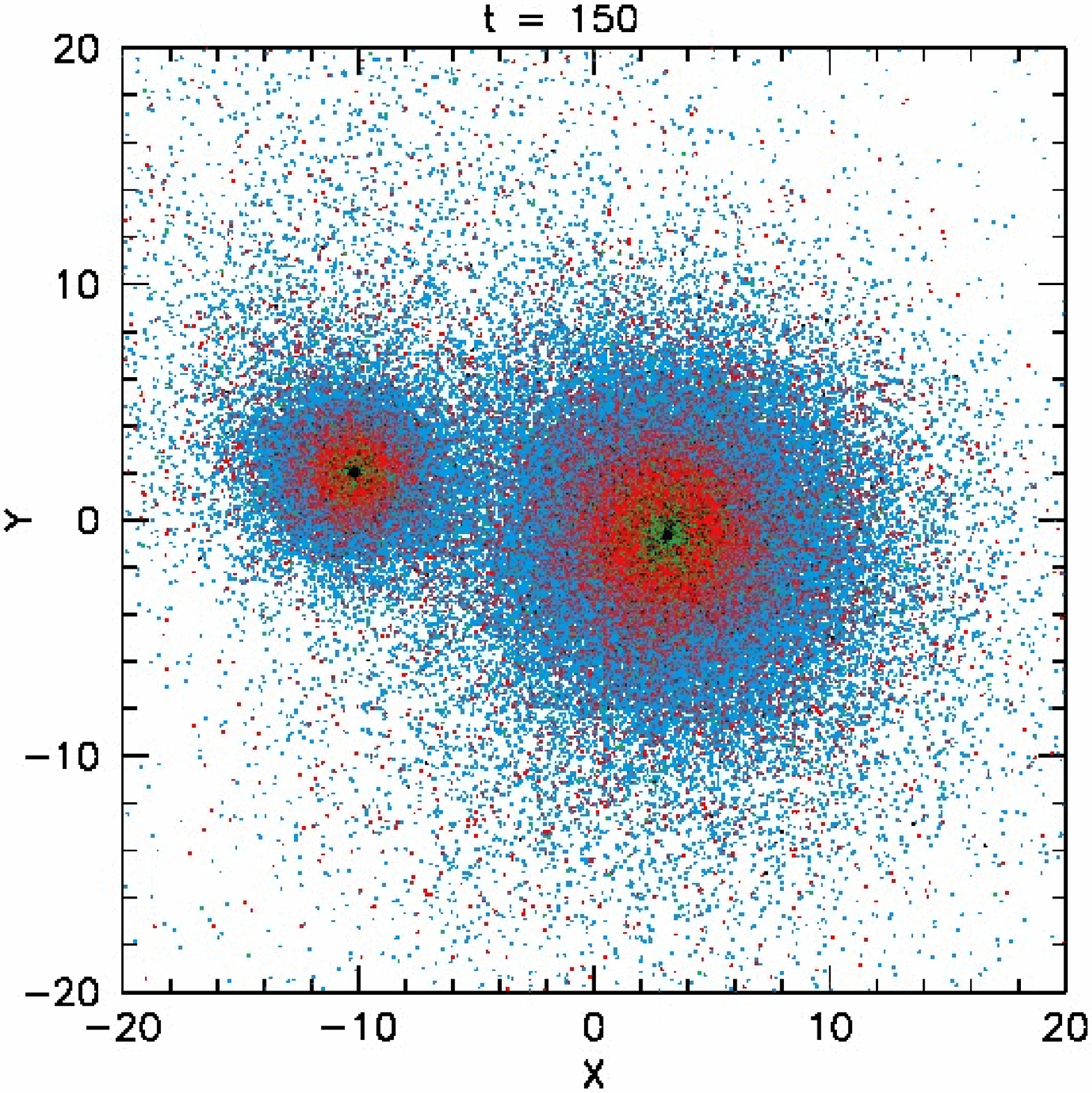}
    \includegraphics[width=5cm]{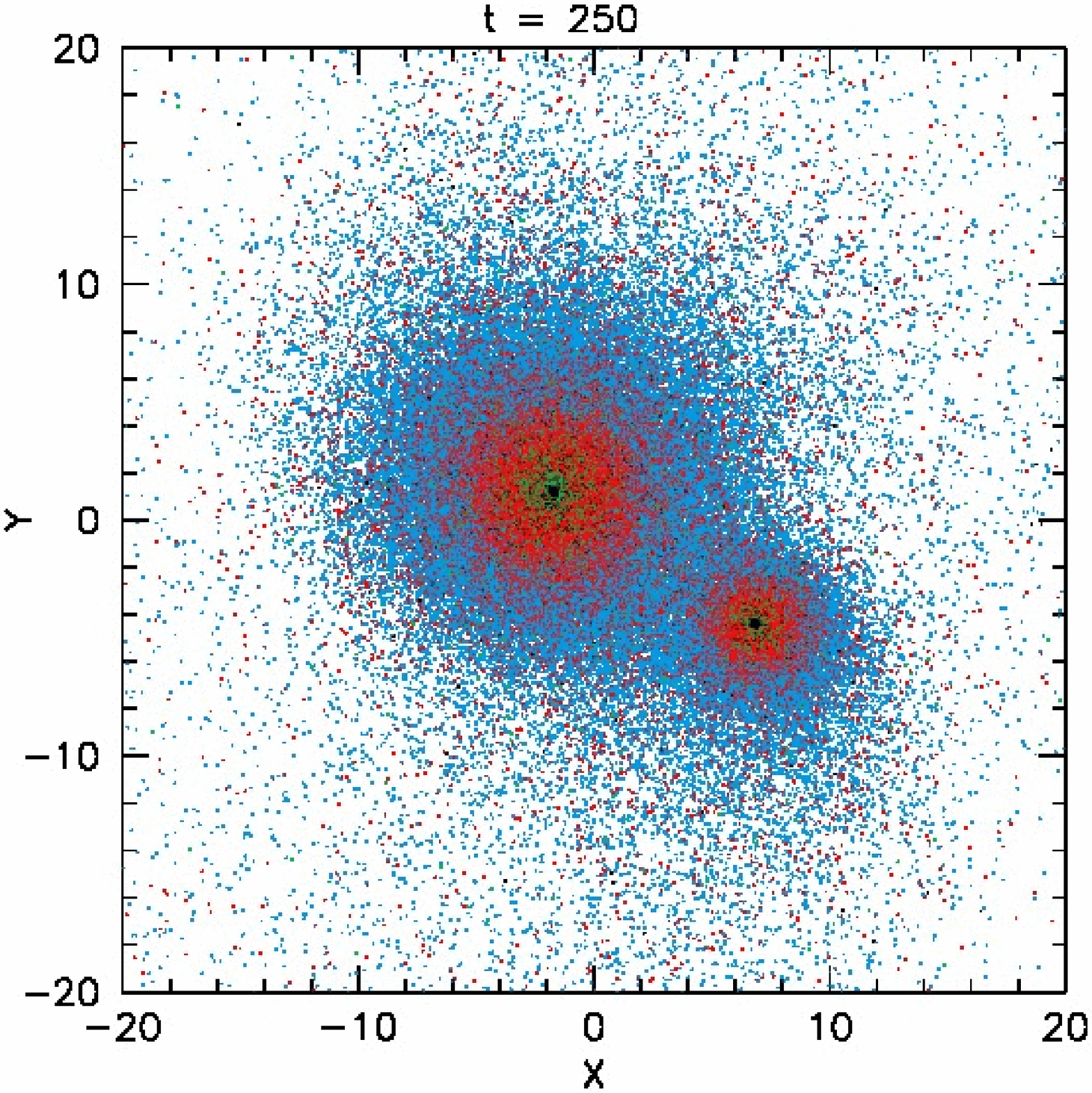}
    \includegraphics[width=5cm]{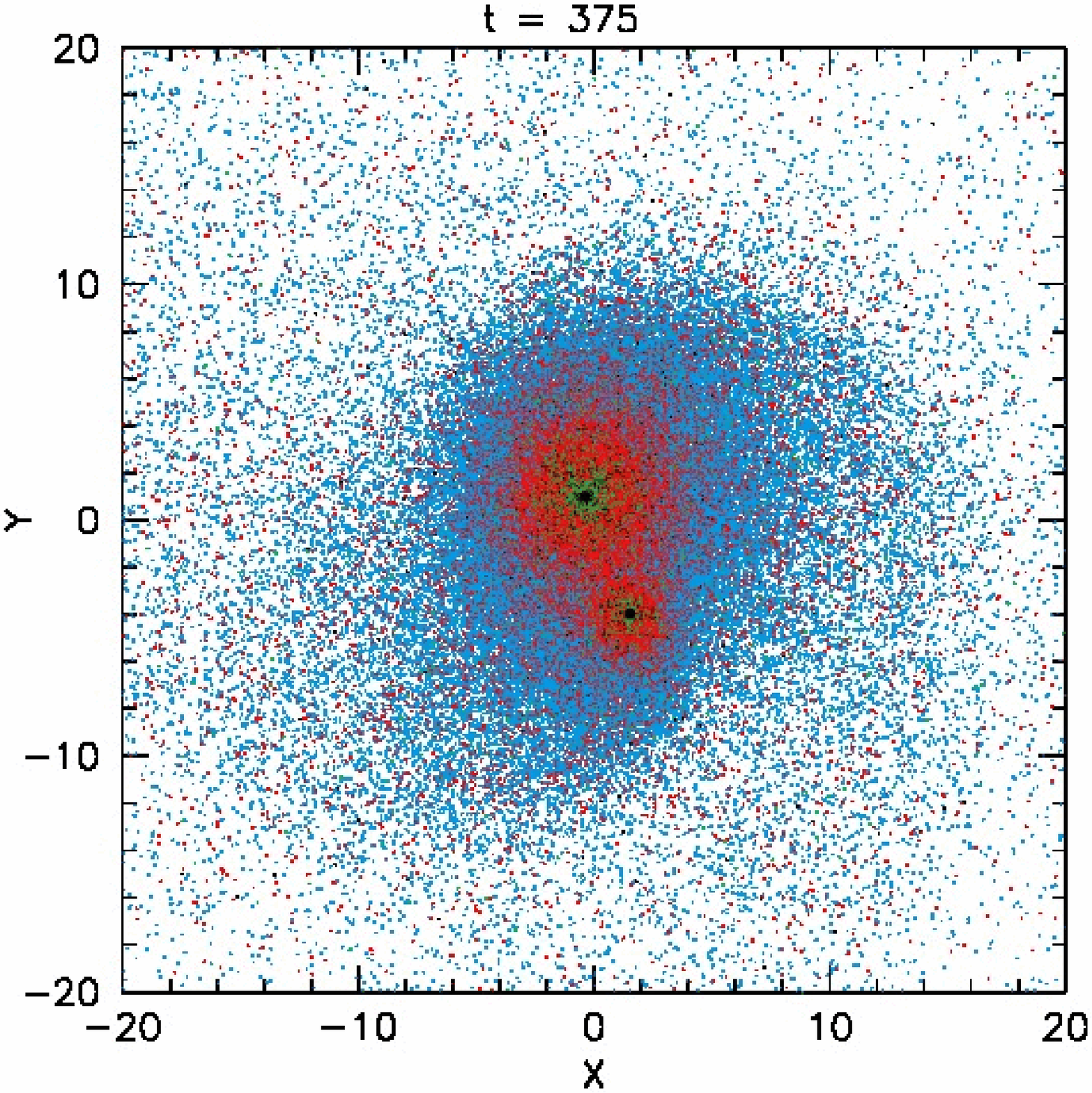}
    \includegraphics[width=5cm]{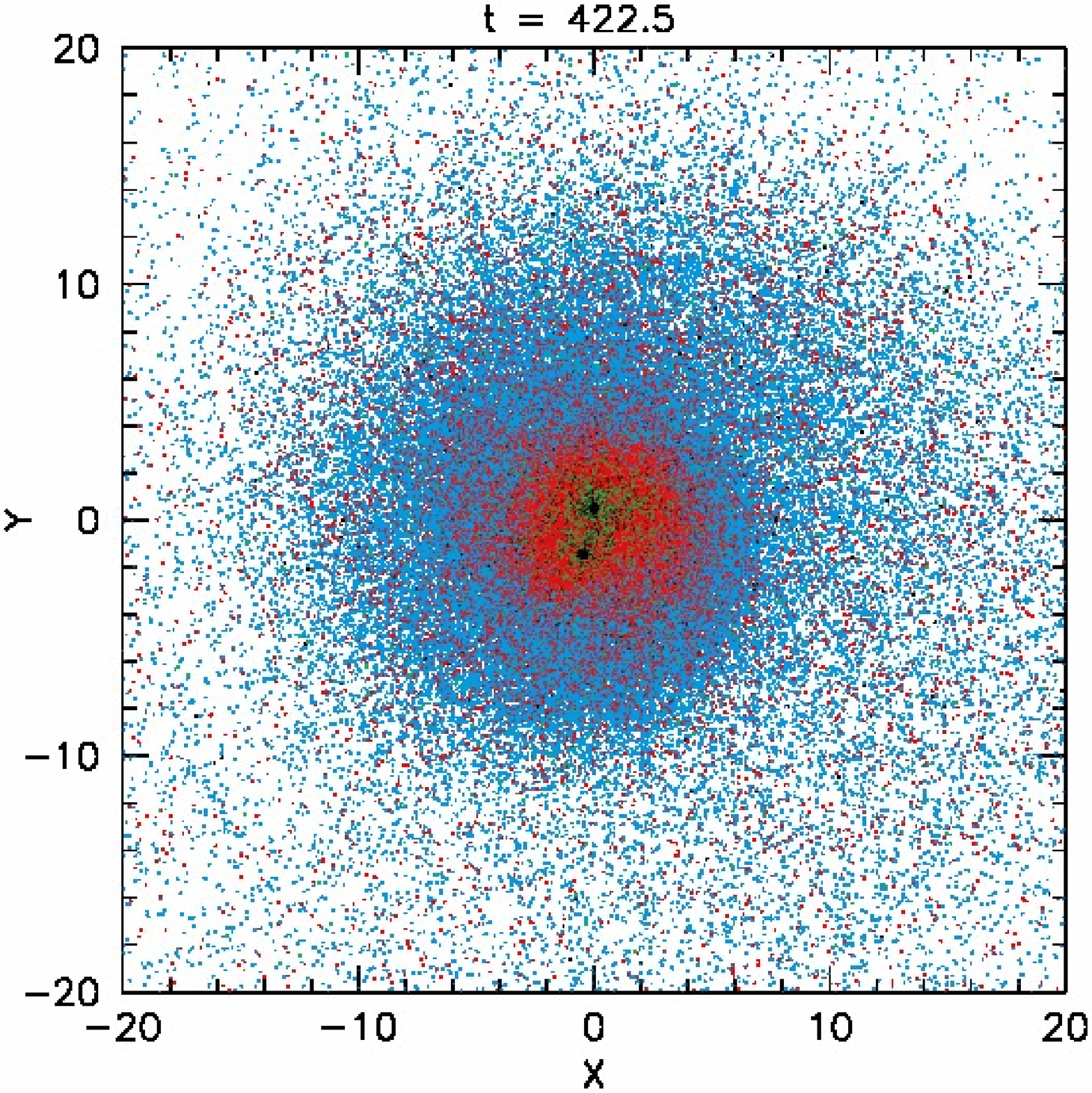}
    \includegraphics[width=5cm]{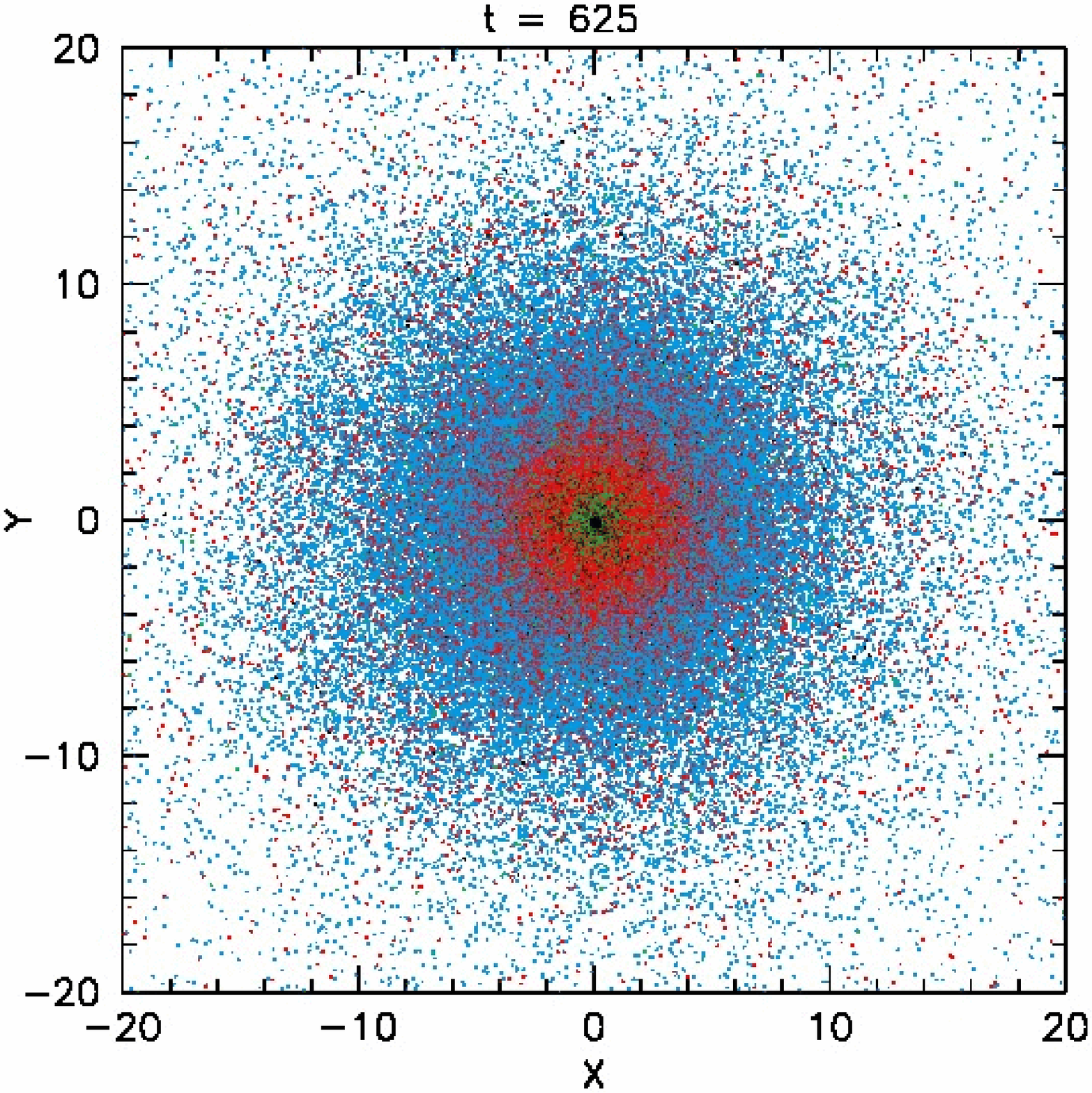}
  \end{center}
  \caption{Snapshots from the early evolution of Model A. The
    different colors refer to the different mass groups: main sequence
    (blue), white dwarfs (red), neutron stars (green), black holes
    (black). The MBHs are indicated by full circles.}
  \label{fig:6framesA}
\end{figure*}
%%% figure 4 %%%
\begin{figure*}
  \begin{center}
    \includegraphics[width=5cm]{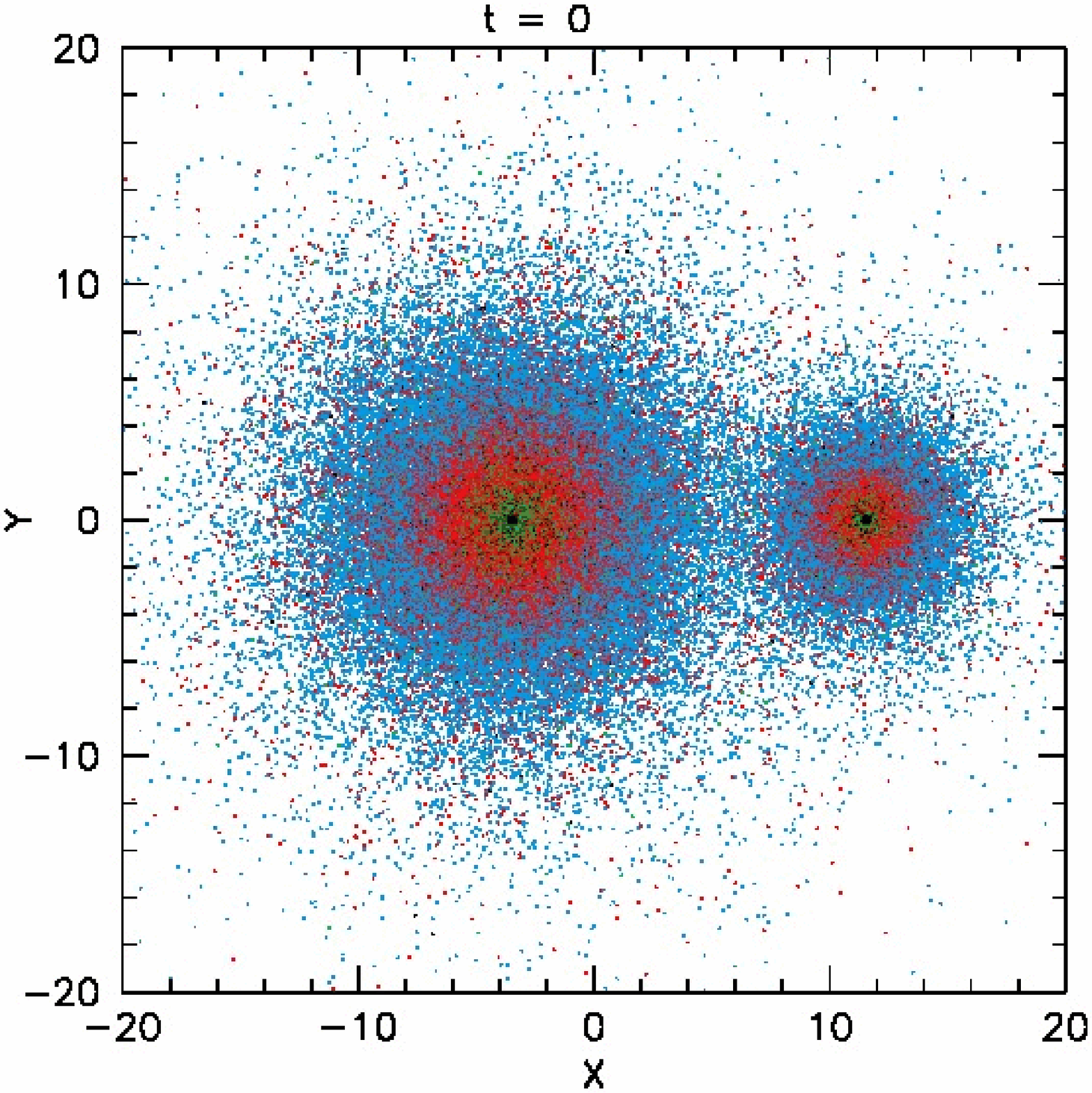}
    \includegraphics[width=5cm]{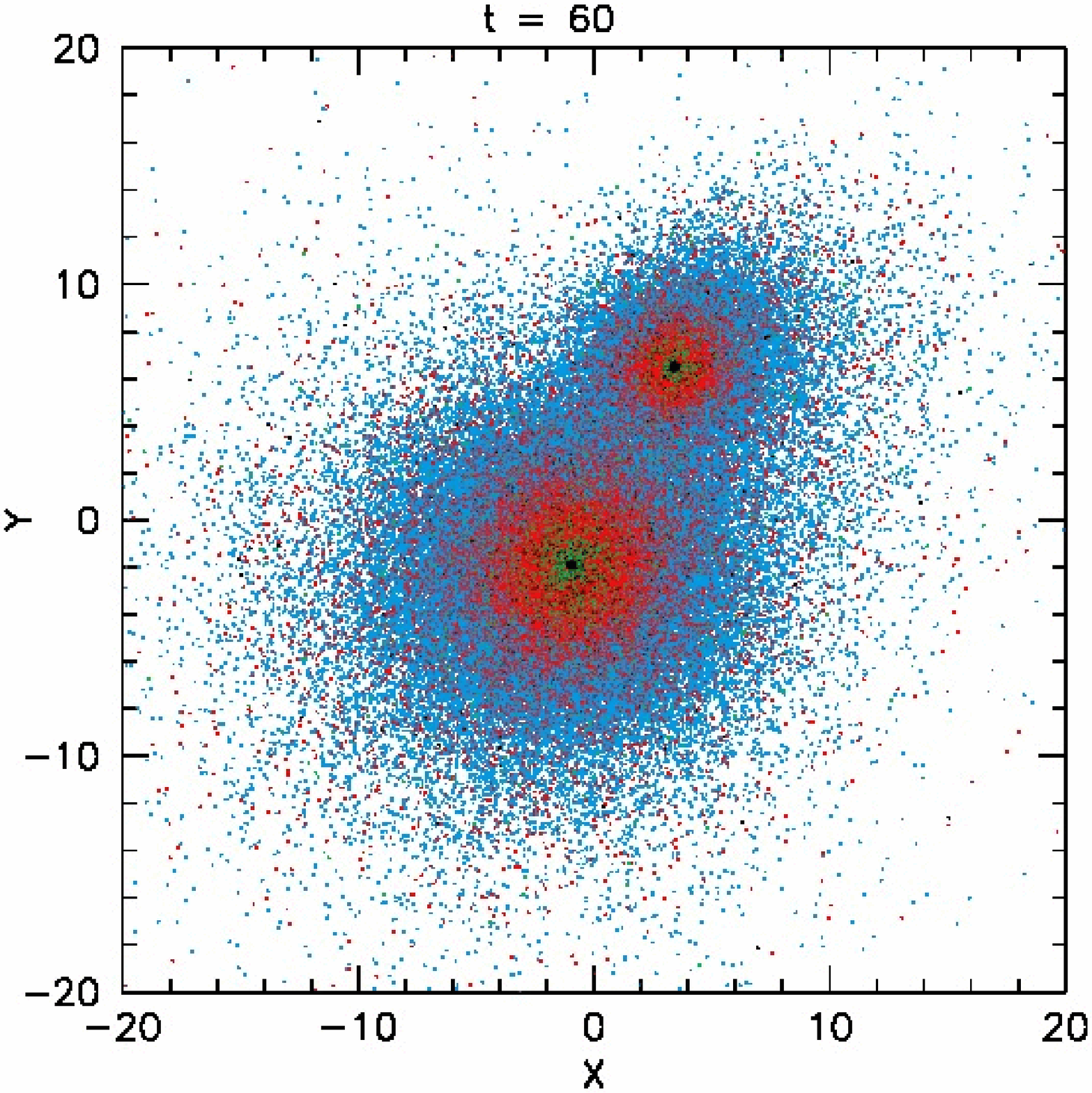}
    \includegraphics[width=5cm]{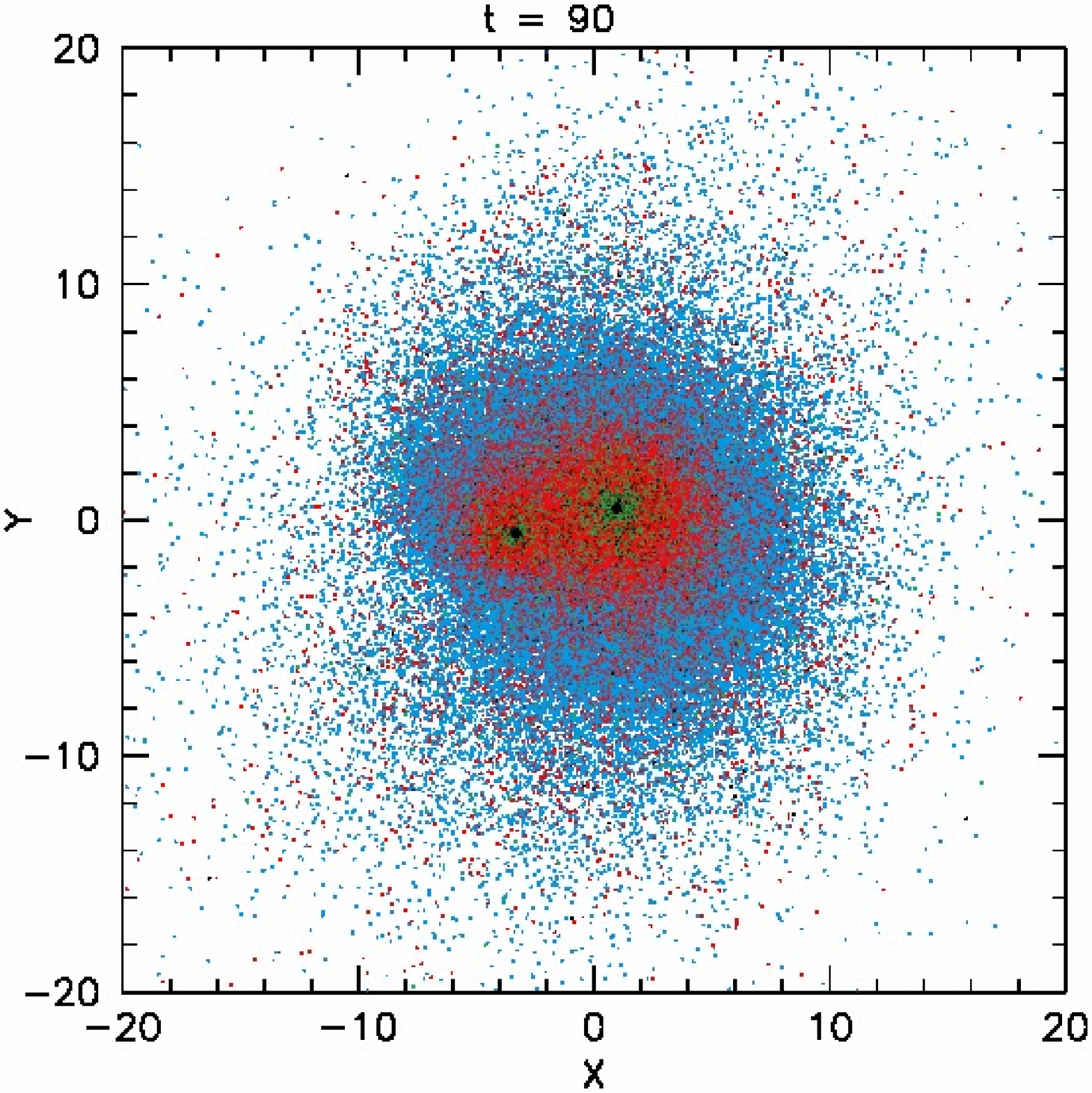}
    \includegraphics[width=5cm]{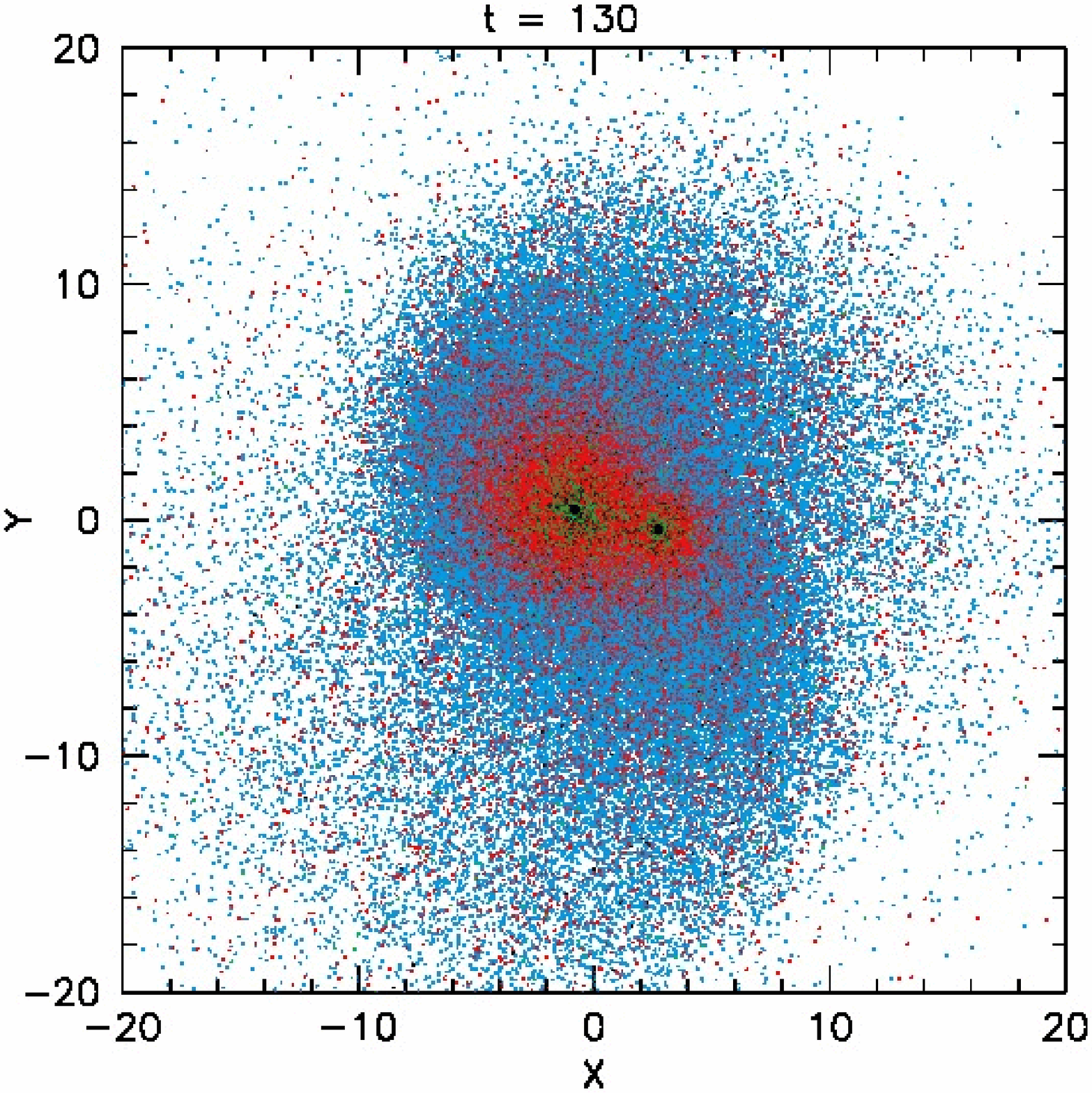}
    \includegraphics[width=5cm]{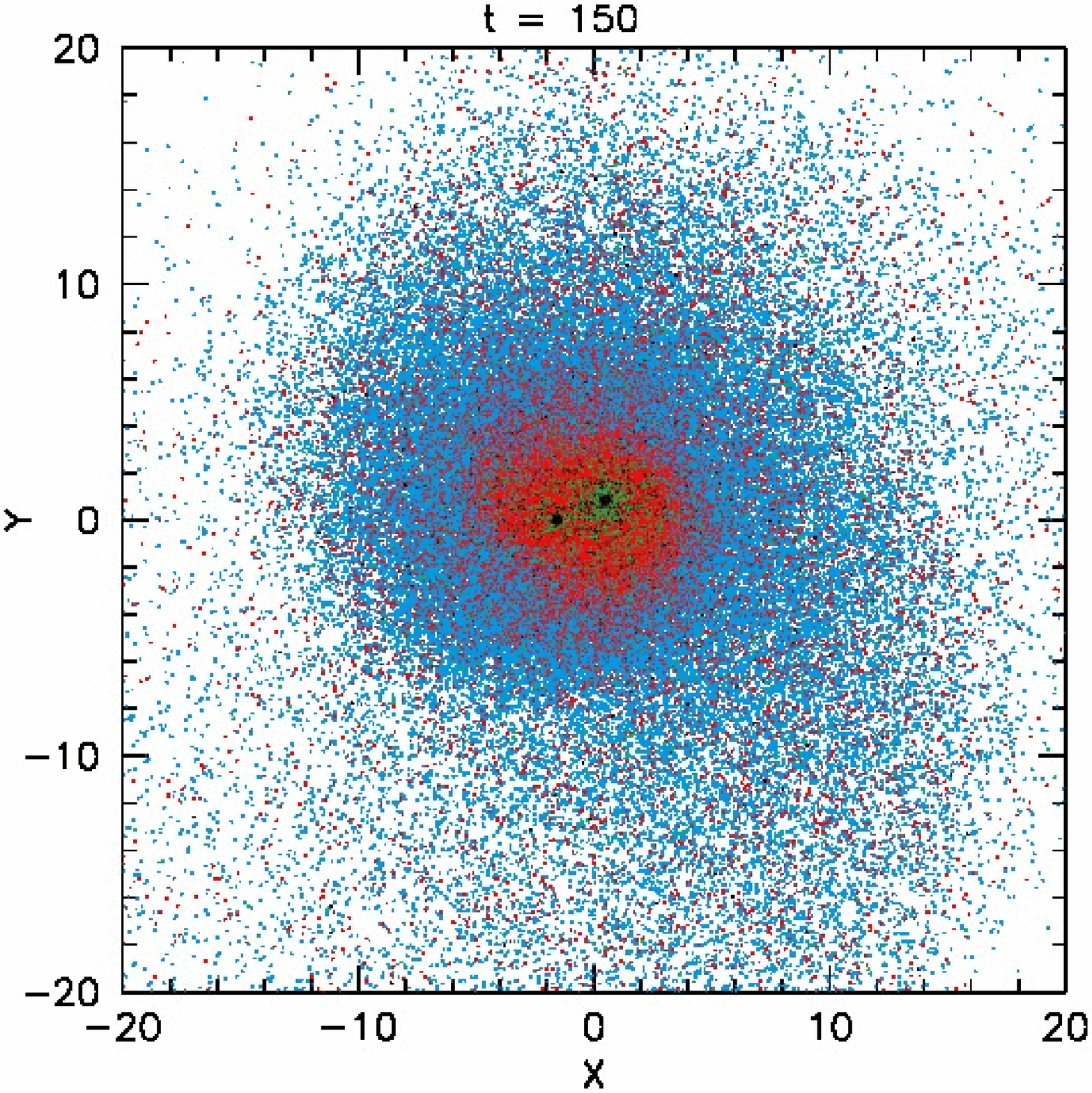}
    \includegraphics[width=5cm]{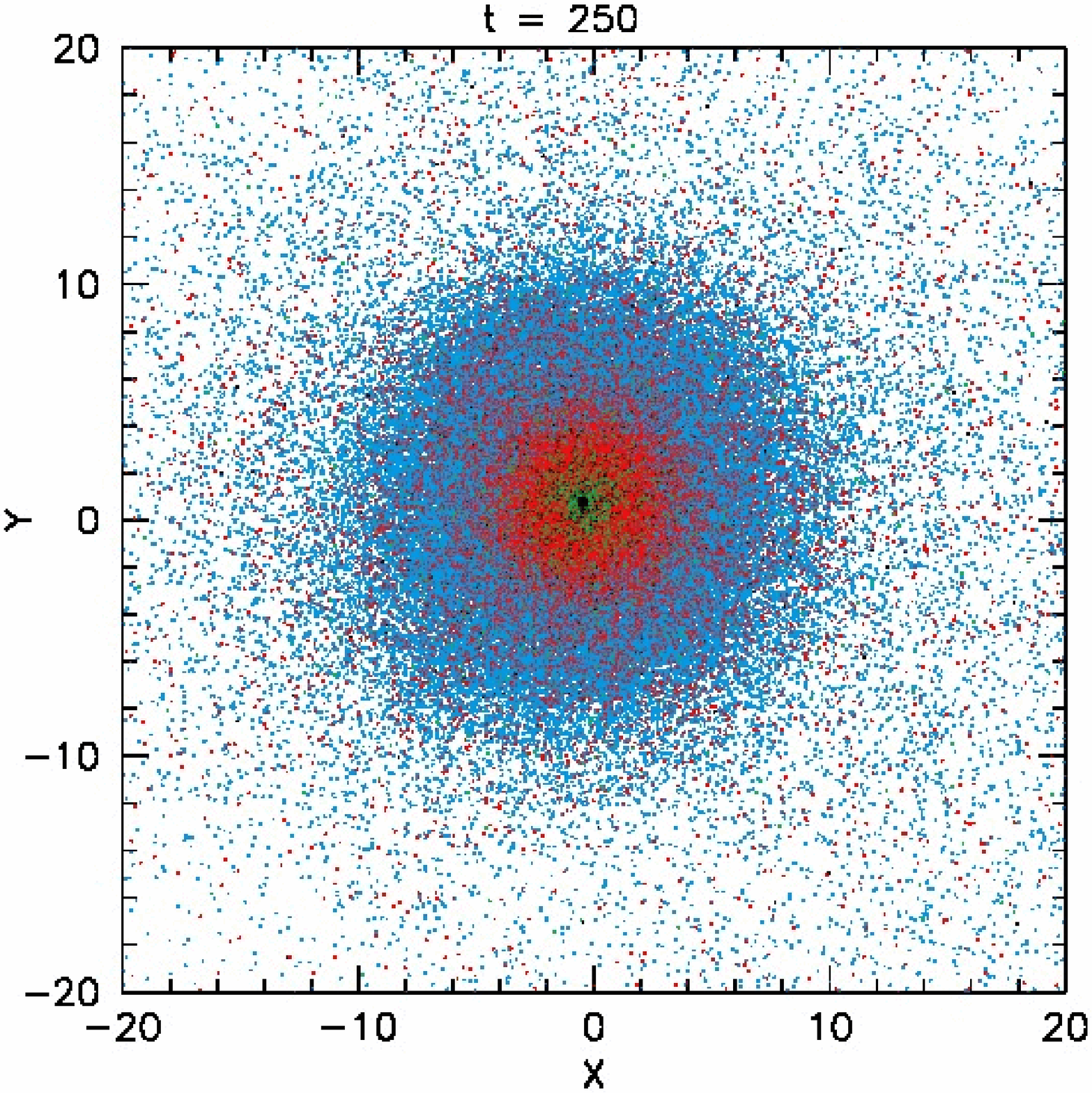}
  \end{center}
  \caption{Snapshots from the early evolution of Model B. Colors are
  as in Figure\,\ref{fig:6framesA}.}
  \label{fig:6framesB}
\end{figure*}

\section{The galaxy merger}
\label{sec:premerger}

\subsection{Evolution of the massive binary}

\begin{figure}
  \begin{center}
    \includegraphics[width=4cm]{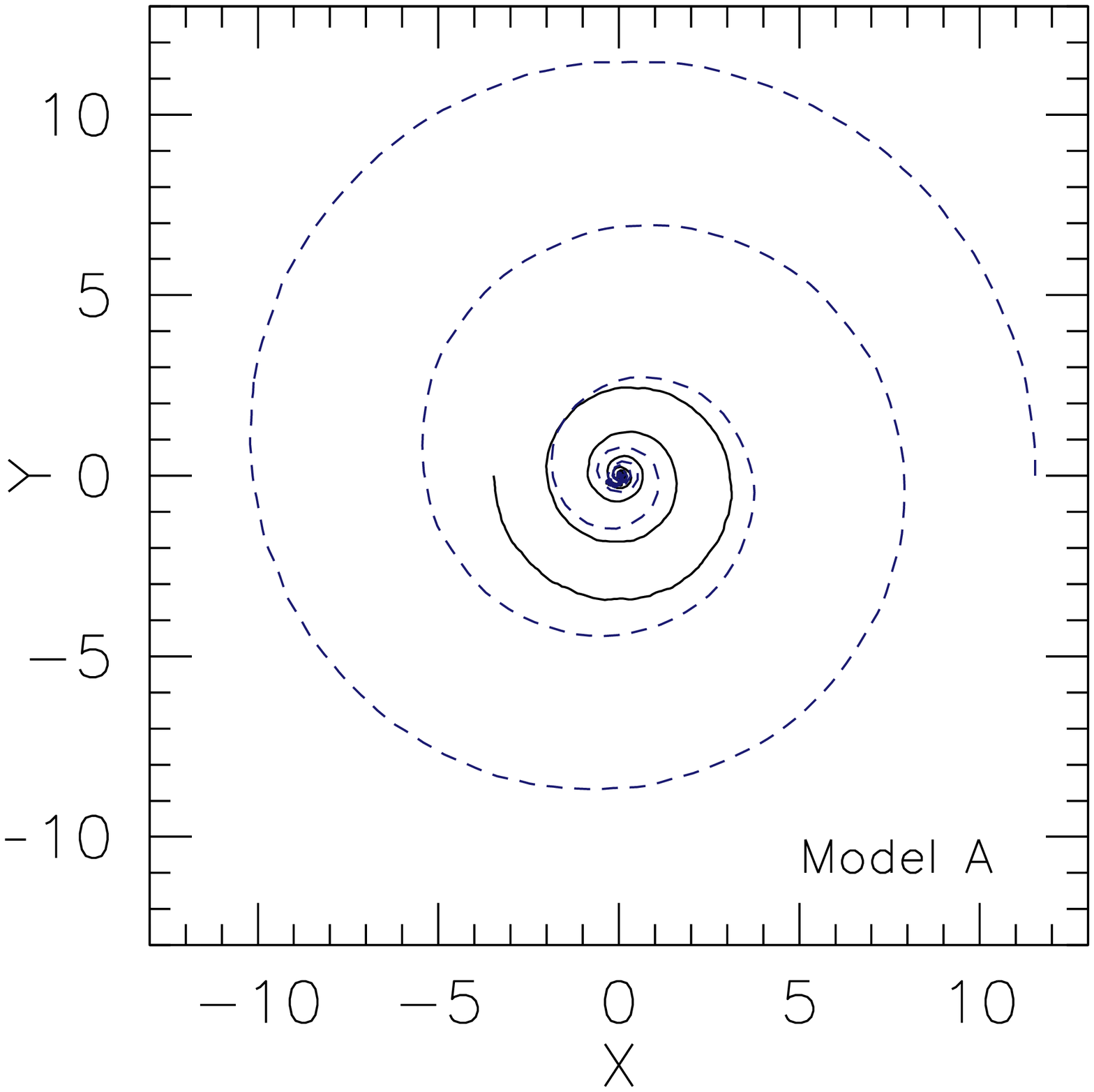}
    \includegraphics[width=4cm]{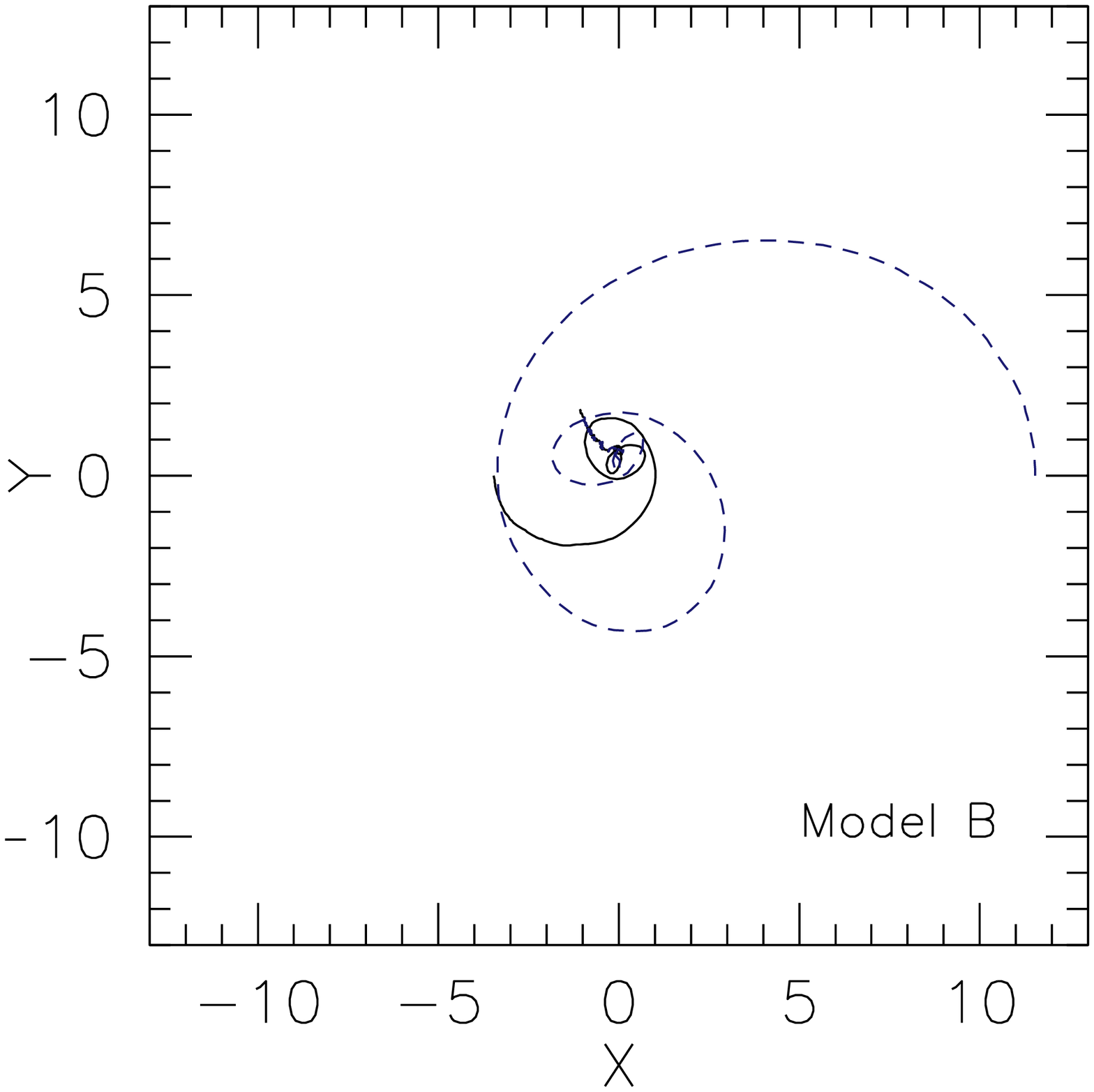}
  \end{center}
  \caption{Trajectories of the MBHs in Model A (left) and B (right)
    during the galaxy merger phase. Dashed line: trajectory of the
    smaller hole. Solid line: trajectory of the larger hole. The
    initial orbit of the galaxy pair lies in the $z = 0$ plane.  }
  \label{fig:traj}
\end{figure}

Figures\,\ref{fig:6framesA} and \ref{fig:6framesB} illustrate the
large-scale evolution of the bulge models described in
Table\,\ref{tab:models1} during the galaxy merger phase.  The
two bulge models start out either on circular (Model A) or eccentric
(Model B) orbits about their common center of mass, with initial
separations roughly one-half the radius of the larger bulge.  Due to its
more eccentric initial orbit, Model B evolves more quickly.  The
trajectories of the MBHs reflect the initial orbits of the parent
systems, as can be seen in Figure\,\ref{fig:traj}.
%%% figure 5 %%%

Evolution of a binary MBH can be divided roughly into four phases.
(1) {\it Formation of a bound pair.} Merger of two galaxies with
central MBHs brings the two MBHs together, in a time comparable with
the galaxy merger time, i.e. a few galaxy crossing times.  (2) {\it
  Formation of a hard binary.} The separation between the two MBHs
decreases very rapidly, due at first to dynamical friction against the
stars, and then to the gravitational slingshot: near stars are ejected
after gaining energy from the massive binary. A core is formed at this
stage, with size roughly equal to the initial separation of the bound
pair.  (3) {\it Binary hardening.} If orbits of stars ejected by the
gravitational slingshot are replenished, the massive binary will
continue to shrink.  Hyper-velocity stars can be produced in this
process. The stellar core will continue to grow.  (4) {\it
  Coalescence.} If replenishment of stellar orbits continues, the
binary separation decreases to a value such that emission of
gravitational waves dominates its evolution, and the two MBHs
coalesce.

Dynamical friction drives the evolution down to a separation $a_f$ at
which the stellar mass enclosed in the binary is of order twice
the mass of the smaller MBH:
\begin{equation}\label{eq:af}
M (< a_f) \approx  2\,M_2\,.
\end{equation}
 This separation is smaller by a factor $\sim M_2/M_1$
than the radius of influence of the larger MBH.

At separations smaller than $a_f$, the binary hardens due to
gravitational slingshot interactions with passing stars.  A binary is
defined as ``hard'' when it reaches a separation
\begin{equation}\label{eq:ah}
a_h \approx \frac{G\,M_2}{4\,\sigma^2}
\end{equation}
called the hard-binary separation;
a binary is ``hard'' when its binding energy per unit mass,
$|E|/(M_1+M_2)$, exceeds $\sigma^2$.

The binary enters the gravitational wave (GW) regime when the
time scale for coalescence due to emission of gravity waves:
\begin{eqnarray}
\label{eq:tgw}
  T_{\rm GW} & = & \frac{5}{256 F(e)} \frac{c^5}{G^3} \frac{a^4}{\mu
    \left(M_1 + M_2 \right)^2} \nonumber\\ & \approx &
  \frac{5.8 \times 10^{11} \yr}{F(e)} \left(\frac{a}{0.1\pc}\right)^4
  \left(\frac{10^7\msun}{\mu}\right)
  \left(\frac{10^8\msun}{M_1+M_2}\right)^2\nonumber\\ 
\end{eqnarray}
becomes shorter than the time for hardening due
to stellar interactions.
Here
\begin{equation}
\label{eq:fe}
  F(e) = \left(1-e^2\right)^{-7/2} \left(1 + \frac{73}{24}e^2 +
  \frac{37}{96} e^4\right) \nonumber
\end{equation}
and $\mu \equiv M_1\,M_2/(M_1+M_2)$ is the reduced mass.

\begin{figure}
  \begin{center}
    \includegraphics[width=8cm]{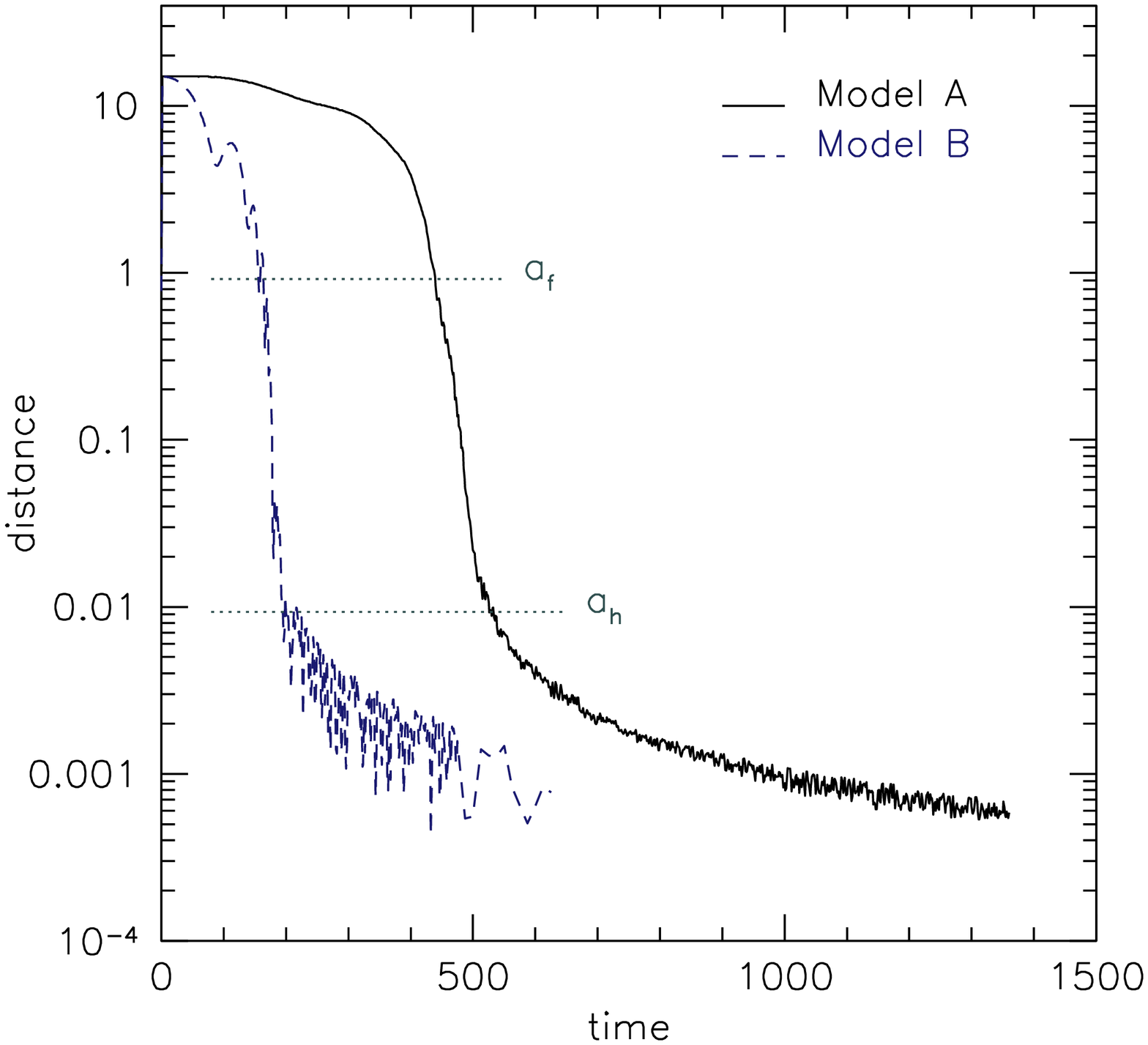}
  \end{center}
  \caption{Separation between the MBH particles as a function of
    time. The horizontal lines indicate $a_f$, the approximate
    separation at which dynamical friction ceases to be efficient
    (equation\,\ref{eq:af}), and $a_h$, the hard-binary separation
    (equation\,\ref{eq:ah}).}
  \label{fig:dist_bh}
\end{figure}

The evolution of the separation between the MBHs
is shown in Figure\,\ref{fig:dist_bh}. 
%%% figure 6 %%%
The first part of this evolution, which is driven by dynamical
friction, ends when the separation reaches $\sim a_f$,
equation\,(\ref{eq:af}); $a_f\approx 0.9$ in model units.  The time to
reach this separation was $t_f \approx 440$ and $\sim 160$,
respectively, for models A and B.  At about the same time,
gravitational slingshot encounters with the stars begin to dominate
the binaries' evolution and the effect on the galaxy structure starts
to become apparent.
%%% figure 7 %%%
\begin{figure}
  \begin{center}
    \includegraphics[width=4cm]{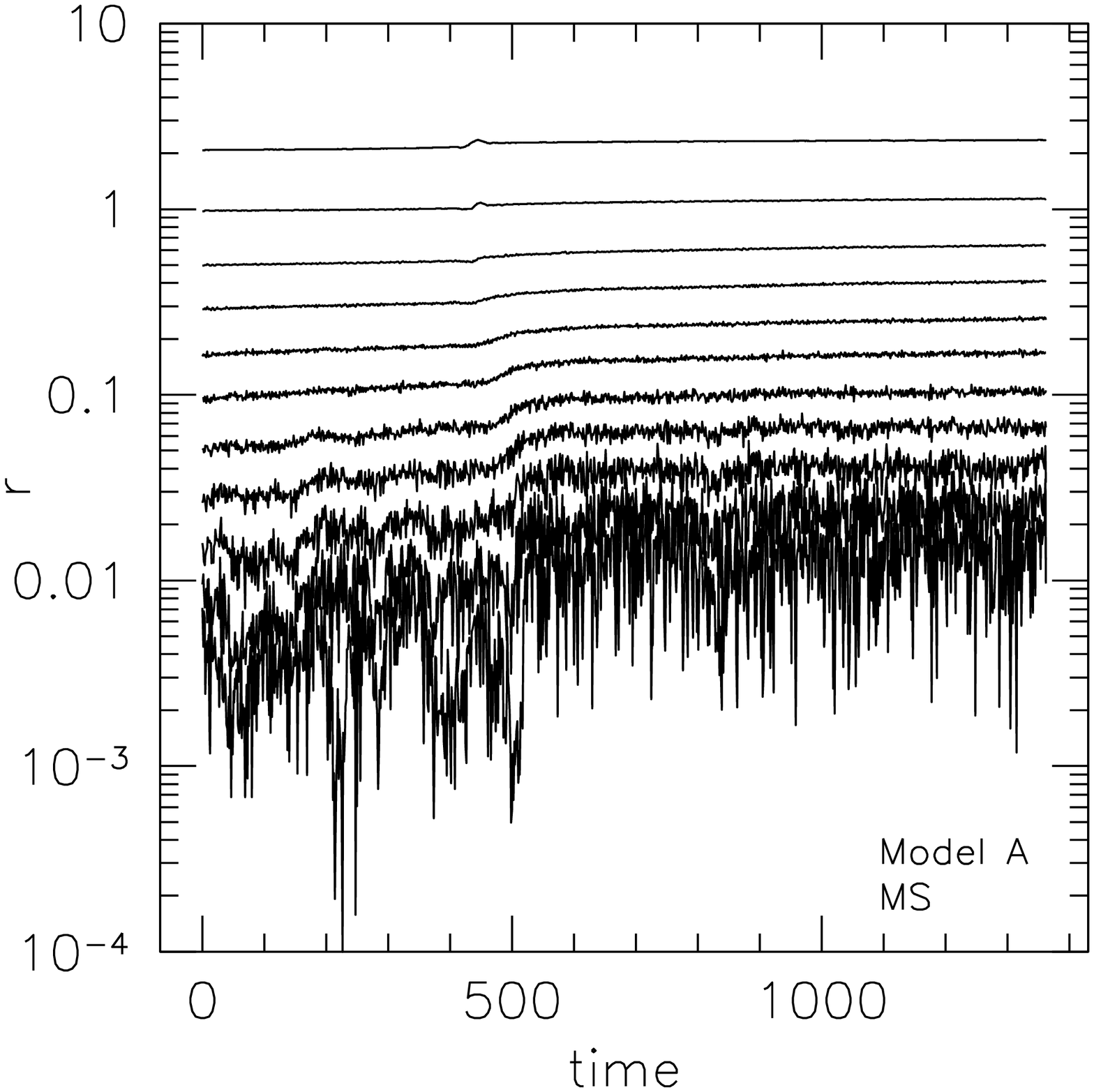}
    \includegraphics[width=4cm]{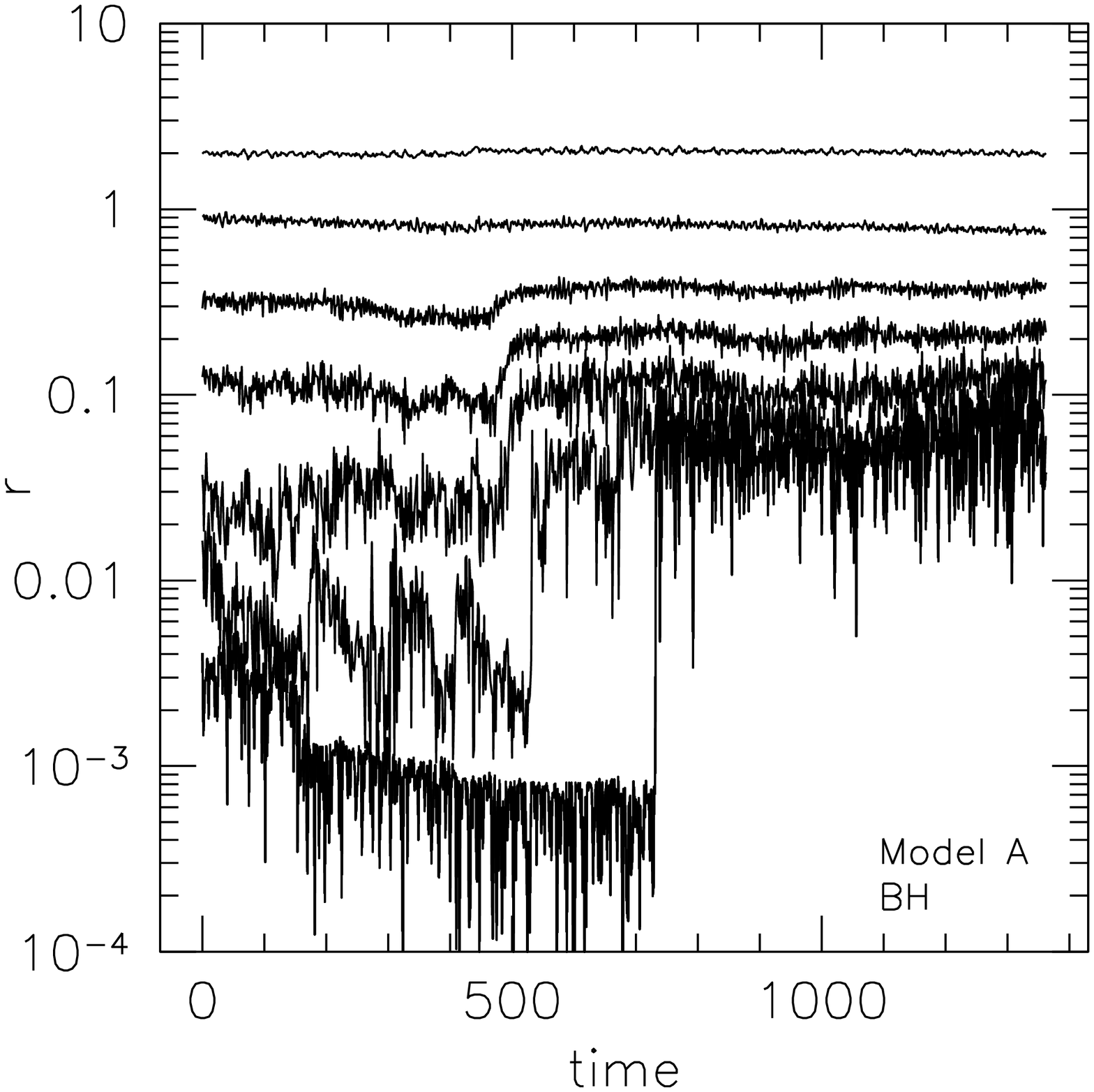}
    \includegraphics[width=4cm]{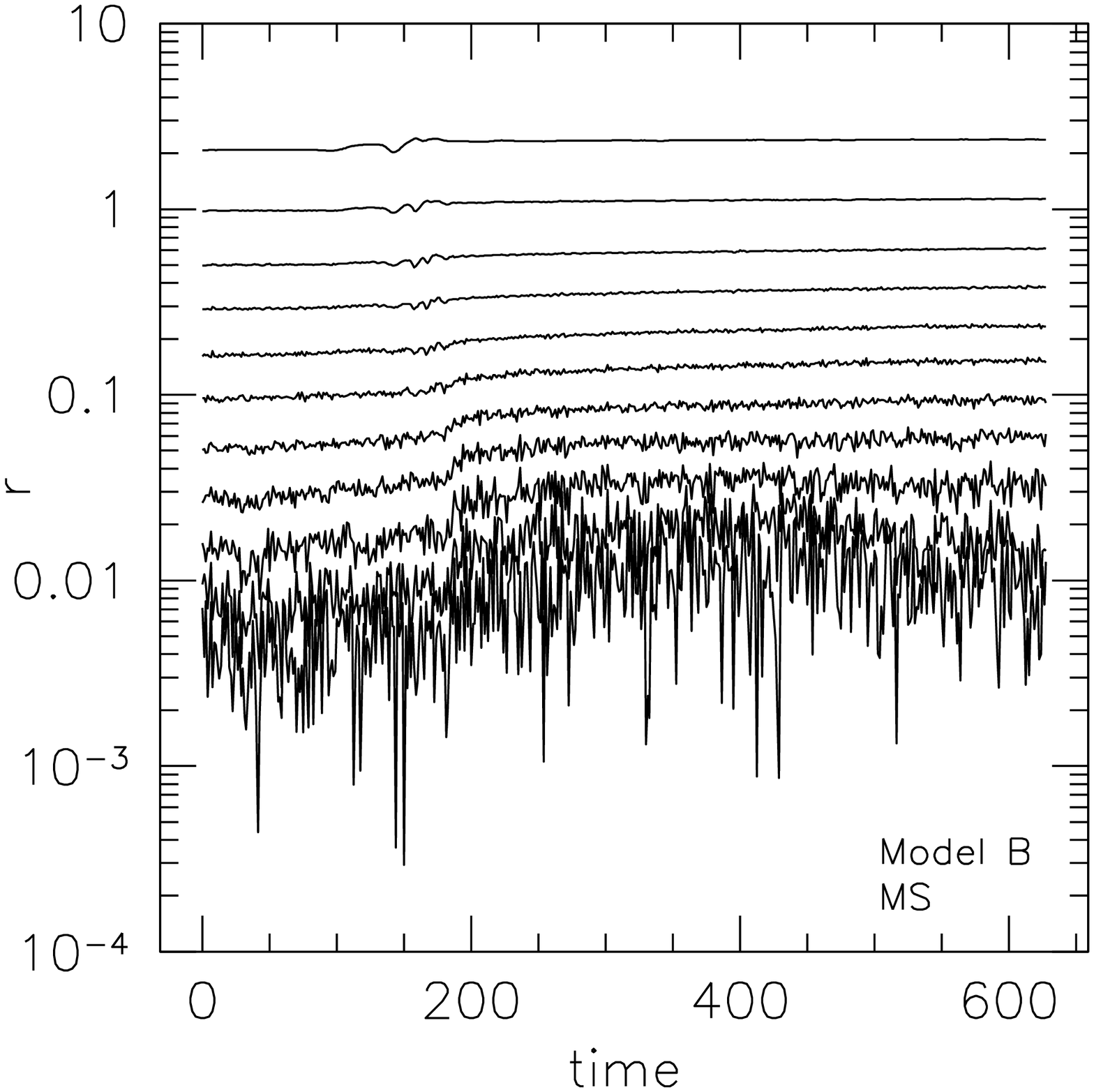}
    \includegraphics[width=4cm]{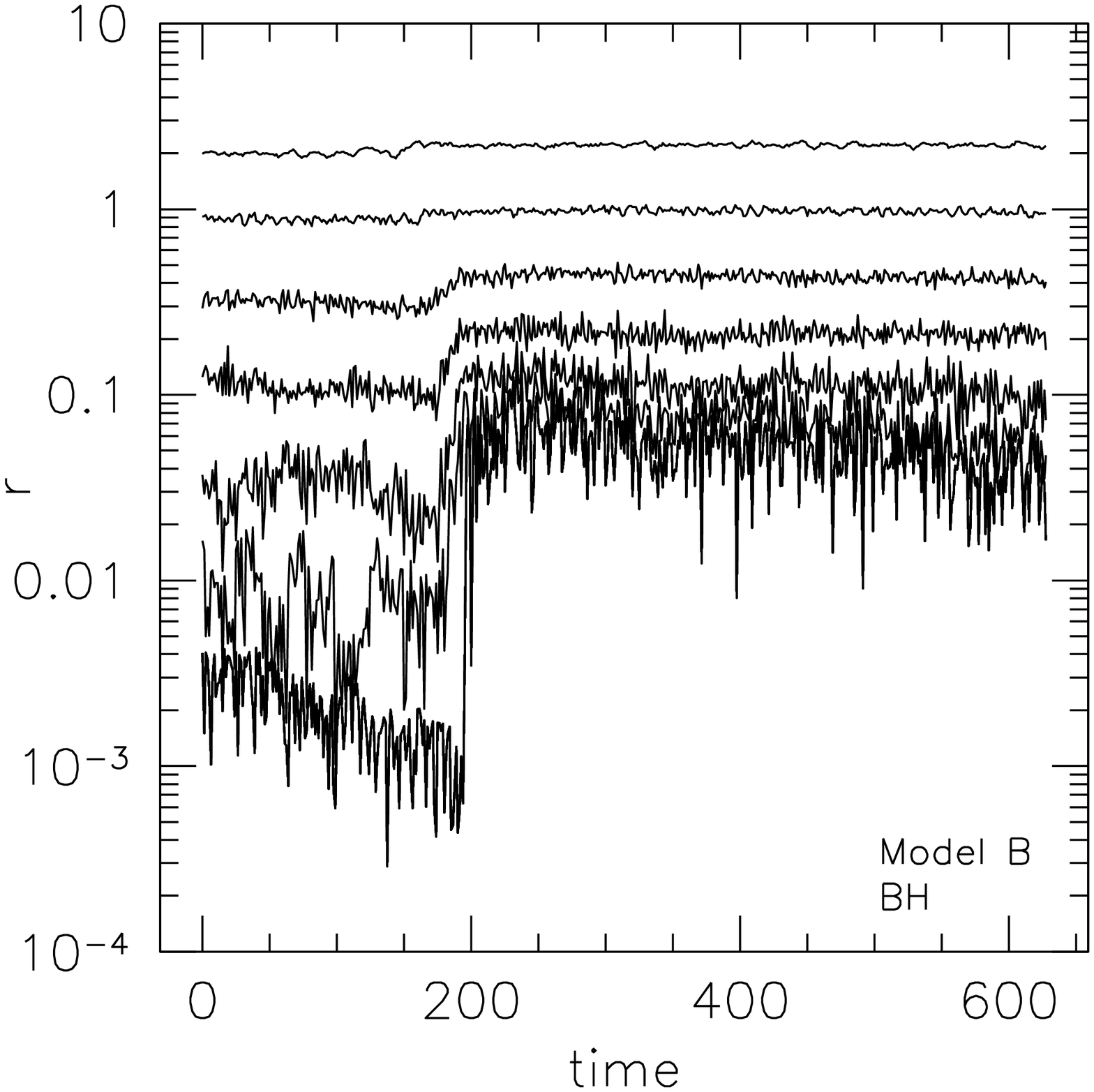}
  \end{center}
  \caption{Evolution of the Lagrange radii of the main sequence stars
    (left) and stellar BHs (right) during the galaxy merger phase, for
    models A (top) and B (bottom). For clarity, only stars initially
    belonging to the larger galaxy are shown. Formation of the binary
    MBH is reflected in the sudden expansion of the central regions,
    at $t\approx 500$ (Model A) and $t\approx 200$ (Model B).}
  \label{fig:lagr1}
\end{figure}
Figure\,\ref{fig:lagr1} clearly shows a decrease in the central
densities of both the main-sequence stars and the stellar BHs at a
time $t\approx t_f$.  Similar expansions are observed in the white
dwarf and neutron star distributions. However, the stellar BHs are
most affected, due to their higher, initial central concentration.

\begin{figure}
  \begin{center}
    \includegraphics[angle=270, width=8cm]{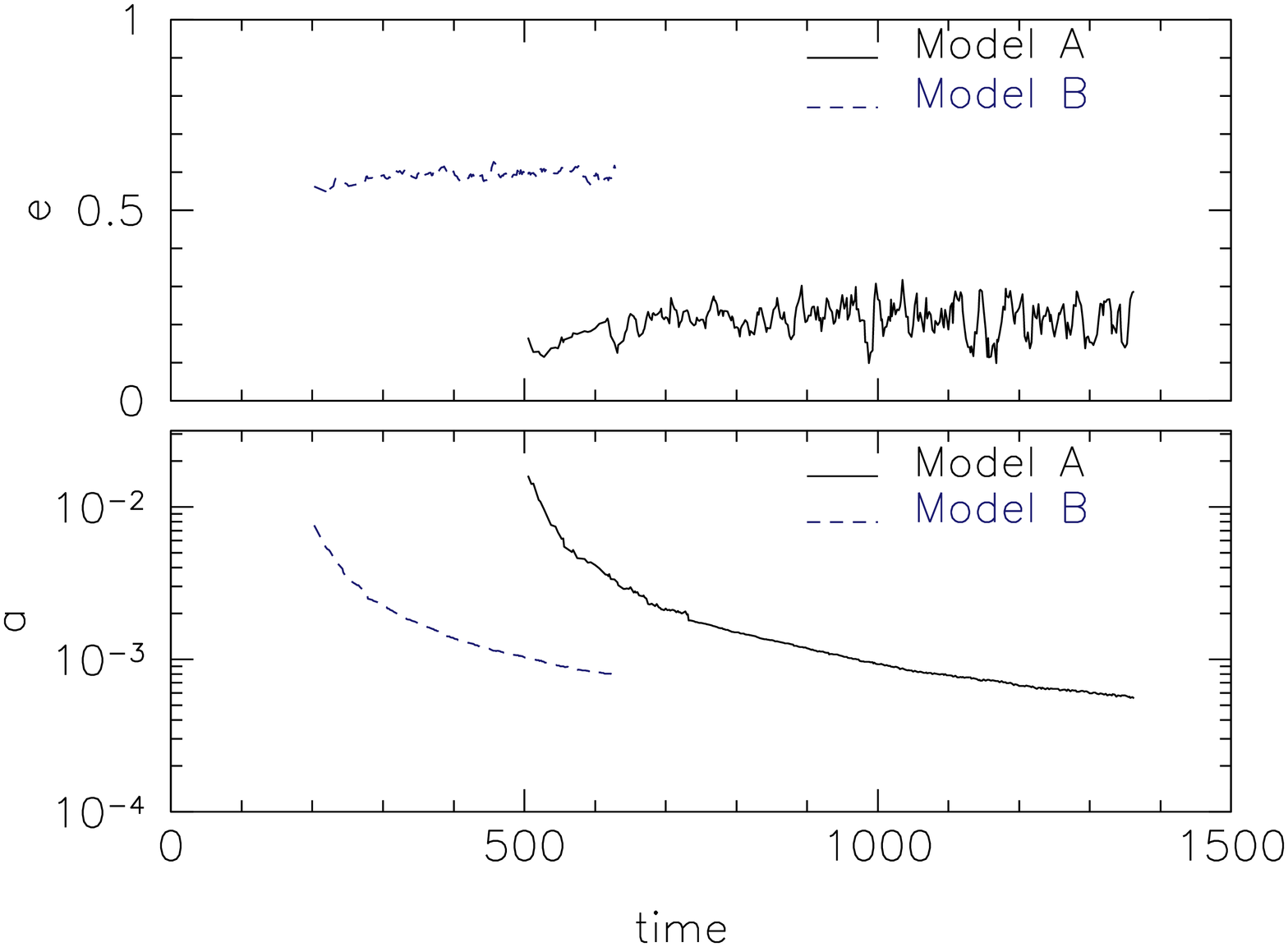}
  \end{center}
  \caption{Evolution of the semi-major axis and eccentricity of the
    MBH binary in Model A and B, starting from the time when
    the MBHs are formally bound. }
  \label{fig:ae}
\end{figure}

The orbital semi-major axis and eccentricity of the massive binary are
shown as functions of time in Figure\,\ref{fig:ae}.  Even in the
circular-orbit merger (Model A), the massive binary forms with a
slightly nonzero eccentricity.  Subsequent evolution of $a$ and $e$ is
driven by interactions with stars on intersecting orbits.
%%% figure 8 %%%

We computed the time-dependent binary hardening rate
\citep{quinlan1996}
\begin{equation}\label{eq:Defs}
s \equiv \frac{d}{dt} \left(\frac{1}{a}\right)
\end{equation}
by fitting a straight line to $a^{-1}(t)$ in small time intervals
between the time of binary formation and the end of the integration.
%%% figure 9 %%%
\begin{figure*}
  \begin{center}
    \includegraphics[width=8cm]{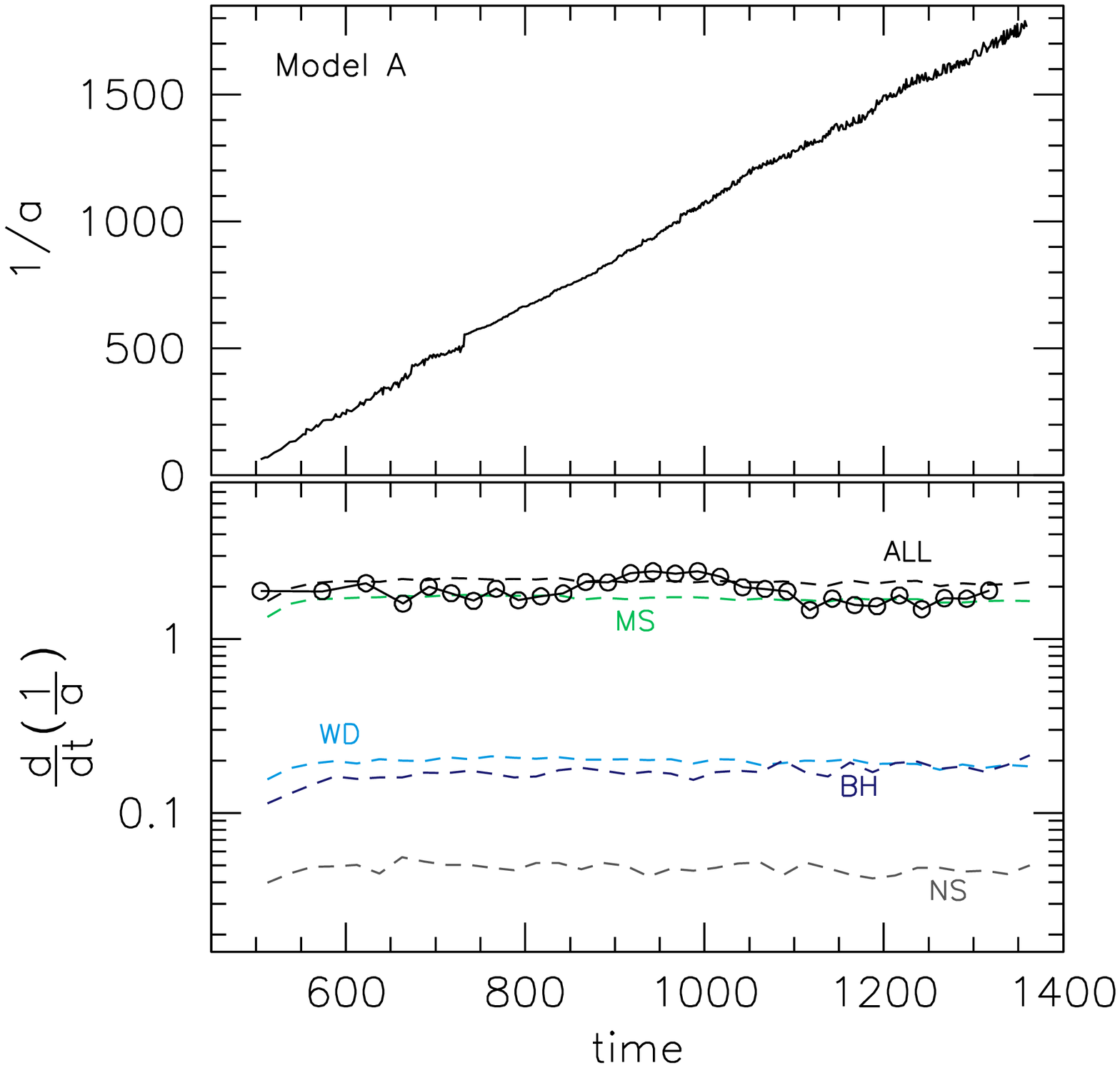}
    \includegraphics[width=8cm]{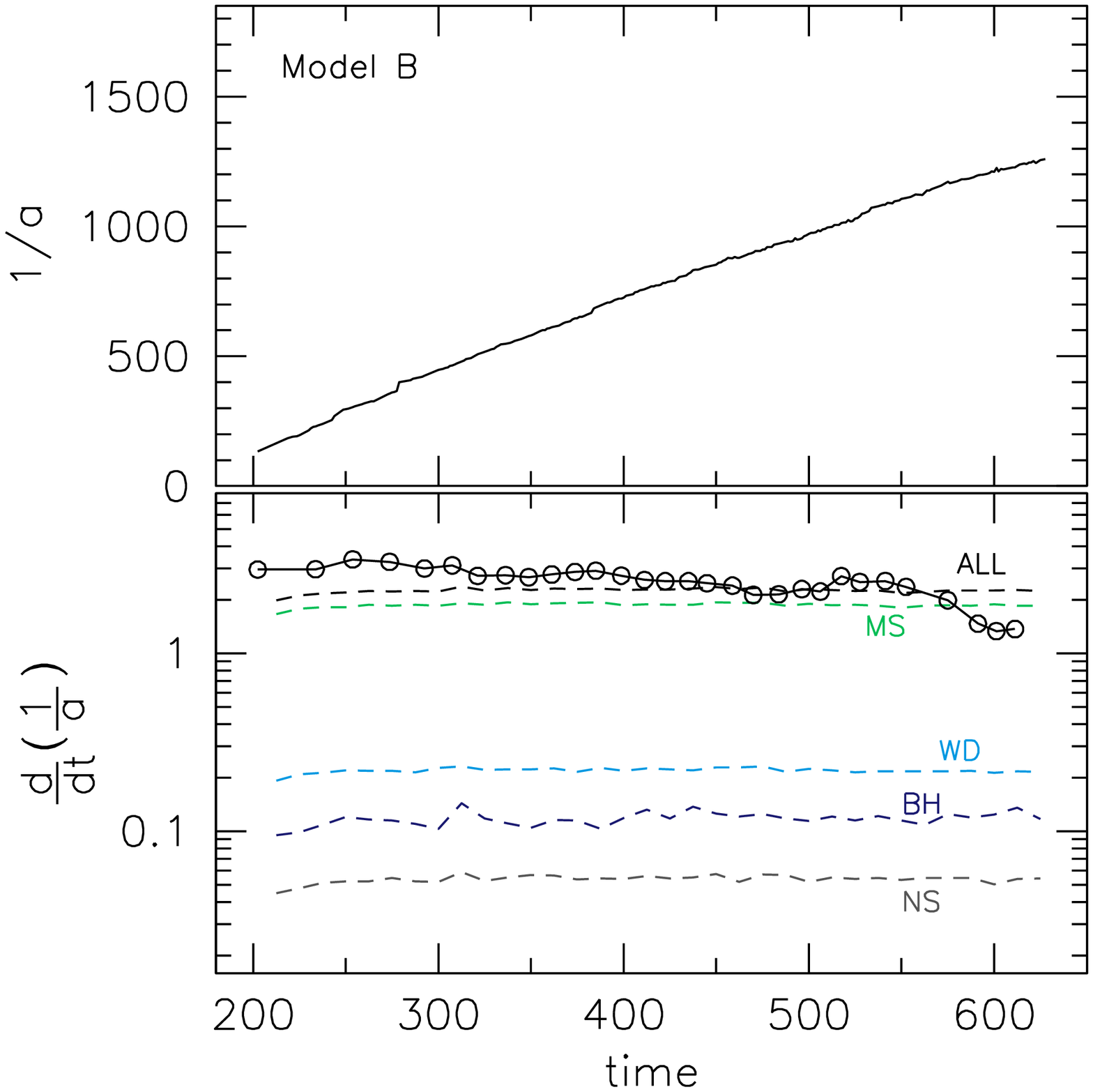}
  \end{center}
  \caption{Evolution of the inverse binary semi-major axis (upper) and
    the binary hardening rate (lower). The points represent the
    $N$-body results while the dashed lines represent the
    semi-analytic estimates for the different mass groups, assuming a
    full loss cone regime and evaluating the density and velocity
    dispersion at $D=0.5$ from the binary center of mass.}
  \label{fig:hard}
\end{figure*}
The results are shown in the bottom panels of Figure\,\ref{fig:hard}
for both models. The top panels show the evolution of $1/a$.  The
hardening rate appears roughly constant in time, which suggests that
the binaries enter the hard phase rather quickly, and then continue
hardening at an approximately constant rate. This behavior is a
defining property of hard binaries, if the stellar background is
unchanging, i.e. unaffected by the binary.

\citet{MikkolaValtonen1992}, \citet{quinlan1996} and
\citet{sesana2006} performed three-body scattering experiments of
circular binaries of varying mass ratio and hardness to derive
estimates of the hardening rate.  They provided fitting formulae for
the dimensionless parameter $H$ which is related to the hardening rate
via
\begin{equation}
H(a) = \frac{\sigma}{G\,\rho} \frac{d}{dt}\left(\frac{1}{a}\right)\,,
\end{equation}
where $\rho$ is the stellar mass density and $\sigma$ is the
one-dimensional velocity dispersion, both assumed constant in space
and time and unaffected by the presence of the binary.  The $H$
parameter obtained from these scattering experiments is a function of
binary mass ratio and hardness; the latter is defined in terms of
$V_\mathrm{bin}/\sigma$, the ratio of binary circular speed to
field-star velocity dispersion.

We wish to compare the $N$-body results for $s$ with the predictions
from the scattering experiments.  Good agreement would imply a ``full
loss cone,'' i.e. that the rate of interaction of the binary with
stars is unaffected by slingshot ejections.  In spherical galaxy
models, binary hardening rates fall much below the predictions from
the scattering experiments, since orbital repopulation is driven by
two-body scattering, which is a slow process for large $N$
\citep{makinofunato2004,ber05,mms07}.

We defined the theoretically-expected hardening rate to be
\begin{equation}\label{eq:sth}
 s_\mathrm{D}(a) \equiv H(a)\left(\frac{\rho}{\sigma}\right)_D
\end{equation}
where $H$ is taken from the published scattering experiments,
and $\rho$ and $\sigma$ are measured in our $N$-body models,
at a distance $D$ from the MBH.
($G=1$ in $N$-body units.)
For $H(a)$ 
we adopted the fitting formula of
\citet{sesana2006}:
\begin{equation}
H = A (1+a/a_0)^{\gamma}, \ \ a_0 = 1.05 \frac{GM_2}{\sigma^2}
\end{equation}
with parameters for a $3:1$ mass ratio: $A=15.82$, $\gamma = -0.95$.
Because $\rho$ and $\sigma$ vary with position (and time) in the
$N$-body models, our computed values of $s$ will depend on $D$.  We
found this dependence to be weak, as long as $D$ is not too different
(i.e. not more than a factor of two) from $\ri$: both $\rho$ and
$\sigma$ are weakly dependent on radius for $r\approx\ri$.  In
Figure\,\ref{fig:hard} we plot values of $s_D$ for $D=0.5$, slightly
larger than estimated influence radii in both models.  We find that
the $N$-body hardening rates, $s$, are quite consistent with the
analytic predictions $s_D$.  The contributions to the binary hardening
from the different mass groups (assumed to scale in proportion to
their mass densities) are shown in the bottom panels of
Figure\,\ref{fig:hard}.  The MS stars appear to be responsible for most
of the binary hardening, followed by the white dwarfs and the stellar
black holes.

These results suggest that the massive binaries in our simulations are
roughly in the ``full-loss-cone'' regime: they harden at a rate that
is consistent with the expected hardening rate of a binary in an
undepleted field of stars.  This is in agreement with the results of
\citet{khan2011} and \citet{preto2011}, who also found that the
stalling that occurs in spherical models is absent in simulations that
start from a stage preceding merger of the two galaxies.  Apparently,
the non-spherical shapes of the merger remnants result in a large
population of stars on ``centrophilic'' orbits: saucer orbits in the
axisymmetric geometry \citep{sridhartouma1999}, pyramid orbits in the
triaxial geometry \citep{merrittvasiliev2010}, etc.  We discuss the
shapes of our models in more detail in Section~\ref{sec:shape}.

Figure\,\ref{fig:ae} shows that the eccentricity of the massive binary
remains roughly constant in both models.  In a nonrotating galaxy,
binary eccentricity is expected to increase gradually with time:
\beq\label{eq:DefK} \frac{de}{dt} =K \frac{d\ln(1/a)}{dt} \eeq where
$K>0$ is a second dimensionless rate, also derivable from scattering
experiments; for instance, \citet{sesana2006} give
\begin{equation}\label{eq:Kofa}
K = A (1+a/a_0)^{\gamma} + B\,,
\end{equation}
and their Table\,\ref{tab:hard} gives values of $A$ and $B$ for a $1:3$
binary mass ratio.  Stars that encounter the binary in a prograde
(co-rotating) sense tend to circularize it; \citet{SGD2011} found that
when the fraction of corotating stars exceeded $\sim 0.7$ (as opposed
to 0.5 for a nonrotating galaxy), binaries tend to circularize.  We
found that our models have roughly this fraction ($\sim 0.7$) of
corotating stars, consistent with the mild eccentricity growth that we
see.

In our simulations, binary hardening rates $s$ are essentially
constant with time.  If we assume that this remains true beyond the
end of our simulations, we can use equations~(\ref{eq:Defs})
and~(\ref{eq:DefK}) to extrapolate the binary elements $(a,e)$ to
arbitrarily later times.  At some point, gravitational wave emission
will dominate the evolution; calculating when this happens requires
assigning physical units to the models.  We considered two
representative scalings, based on assumed MBH masses of $4.0\times
10^6\msun$ and $10^8\msun$ respectively.  Scaling factors for length
were set at $10\pc$ and $40\pc$ respectively, yielding physical
values of the influence radii of $\sim 3\pc$ and $\sim 12\pc$.  As
discussed in Section~\ref{sec:scaling}, the unit of time is determined
in this (full loss cone) regime by orbital periods, not relaxation
times; hence the scaling factor for time, $[T]$, is related to the
scaling factors for mass and length by $[T] = \sqrt{[L]^3/(G[M])}$ and
the binary hardening rate in physical units is $\left[L\right]^{-1}
\left[T\right]^{-1}$ times the $N$-body hardening rate.
%%% figure 10 %%%
\begin{figure*}
  \begin{center}
    \includegraphics[angle=270, width=12cm]{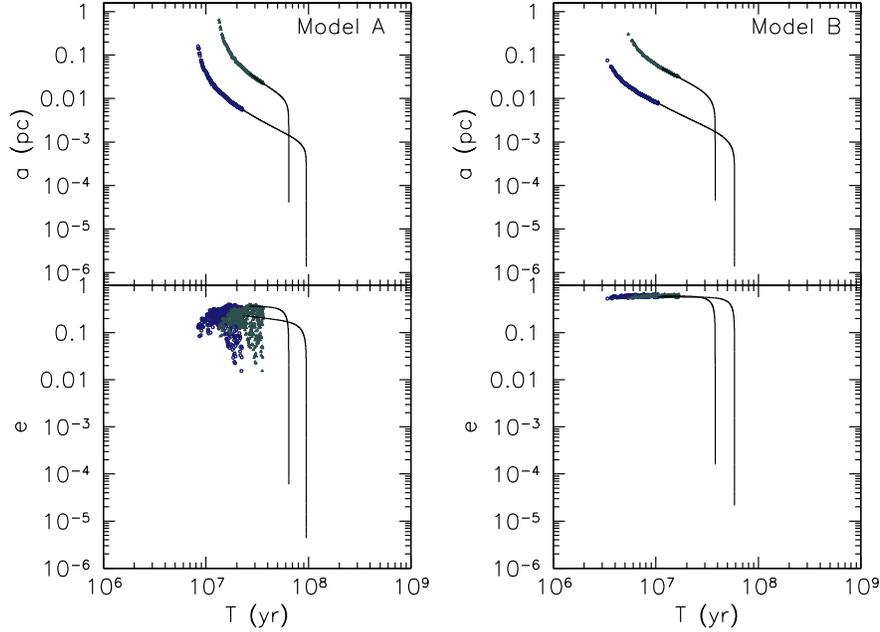}
  \end{center}
  \caption{Evolution of the binary elements in the $N$-body phase
    (symbols) and in the semi-analytic phase (lines). For each model,
    two physical scalings are adopted, as discussed in the text.}
  \label{fig:harden}
\end{figure*}

Given a value for $s$, and scale factors for mass and length, the
evolution of the binary semi-major axis at late times is determined by
\begin{equation}
\frac{da}{dt} =  \frac{da}{dt}\bigg|_{\rm SI} +
\frac{da}{dt}\bigg|_{\rm GW} = -s\,a^2(t) + \frac{da}{dt}\bigg|_{\rm GW}
\end{equation}
where the first term on the right hand side represents hardening due
to interactions with stars, and the second term represents energy
lost to gravitational waves.  The rate of the latter process depends
on $e$ as well as $a$.  In extrapolating $e$ beyond the end of the
$N$-body integrations, we ignored the effect of the galaxies' rotation
on the binaries' eccentricity growth and simply applied
equations~(\ref{eq:DefK}) and~(\ref{eq:Kofa}). 
The rate of change of the orbital elements due to GW emission is
\citep{peters64}
\begin{subequations}
\label{eq:peters64}
\begin{eqnarray}
\frac{da}{dt}\bigg|_{\rm GW} = -\frac{64}{5} \beta \frac{F(e)}{a^3}\\
\frac{de}{dt}\bigg|_{\rm GW} = -\frac{304}{15}\beta \frac{e G(e)}{a^4}
\end{eqnarray}
\end{subequations}
where $F(e)$ is given in equation\,(\ref{eq:fe}),
\begin{equation}
 \label{eq:ge}
G(e) = \left(1-e^2\right)^{-5/2} \left(1 + \frac{121}{304}e^2 \right)
\end{equation}
and 
\begin{equation}
\beta = \frac{G^3}{c^5} M_1 M_2 \left(M_1+M_2\right)\,.\nonumber
\end{equation}

\begin{table}
\begin{center}
\caption{Hardening phase parameters}
\label{tab:hard}
\begin{tabular}{ccccccc}
\hline 
Model & $M_1$ & $\tnb$ & $\thd$ & $a_f$ & $e_f$ & $e_g$\\
      & $(\msun)$ & $(\yr)$ & $(\yr)$ & $(\mpc)$ & & \\
\hline 
A  &  $4\times 10^6$ &  $2.3\times 10^7$ & $7.3\times 10^7$ &  5.6 & 0.23 & $8\times 10^{-6}$ \\
   &  $1\times 10^8$ &  $2.7\times 10^7$ & $3.7\times 10^7$ & 37.2 & 0.37 & $8\times 10^{-5}$ \\  
B  &  $4\times 10^6$ & $1.0\times 10^7$ & $4.8\times 10^7$ &  8.0 & 0.58 & $4\times 10^{-5}$ \\
   &  $1\times 10^8$ & $1.1\times 10^7$ & $2.6\times 10^7$ & 50.2 & 0.60 & $2\times 10^{-4}$ \\
\hline
\end{tabular}
\end{center}
\end{table}

We numerically solved the coupled equations for the evolution of the
semi-major axis and eccentricity, starting from values at the end of
the $N$-body phase or somewhat earlier, for each of the two sets of
scaling factors.  The results are shown in Figure\,\ref{fig:harden} and
Table\,\ref{tab:hard}.  In this table, the time $\tnb$ represents the
time spent in the $N$-body hardening phase while $\thd$ represents the
time from the end of the $N$-body integration until coalescence; the
latter time includes both a stellar-interaction driven, and a GW
dominated, regime. The total evolution time from the beginning of the
galaxy merger to full coalescence is given by the sum of the two
times.  For a Milky Way type galaxy undergoing a $3:1$ merger, this
time is of the order of $10^8\yr$ for an initial circular orbit and
$6\times10^7\yr$ for an initially moderately eccentric orbit.  The
table also lists the values of the semi-major axis $a_f$ and
eccentricity $e_f$ at the transition between the simulated $N$-body
phase and the semi-analytical phase, as well as the value of the
eccentricity when the separation reaches $10 r_g = 10\,GM_1 / c^2$.
At smaller separations, equations like~(\ref{eq:peters64}) are not
valid.

\subsection{Core formation}
\label{sec:mdef}

%%% figure 11 %%%
\begin{figure}
  \begin{center}
    \includegraphics[width=8cm]{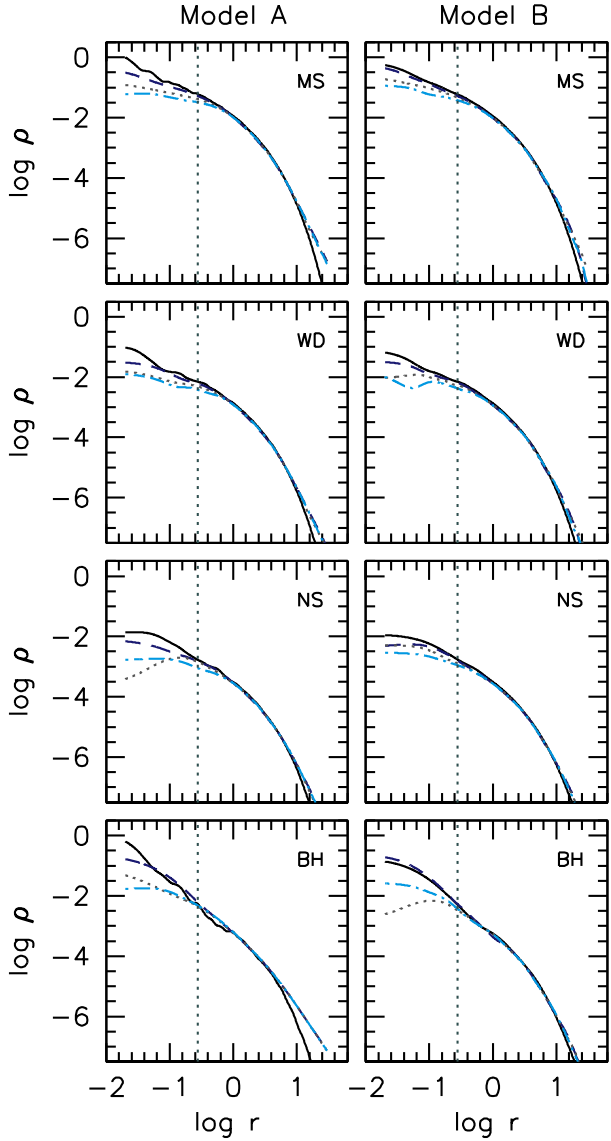}
  \end{center}
  \caption{Spatial density profiles of the different mass groups in
    the larger galaxy. For each species, different lines are for
    different times during the merger phase: at the start of the
    simulation (solid lines), at the time when $a\sim a_f$ (dashed
    lines), at the time when $a \sim a_h$ (dotted lines) and at a late
    time when $a \sim 0.1 a_h$ (dashed-dotted lines).  Vertical lines
    indicate $\ri$, defined as described in the text.}
  \label{fig:density}
\end{figure}

The pre-merger galaxies had central density profiles that were well
approximated as power laws with respect to radius.  Following the
merger, the density on scales $r\lap\rh$ is strongly modified
(lowered) by the action of the massive binary.  The process of ``core
scouring'' is illustrated in Figure\,\ref{fig:density}, which compares
the spatial density profiles of the different species in the larger
galaxy at four different times.  For this plot, only particles that
were originally associated with the larger galaxy were used.  By the
time the binary has become hard, a mass deficit \citep{mm02} is
clearly visible in all components on scales significantly larger than
$\ri$.  Model A shows substantially more evolution of the central
density at early times; at late times the cores in the two models are
more similar.

We estimated core sizes in two ways.  First, we fit core-S\'{e}rsic
profiles to our models at the end of the merger phase, and adopted the
resulting values of the break radius $R_b$ as estimates of the core
radius.  The projected density profiles of galaxies are globally well
fit by the S\'{e}rsic (1968) law:
\begin{equation}
\label{eq:sersic}
I(R) = I(0)~{\rm exp}\left\{-b \left(R/R_e\right)^{1/n} \right\}\,
\end{equation}
where $I(0)$ is the central intensity, $R_e$ is the effective
half-light radius, and $n$ is a parameter controlling the curvature in
a log-log plot. The term $b$ is not a parameter but a function of $n$
\citep[see e.g.][]{tg05}.  Deviations from this law appear close to
the center, where bright galaxies typically show central light
deficits with respect to the inward extrapolation of the S\'{e}rsic
law while faint galaxies show central light excesses
\citep[e.g.][]{GG2003,ferr2006,cote2007}.  Adding an inner power-law with
slope $\gamma$ \citep{graham2003} yields
\begin{equation}
\label{eq:trujillo}
I(R) = I^{\prime}\,
\left[1+\left(R_b/R\right)^{\alpha}\right]^{\gamma/\alpha}{\rm
  exp}\left\{ -b\, \left(R^\alpha +
R_b^\alpha\right)/R_e^\alpha\right\}^{1/\left(n \alpha\right)}
\end{equation}
with 
\begin{equation}
I^{\prime} = I_b\,2^{-\gamma/\alpha} {\rm exp}\left[ b\,
  2^{1/\left(\alpha n\right)} \left(R_b/R_e\right)^{1/n} \right] \,,
\end{equation}
the so-called core-S\'{e}rsic law.  Here, $\alpha$ is an additional
parameter regulating the transition from the inner power-law to the
outer S\'{e}rsic law and $R_b$, the break radius, marks the distance
where the profile changes from one regime to the other.

We also considered a second definition of the core radius
\citep{King1962}, as the projected radius $r_c$ at which the surface
density falls to one-half its central value.  Here, ``central'' was
taken to be the value at a projected radius of $0.1\ri$.

\begin{table}
  \begin{center}
    \caption{Core radii}
    \label{tab:merger}
    \begin{tabular}{cccccc}
      \hline 
      Model & $\ri$ & $R_b$ (MS) & $r_c$ (MS) & $R_b$ (BH)  & $r_c$ (BH) \\
      \hline 
      A &  0.38  &  2.4   & 0.8  & 5.1  &  0.3 \\
      B &  0.36  &  0.9   & 0.6  & 3.5  &  0.2 \\
      \hline
    \end{tabular}
  \end{center}
\end{table} 
Both estimates, computed at the time when $a\sim 0.1\,a_h$, are listed in
Table\,\ref{tab:merger}, separately for the MS stars and the
stellar-mass BHs.  These radii were computed using {\it all} the
particles from the respective mass groups, without regard to the
galaxy with which they were originally associated. We also give $\ri$,
computed using equation\,(\ref{eq:DefrM}); we set $M_\bullet=M_1+M_2$
in that equation, and combined together {\it all} the stellar species
from {\it both} galaxies when computing $M_\star$.
%%% figure 12 %%%
\begin{figure}
  \begin{center}
    \includegraphics[width=8cm]{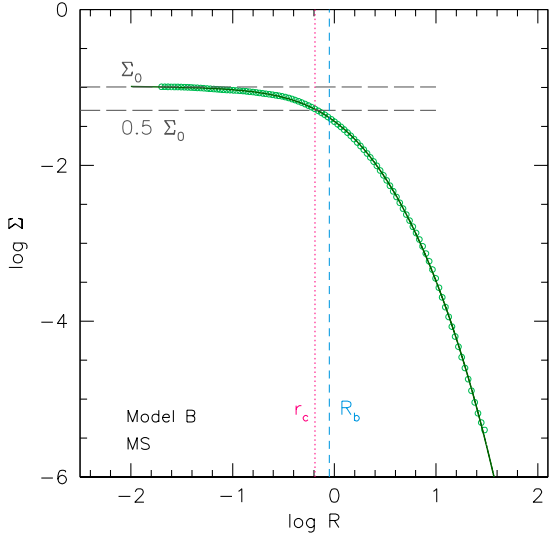}
  \end{center}
  \caption{Projected density profile for MS stars in Model B, at the
    time when $a\sim 0.1\,a_h$ (points), with superimposed the best
    fitting core-S\'{e}rsic model (solid line). The vertical dotted line
    indicates $r_c$ while the vertical dashed line indicates
    $R_b$. The horizontal lines indicate the value of the central
    density $\Sigma_0$ and half the central density.}
  \label{fig:rcore}
\end{figure} 

In the case of the dominant (MS) mass component,
Table\,\ref{tab:merger} shows that $r_c\approx 2\ri$ for both Models A
and B.  However, $R_b$ is somewhat larger than $r_c$.  The reason is
that, due to the flatness of the profile, the radius at which the
inner profile transitions to the outer profile (i.e. the break radius)
is much larger than the radius at which the density falls to half its
central value (i.e. the core radius). This is illustrated in
Figure\,\ref{fig:rcore} for Model B.

In what follows, we will adopt $r_c$ as the more robust measure of the
core radius.  We emphasize that the size of the core in the MS stars
is larger than the influence radius of the massive binary in both of
our models, whichever definition of ``core radius'' or ``influence
radius'' is used.  One important consequence, discussed in more detail
below, is that regrowth of a Bahcall-Wolf cusp at radii $\lap \ri$
following coalescence of the two MBHs leaves the core structure
essentially unchanged.

\subsection{Stellar ejections}
\label{sec:ejections}

\begin{table*}
\begin{center}
\caption{Fraction of unbound stars}
\label{tab:escapers}
\begin{tabular}{cccccccccc}
\hline 
Model & $f_{\rm e}$  & $f_{\rm e1}$ & $f_{\rm e2}$ & $f_{\rm eMS}$  &
$f_{\rm eWD}$  & $f_{\rm eNS}$  & $f_{\rm eBH}$ & $M_{\rm eTS} /
M_{\rm gal}$ & $M_{\rm eGS} / \msbh$ \\
\hline 
A   &  0.021  & 0.005  &  0.016  &  0.021  &  0.020  &  0.024  &  0.024  &  0.018 &  0.55\\
B   &  0.014  & 0.001  &  0.013  &  0.014  &  0.012  &  0.016  &  0.018  &  0.012 &  0.45\\
\hline
\end{tabular}
\end{center}
\end{table*}

Stars can become unbound during a galaxy merger due to both tidal
stripping in the early phases of the merger and gravitational
slingshot interactions with the binary MBHs when this is hard.  Unbound
stars at the end of the merger phase are a result of both mechanisms.
We determined the number of unbound stars in the two merger
simulations by selecting stars with velocity in excess of the local
escape speed from the system.  At any given time, the escape velocity
at a distance $r$ from the binary center of mass is $V_{\rm esc}(r) =
\sqrt{-2\Phi(r)}$, where the gravitational potential $\Phi$ can be
expressed as \citep[equation\,2-22]{bt87}
\begin{equation}
\Phi(r) =  \Phi(r' < r) + \Phi(r' > r)
\end{equation}
with
\begin{equation}
\Phi(r' < r)  = -4\pi G \, \frac{1}{r} \int_0^r \rho(r') r'^2 dr' \approx
\frac{-G M(<r)}{r}
\end{equation}
having defined $M(<r)$ the mass enclosed within a sphere of radius
$r$, and
\begin{equation}
\Phi(r' > r)  = -4\pi G \, \int_r^{\infty} \rho(r') r' dr' \approx
-G \sum_{i=1}^N \frac{m_i}{r_i},  \mbox{ for } r_i > r \,.
\end{equation}

Table\,\ref{tab:escapers} reports the total fraction of unbound stars
$f_e$ at the time when $a \sim 0.1 a_h$, the fraction of unbound stars
originally belonging to the first and second galaxy $f_{e1}$,
$f_{e2}$, and the fraction of unbound MS, WD, NS and BHs $f_{MS}$,
$f_{WD}$, $f_{NS}$, $f_{BH}$, normalized to the total number of stars
in their mass group.  We find that about 2\% of all stars are ejected
by the end of the merger phase. Model A, which has a more gradual
evolution, produces more unbound stars than Model B.  Stars from the
different mass groups show roughly equal probabilities of being
ejected.

We distinguished stars ejected via tidal stripping (TS) versus
gravitational slingshot (GS) based on the time they become unbound: if
a star becomes unbound before the binary has become hard we consider
it ejected by TS whereas if it becomes unbound after the binary has
become hard we consider it ejected by GS. Based on this selection, we
find that TS is responsible for the ejection of about 90\% of the
escapers. The smaller galaxy is the most susceptible to TS, as shown
by the fact that $f_{e2} \gg f_{e1}$.  Table\,\ref{tab:escapers} also
lists the mass in stars unbound by TS in units of the total stellar
mass $\mg$ and by the mass in stars unbound by GS in units of the
binary mass.  High velocity escapers, with velocities as high as
several times the central escape speed, are mainly produced by
GS. These stars are analogs of the hypervelocity stars detected in the
halo of the Milky Way \citep[e.g.][]{brown05, brown06, GPS05,
  GPZ07, gva2009}. Stars unbound by TS tend to have lower velocities, and would
be less numerous in a real galaxy, due to the deeper potential well.

\section{Post-merger evolution}
\label{sec:postmerger}

After the two MBH particles were combined into one, we continued the
integrations of the $N$-body models for several relaxation times.  The
relaxation time of a multi-component system is ill-defined.  We are
primarily interested in the time scale for collisional evolution of
the dominant population, the MS stars.  As a simple estimate of the
relaxation time, we used equation\,(\ref{eq:tr_spitz}), setting $m$ to
the mass $m_{\rm MS}$ of a MS particle, and replacing the number
density $n$ by a simple summation, $n_t = n_{MS} + n_{WD} + n_{NS} +
n_{BH}$.  Other reasonable definitions of $\tr$ were found to give
nearly identical numerical values, a consequence of the fact that the
MS stars dominate the total numbers at the radii of interest, and the
fact that the masses of the three major groups (MS, NS, WD) are very
similar.  The resulting estimates of $\tr$, computed at the MBH
influence radius $\ri$ at the beginning of the post-merger phase, are
given in Table\,\ref{tab:trelax}.  For the Coulomb logarithm we used
$\ln \Lambda = \ln(r_h \sigma^2/2 G m_{\rm MS}) = \ln (\msbh / 2m_{\rm
  MS})$, with $r_h = G \msbh /\sigma^2$ and $\msbh$ the mass of the
merged MBHs.  

Ignoring the influence of the other components, a cusp in the MS stars
is expected to re-form in a time of roughly $\tr(\ri)$, at radii
$r\lap 0.2\ri$, after being destroyed by the massive binary
\citep[e.g.][]{MS06}.  In the case of the BHs, their central density should
increase more rapidly, by a factor $\sim m_{BH}/m_{MS}=10$, as they
segregate spatially with respect to the lighter components.  If the
density in the heavier (BH) component should ever approach locally the
density in the lighter components, heating of the light particles by
the heavy particles will occur, causing the density of the former to
decrease.
%%% figure 13 %%%
\begin{figure*}
  \begin{center}
    \includegraphics[width=6.5cm]{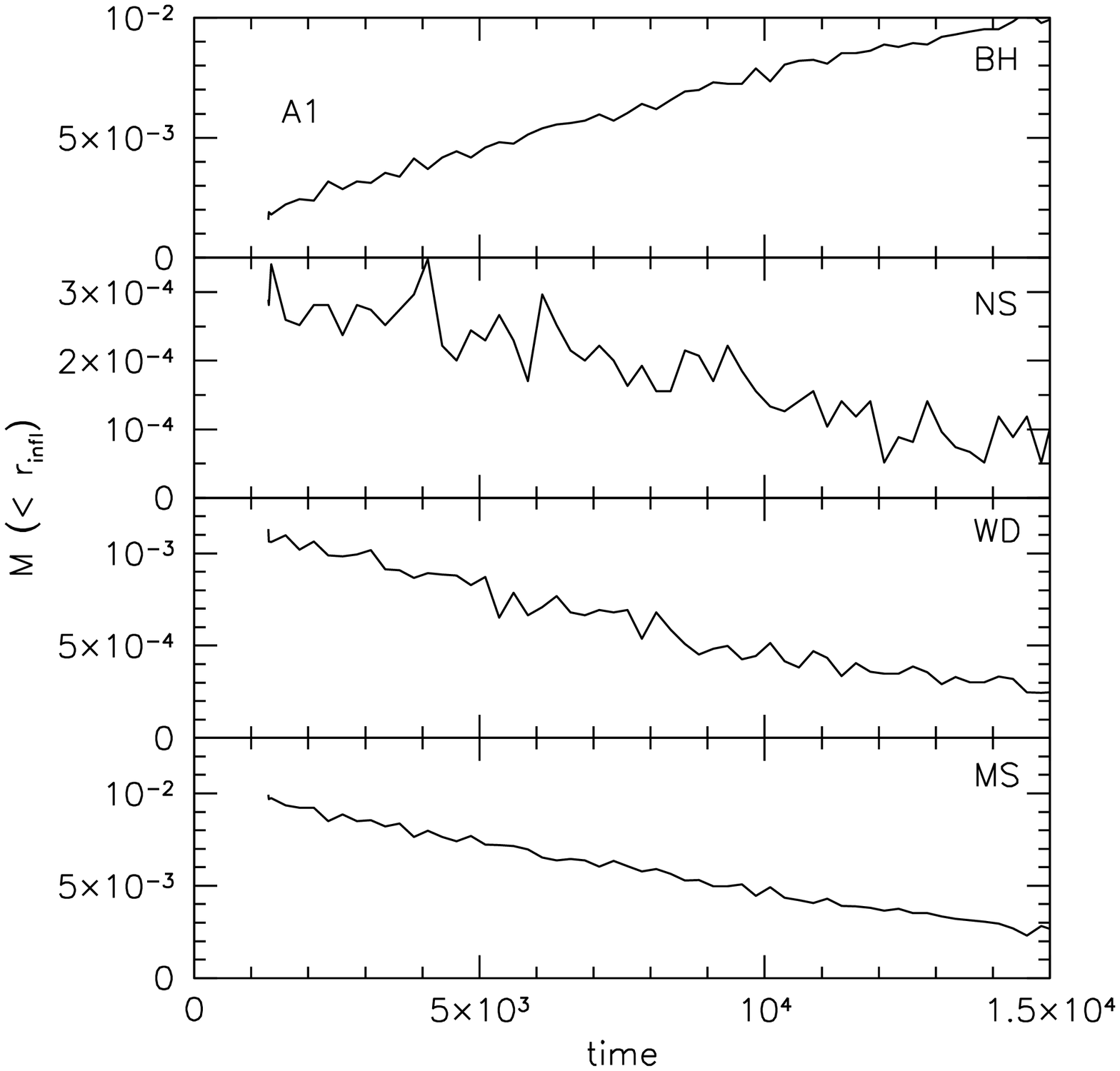}
    \includegraphics[width=6.5cm]{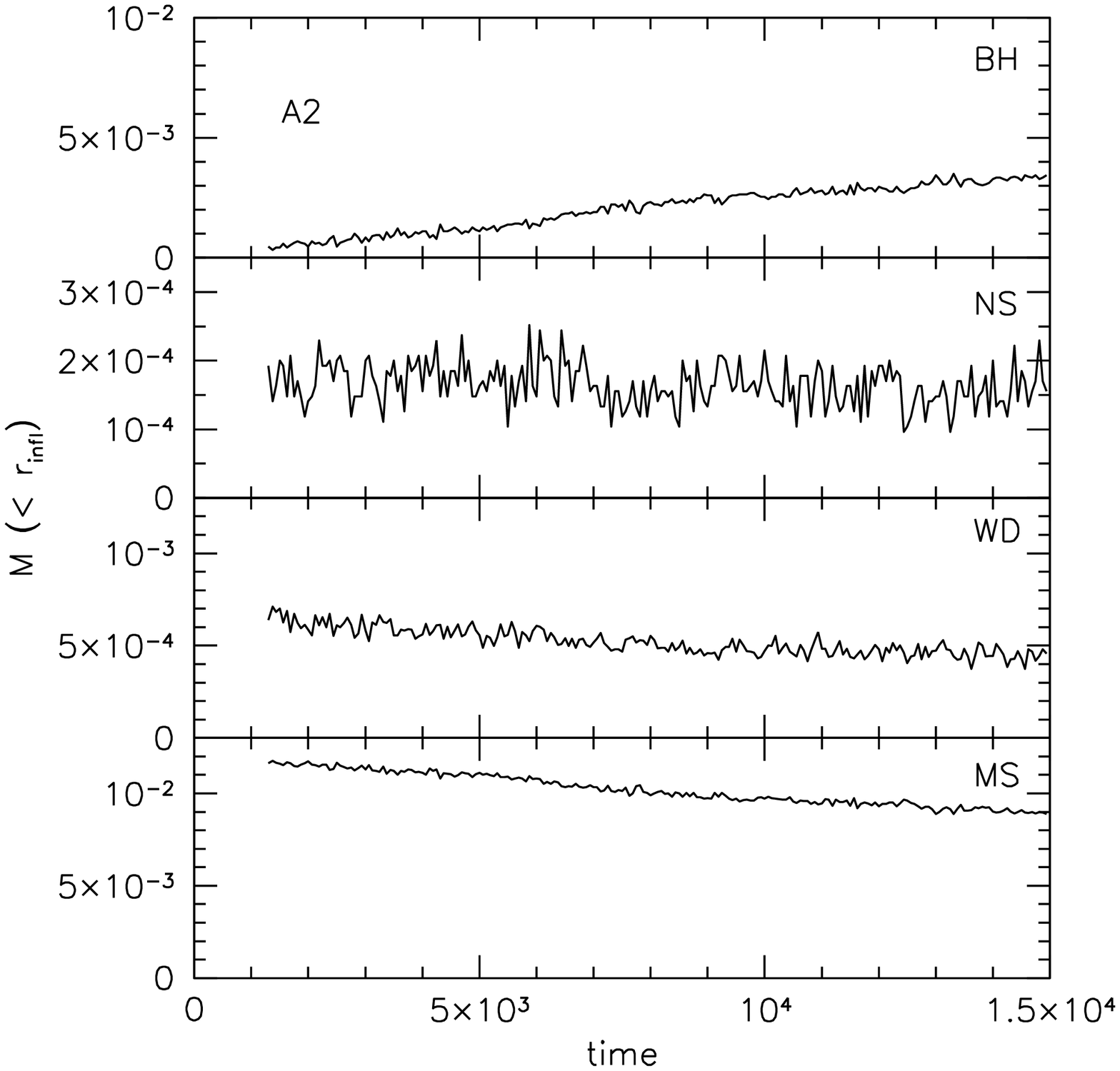}
    \includegraphics[width=6.5cm]{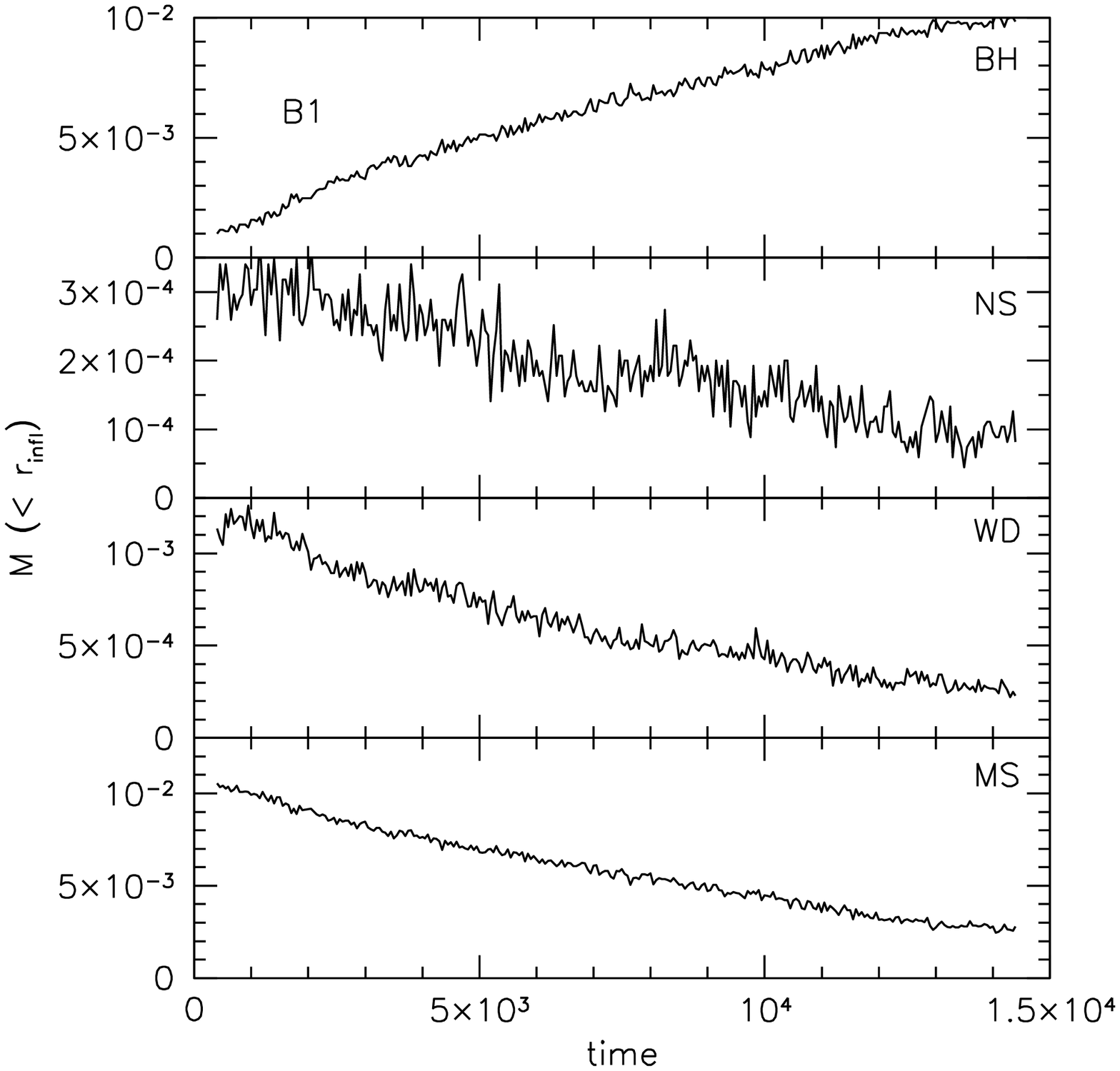}
    \includegraphics[width=6.5cm]{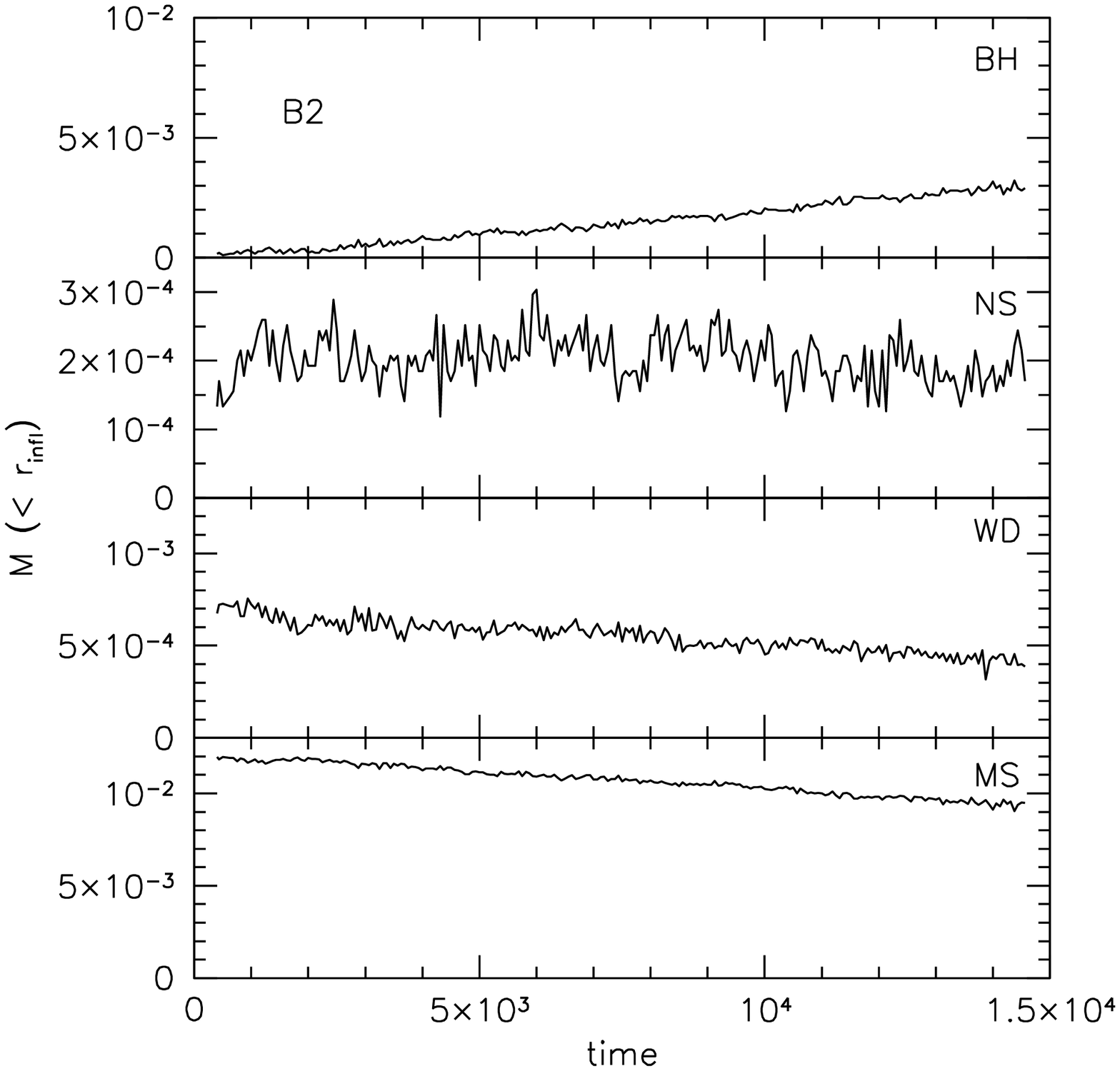}
  \end{center}
  \caption{Time evolution of the stellar mass enclosed in the MBH
    influence sphere, for each species, in the post-merger models.}
  \label{fig:massp}
\end{figure*}

\begin{table}
  \begin{center}
    \caption{Central properties at the start of the post-merger phase}
    \label{tab:trelax}
    \begin{tabular}{ccccc}
      \hline 
      Model & $\rh$ & $\ri$ & $\ln \Lambda$ & $\tr(\ri)$ \\
      \hline 
      A1 & 0.09 & 0.40 & 6.4 & $4.7\times10^3$\\ 
      A2 & 0.09 & 0.43 & 6.4 & $5.8\times10^3$\\
      B1 & 0.09 & 0.35 & 6.4 & $3.8\times10^3$\\
      B2 & 0.09 & 0.37 & 6.4 & $3.7\times10^3$\\
      \hline
    \end{tabular}
  \end{center}
\end{table}

%%% figure 14 %%%
\begin{figure*}
  \begin{center}
    \includegraphics[width=8cm]{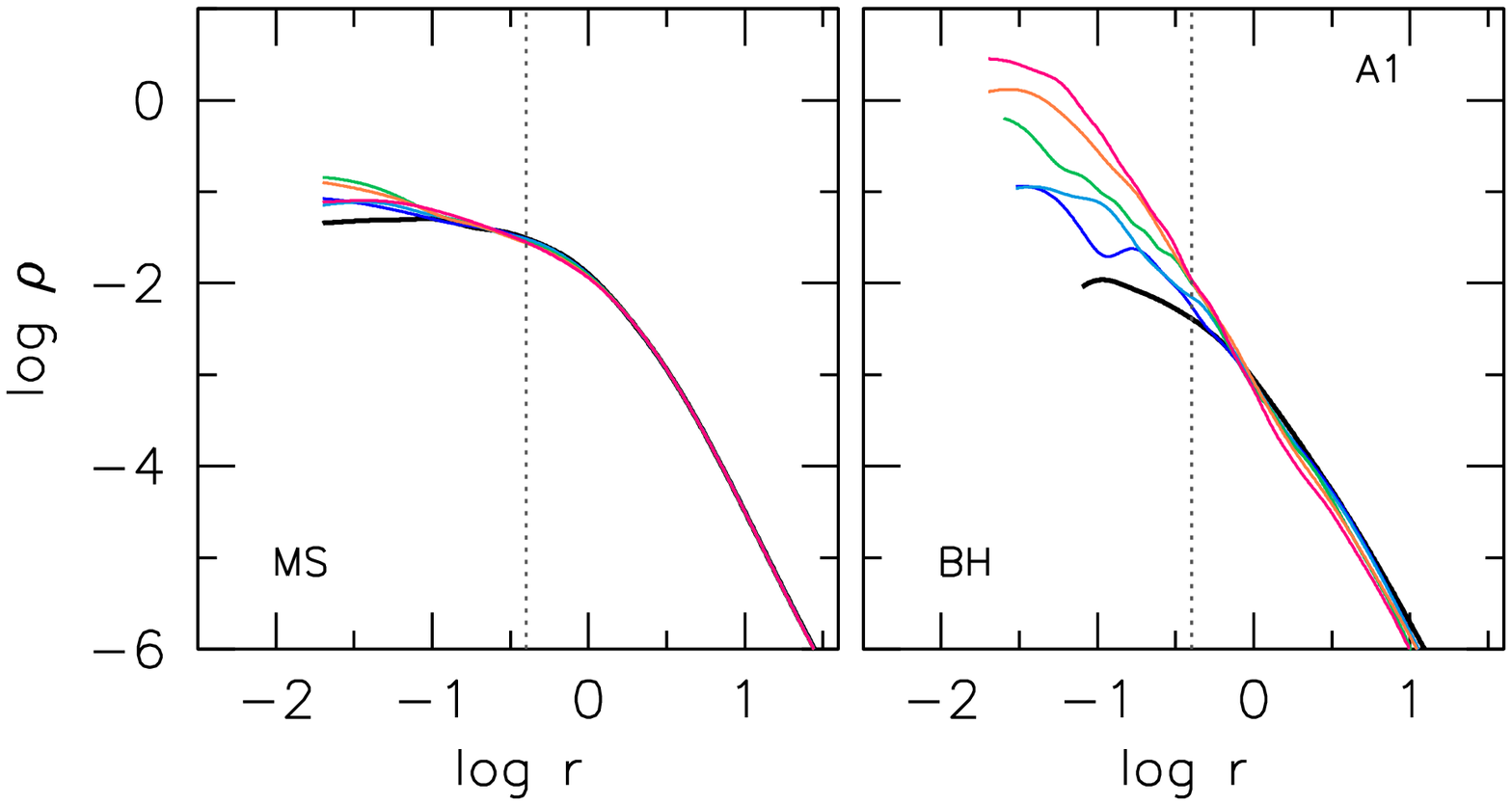}
    \includegraphics[width=8cm]{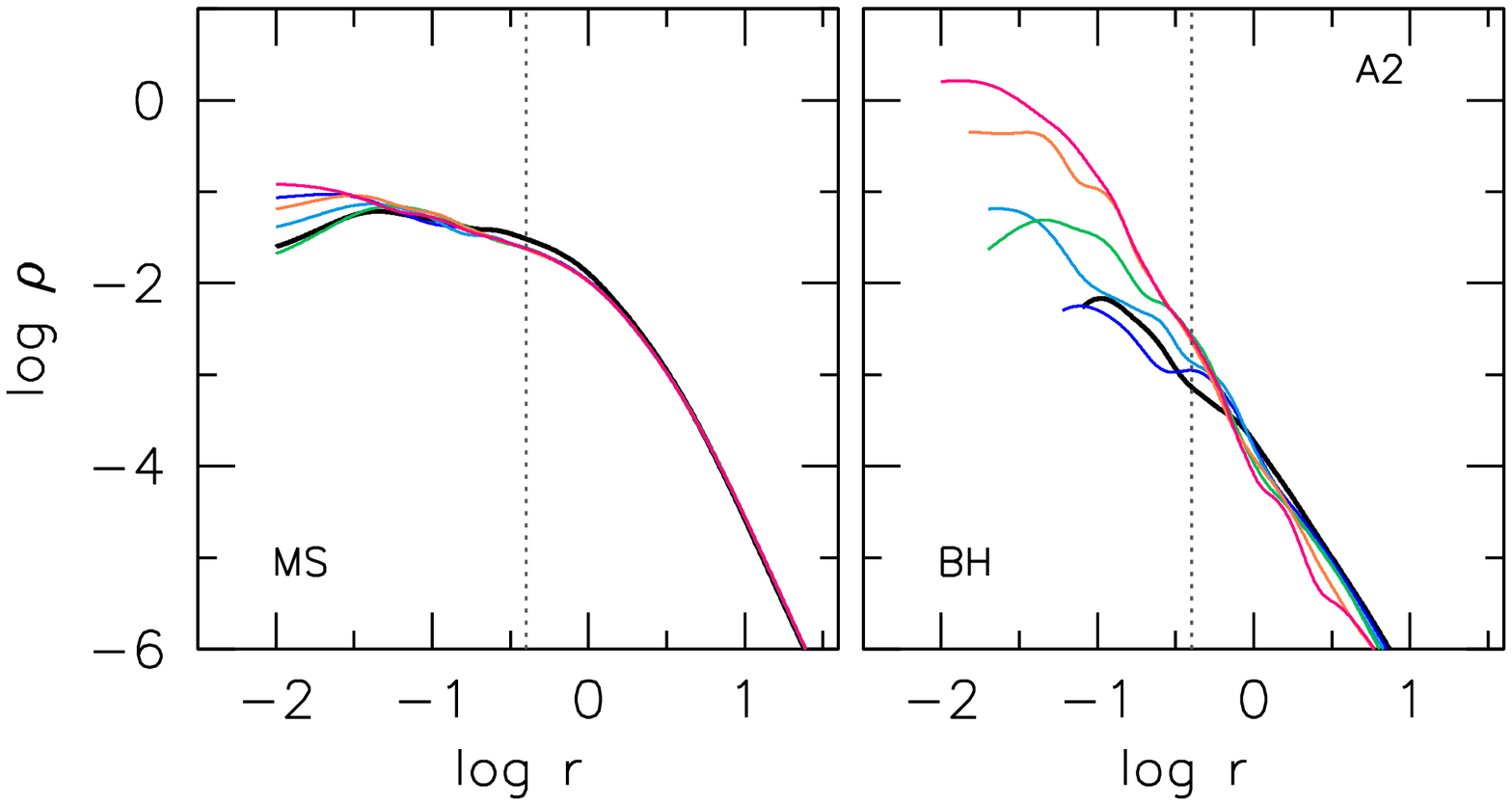}
    \includegraphics[width=8cm]{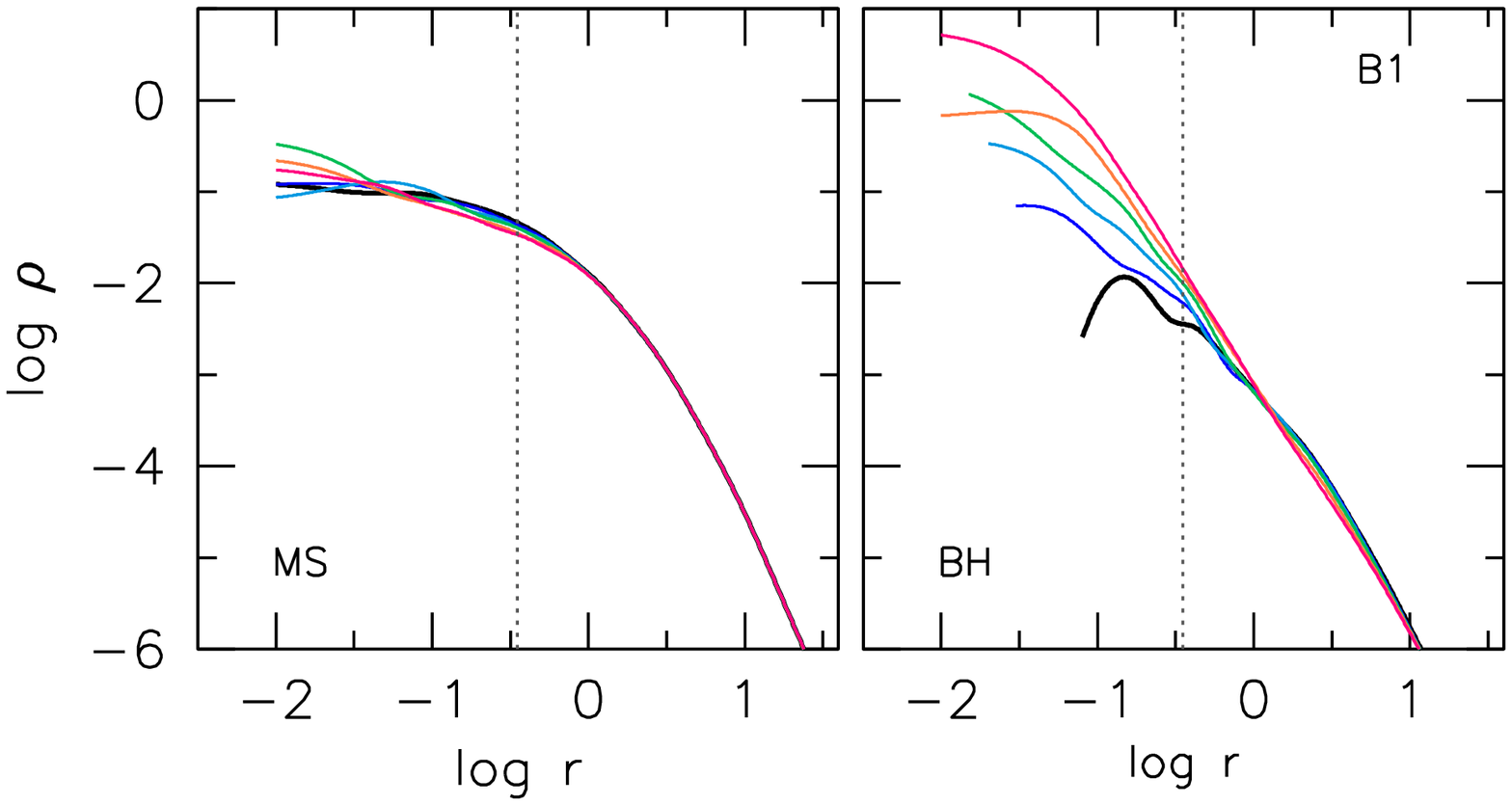}
    \includegraphics[width=8cm]{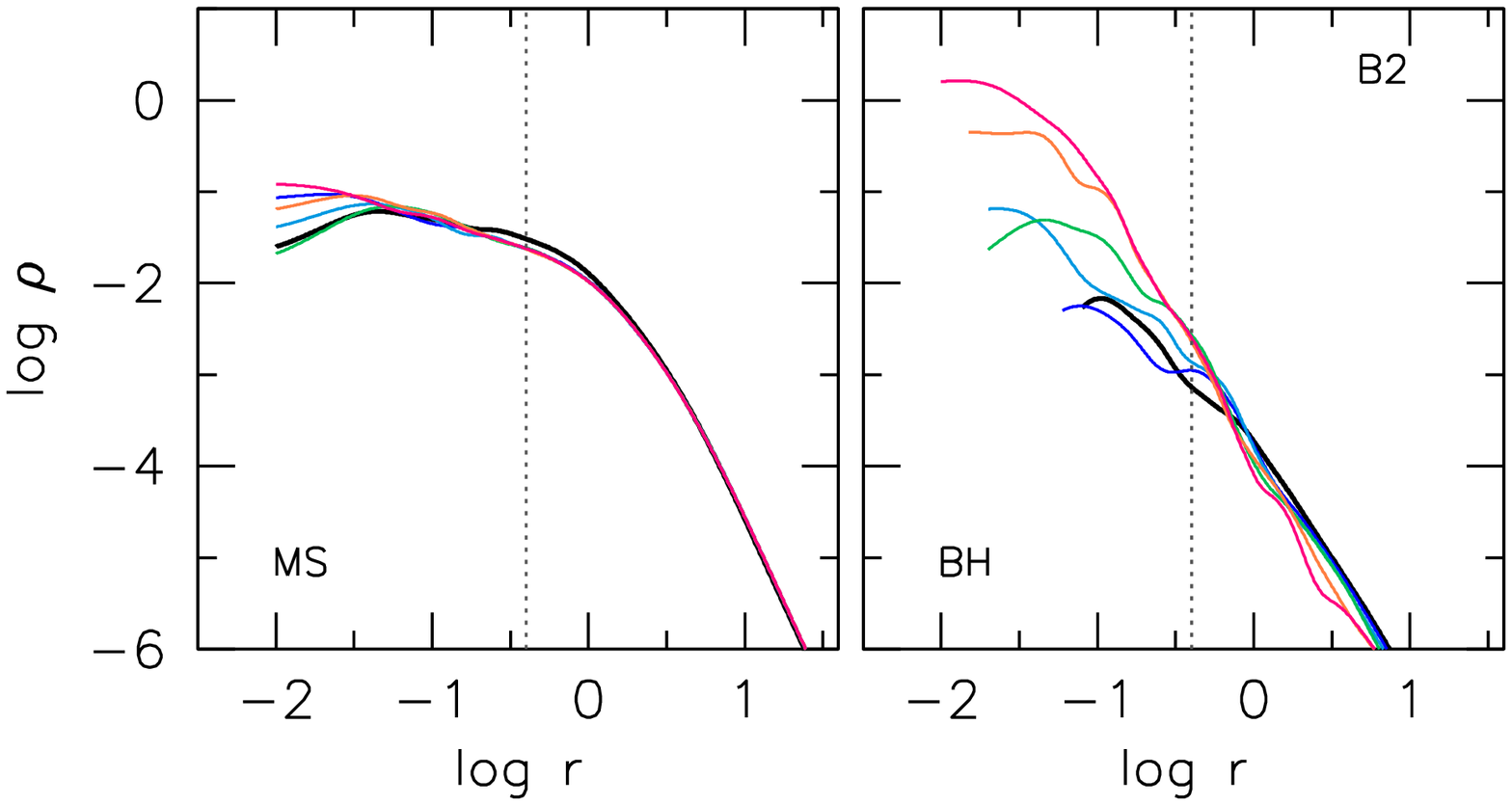}
  \end{center}
  \caption{Density profiles for MS stars and BHs at different times
    (0.2,0.5,1,2,3) $\tr$ during the post-merger phase. Thick solid
    lines indicate profiles at the beginning of the post-merger phase.
    Dotted lines indicate $\ri$.}
  \label{fig:profiles}
\end{figure*}

Figure\,\ref{fig:massp} shows the evolution of the enclosed mass in each
species in the post-merger integrations.  Within the influence sphere, the
general trend is for the mass in the lighter components to decrease
with time, while the mass in BHs increases.  The mass in NSs decreases
in models A1 and B1 while it appears nearly constant in models A2 and
B2.  In terms of mass density, Figure\,\ref{fig:profiles} shows that the BHs
quickly (on a time scale $\lap \tr(\ri)$) form the expected, steep
density profile, $\rho\sim r^{-2}$.  However, the MS stars maintain a
core-like profile.  Only at very small radii, $r\lesssim 0.2 \ri$,
does evolution toward a cusp appear to occur in the lighter species.

These results are reasonable.  In single-component models, a
Bahcall-Wolf cusp only extends outward to a fraction of $\ri$; one
would not expect the pre-existing MS core to disappear, since it extends
well beyond $\ri$, where the relaxation time is considerably longer
than its value at $\ri$.
Furthermore, heating by the heavier BHs should cause the mean
density of the lighter species to decrease with time, as observed.

However, at first blush, the results shown in
Figures\,\ref{fig:massp} and~\ref{fig:profiles} seem to contradict the
results described by \citet{PAS2010}, who used Fokker-Planck and
$N$-body integrations to follow the evolution of models with two mass
species and a MBH.  Those authors stated that ``mass
segregation...speeds up cusp growth [in the lighter component] by
factors ranging from 4 to 10 in comparison with the single-mass
case.''  
\citet{PAS2010} concluded that relaxation to a mass segregated, 
collisional steady state takes place in a time much less than the relaxation
time at the influence radius, hence that collisionally-relaxed models 
like that of \citet{HA06} should be a good description of nuclei like that
of the  Milky Way (in spite of the fact that the Milky Way is not observed
to contain a Bahcall-Wolf cusp in the stars).

The initial conditions adopted by \citet{PAS2010} were rather
different than in our post-merger models: our models have bona-fide
cores, while their initial models had density cusps, $\rho\sim
r^{-\gamma}$ with $\gamma=(1/2,1)$.  Nevertheless, the
``acceleration'' that they describe might be expected to occur also in
our models.

To understand the nature of this apparent discrepancy, we carried out
a number of separate $N$-body experiments, as well as Fokker-Planck
integrations.  The latter are presented in the next sub-section.
%%% figure 15 %%%
\begin{figure*}
  \begin{center}
    \includegraphics[width=16cm]{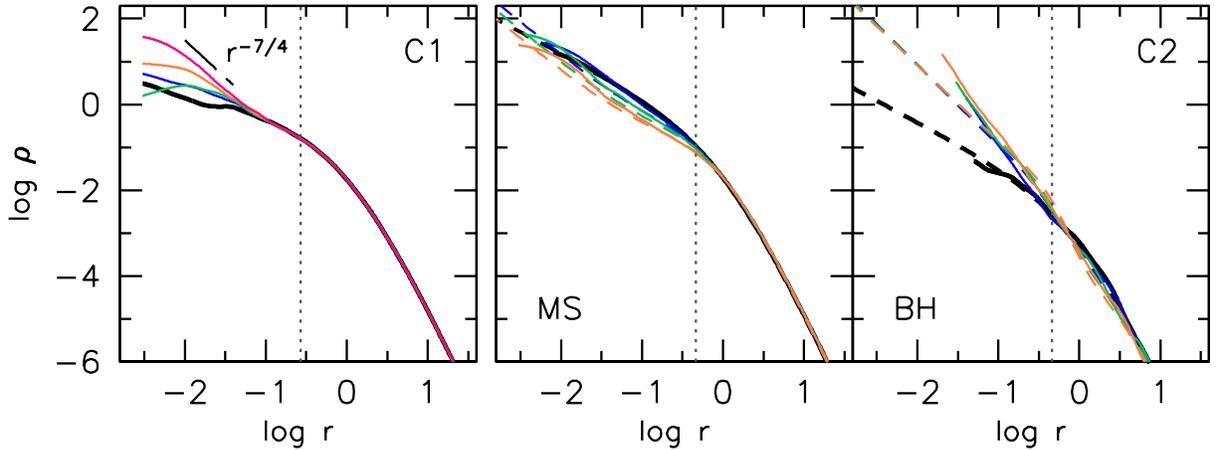}
  \end{center}
  \caption{Left: Density profiles in the single mass model C1 at
    different times: (0, 0.25, 0.5, 1, 2) $\tr(\ri)$. The thick line
    indicates the initial profile.  Right: Density profiles in the
    two-component model C2 at different times: $(0, 0.25, 0.5, 1)
    \tr(\ri)$.  Solid lines are from the $N$-body integrations while
    dashed lines are FP models. Vertical dotted lines indicate $\ri$
    in all panels.}
  \label{fig:C2}
\end{figure*}

We first checked whether our $N$-body code would reproduce the rate of
cusp formation observed by other authors in single-component models.
The left panel of Figure\,\ref{fig:C2} shows the evolution of model C1,
which had the same initial density profile and MBH mass as ``Run 1''
of \citet{preto2004}, and $N=131072$.  A $\rho(r) \sim r^{-7/4}$
density cusp forms in roughly one relaxation time at radii $r\lap
0.5\ri$, and the time dependence of the density is in excellent
agreement with what was found in that paper and in other $N$-body
studies \citep[e.g.][]{BME2004a}.  Additional experiments, varying the
softening length, time step and particle number, convinced us of the
robustness of this result.

The right panels of Figure\,\ref{fig:C2} show the results of
integrating a two-component model, C2, with the same initial
conditions as ``Run 1'' of \citet{PAS2010}, and $N=131072$.  Cusp
growth is clearly observed only in the heavier component, similar to
what we described above in the $N$-body integrations of the
four-component A and B.  Again in this case, additional experiments
using different integration parameters confirmed the results.

\subsection{Fokker-Planck models}
\label{sec:FP}

We carried out integrations of the isotropic Fokker-Planck
equation describing galaxies containing two stellar mass groups and a
MBH.  We used these integrations to address two questions.  (1) What
is the nature of the ``accelerated cusp growth'' that Preto \&
Amaro-Seoane observed in their Fokker-Planck integrations?  (2) Is the
evolution that we observe in our multi-component $N$-body models
consistent with the predictions of Fokker-Planck models?

We begin by summarizing the evolution equations.\footnote{These
  equations differ from the similar equations given by
  \citet{PAS2010}; the latter appear to be missing some multiplicative
  factors, as well as having an incorrect dependence of the diffusion
  coefficients on the $m_j$.  However, we believe that their numerical
  implementation was based on the equations in their correct form.  }

Let $f_i(E,t)$ be the phase-space number density of the $i$-th species
at time $t$ and (binding) energy $E$, where $E=-v^2/2+\psi$ and 
$\psi=-\Phi$ with $\Phi$ the gravitational potential.  
Let $i=1$ denote main-sequence stars,
of mass $m_1$, while $i=2$ denotes stellar-mass BHs, of mass $m_2$;
we set $m_2/m_1 = 10$ as in the $N$-body integrations.  The evolution
equation for the $i$-th component is
\begin{subequations}\label{Equation:FP}
\begin{eqnarray}
&&4\pi^2 p(E)\frac{\partial f_i}{\partial t} = -\frac{\partial F_i}{\partial E} \\
&&F_i = \sum_{j=1,2}\left(-{D_{EE}}_{ij}\frac{\partial f_i}{\partial E}
-{D_E}_{ij}f_i\right)\\\label{Equation:FPc}
&&{D_{EE}}_{ij} = 16\pi^3\Gamma m_j^2\bigg[q(E)\int_0^Ef_j(E',t)dE' +
\int_E^{\infty}f_j(E',t)q(E')dE'\bigg]\nonumber\\
&&\\
&&{D_E}_{ij} = -16\pi^3\Gamma m_im_j\int_E^\infty f_j(E',t)p(E')dE'
\end{eqnarray}
\end{subequations}
\citep{DEGN}.
Here, $p(E)$ and $q(E)$ are given by
\begin{subequations}
\begin{eqnarray}
p(E) &=& 4\int_0^{\psi^{-1}(E)} r^2 v(E,r) dr,\\
q(E) &=& \frac43\int_0^{\psi^{-1}(E)} r^2 v^3(E,r) dr
\end{eqnarray}
\end{subequations}
with $\psi^{-1}(E)$ the inverse of the potential function, $v(E,r) =
\sqrt{2\left[\psi(r)-E\right]}$, and $\Gamma=4\pi G^2\ln\Lambda$.  To
simplify the calculations, we made the assumption (as in numerous
earlier studies e.g. \citealt{preto2004, mhb07, PAS2010}) that the
gravitational potential, \beq \psi(r) = \frac{GM_\bullet}{r} +
\psi_\star(r), \eeq the sum of the stellar and MBH potentials, was
constant with time, and given by its value at $t=0$.  This is a
reasonable approximation at all radii: even if the stellar density
evolves significantly at $r\lap\ri$, the potential at these radii is
dominated by the MBH and is nearly unchanging.  At $r\gap\ri$ there is
no significant evolution in the density and the approximation is again
valid.
%%% figure 16 %%%
\begin{figure*}
  \begin{center}
    \includegraphics[angle=0,width=12.cm]{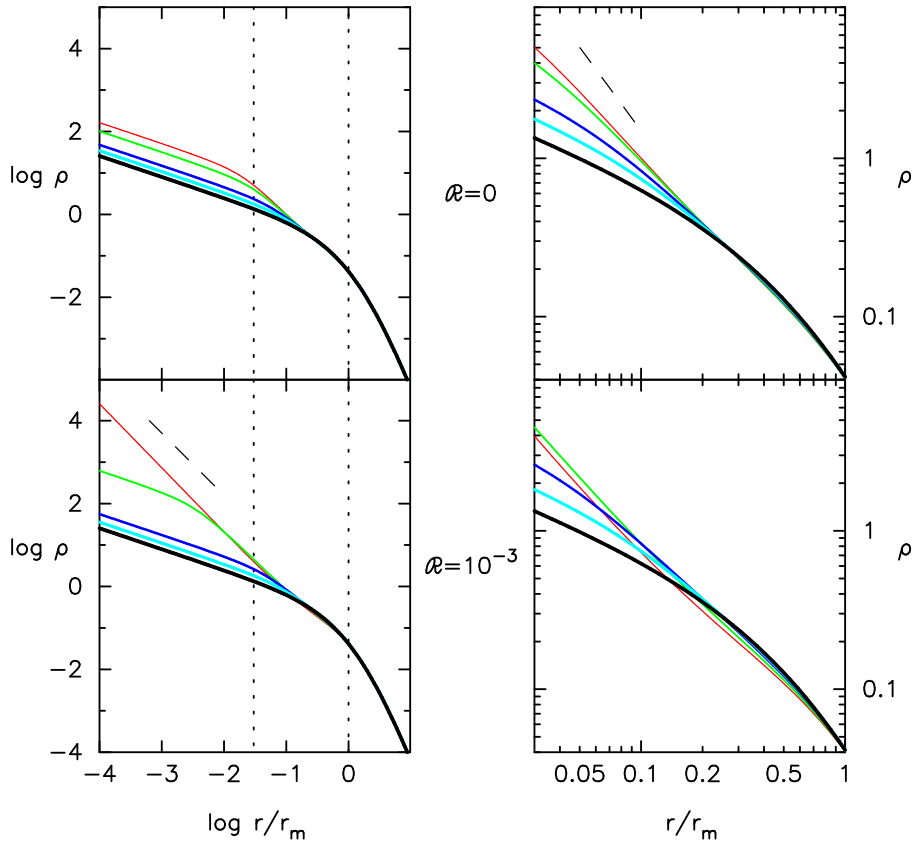}
  \end{center}
  \caption{Density profiles of the lighter (MS) component in
    Fokker-Planck integrations of two, two-component models; the
    heavier component, the BHs, are assumed to have masses ten times
    the mass of a MS star.  The initial conditions differ only in
    terms of the fraction of heavy particles:
    ${\cal R}=N_\mathrm{BH}/N_\mathrm{MS}=0$ (top) and $10^{-3}$
    (bottom).  Both components have $\gamma=0.5$ initially, a
    ``core,'' and $M_\bullet/\mg=0.05$.  The model with
    ${\cal R}=10^{-3}$ is the same model plotted in Figure 3 of
    \citet{PAS2010} , as an illustration of ``accelerated cusp
    growth;'' times shown are also the same as in that figure, i.e.,
    $t=(0,0.05,0.1, 0.2,0.25)$ in units of the relaxation time at the
    influence radius.  Panels on the left show the MS density profile
    at low spatial resolution, while panels on the right focus in on
    the region nearer $\ri$; the dotted lines on the left
    delineate the region plotted on the right and the dashed lines
    have logarithmic slopes of $-7/4$ (right) and $-3/2$ (left).  The
    accelerated cusp growth described by \citet{PAS2010} is seen to be
    present only at small radii, $r\lap 0.05\ri$; it is
    due to scattering of the MS stars by the BHs, which causes the
    initial ``hole'' in phase space to rapidly fill in.  At radii
    $r\gap 0.05 \ri$, adding the BHs has the opposite
    effect, resulting in a lower density of the MS component at all
    times.
  \label{fig:FP_g0.5}}
\end{figure*}

The coupled equations~(\ref{Equation:FP}) were solved by standard
techniques, starting from initial conditions in which the two
components had configuration-space densities 
\beq \rho_i(r) =
\rho_i(0)\left(\frac{r}{r_0}\right)^{-\gamma}
\left(1+\frac{r}{r_0}\right)^{\gamma-4} \eeq with the same
$(r_0,\gamma)$.  This is the same initial mass distribution adopted by
\citet{PAS2010}, who set $\gamma=(1/2,1)$.  Parameters defining the
initial conditions were $\gamma$; $M_\bullet/\mg$, the
ratio of MBH mass to the total mass in components 1 and 2; and the
number density ratio \beq {\cal R} \equiv
\frac{N_\mathrm{BH}}{N_\mathrm{MS}} .
\eeq As unit of time we adopted the relaxation time defined above,
evaluated at the MBH influence radius $\ri$, with $\ri$ defined as in
the $N$-body models; note that the value of $\log\Lambda$ becomes
irrelevant when the time is expressed in this way.

%%% figure 17 %%%
\begin{figure}
  \begin{center}
    \includegraphics[angle=0,width=8cm]{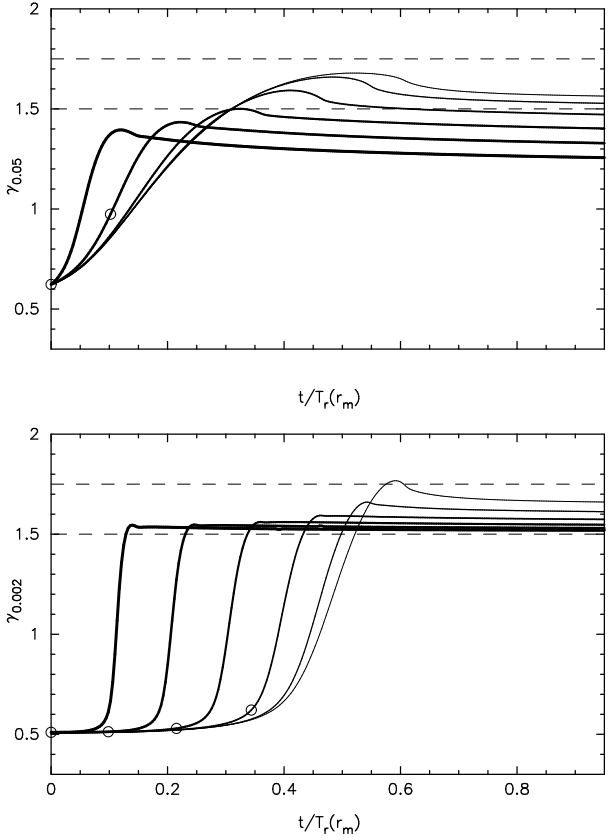}
  \end{center}
  \caption{Evolution of the local slope, $\gamma\equiv -d\log\rho/d\log r$,
    of the MS density profile, in two-component (MS + BH) Fokker-Planck models
    like those in Figure\,\ref{fig:FP_g0.5}.
    The curves differ in terms of the heavy-particle number fraction,
    $\log_{10}{\cal R}=\log_{10}(N_\mathrm{BH}/N_\mathrm{MS})=
    -7,-6,-5,-4,-3,-2$; increasing line width corresponds to
    increasing ${\cal R}$.
    Upper panel shows slopes at $r=0.05 \ri$, bottom
    panel at $r=0.002 \ri$.
    Open circles indicate the times at which 
    $\rho_\mathrm{BH}=0.1\rho_\mathrm{MS}$ at the respective radii.
    Adding a heavy (BH) component has relatively little effect on
    the growth of a Bahcall-Wolf cusp in the MS component (top panel),
    however it greatly affects the form of the density profile
    at $r\lap 0.01 \ri$ (bottom panel), 
    for the reasons discussed in the text.
    The latter phenomenon is the ``accelerated cusp growth''
    described by \citet{PAS2010}.}
  \label{fig:FP_g0.5_slopes}
\end{figure}

The right panels of Figure\,\ref{fig:C2} show results from a
Fokker-Planck integration of the same initial conditions used for the
$N$-body model C2 described above.  The agreement is very good.

Figure\,\ref{fig:FP_g0.5} shows integrations of two models both with
$\gamma=0.5$ and $M_\bullet/\mg=0.05$.  The models differ
in the fraction of heavy objects: ${\cal R}=0$ and
${\cal R}=10^{-3}$.  The model with ${\cal R}=10^{-3}$ is the
same model plotted by \citet{PAS2010} in their Figure 3.  We have
plotted only the density of the lighter (MS) component.

The superficial appearance of these plots depends strongly on the
radial range plotted.  If the range is sufficiently large,
extending to radii $\ll \ri$, a MS cusp with slope
$d\log\rho/d\log r\approx -3/2$ catches the eye in the model with BHs.
This is the feature emphasized by \citet{PAS2010}.  Viewed more
closely, in the region $0.02\lap r/\ri \lap 1$, the plots tell a
different story: the model including BHs exhibits a {\it lower}
density in MS stars at all times.  Excepting at very small radii,
addition of the BHs has the expected effect of reducing the density of
the MS stars.

\citet{PAS2010} attributed the ``accelerated cusp growth'' in the
lighter component to ``mass segregation,'' without stating explicitly
the connection between the two phenomena.  
In fact, as we now argue, the mechanism 
driving the evolution of the lighter component at small radii is
not mass segregation; it is scattering of the light component by
the heavy component \citep{mhb07,AH2009}.

Consider a two-component system; as before, species no. 1 is
the light (MS) component and species no. 2 is the heavy (BH) component.
The four diffusion coefficients that appear in the Fokker-Planck equation
for the light (MS) component scale with $m_i$ and $f_i$ as
\begin{eqnarray}
D_{EE_{11}} &\simeq& m_1^2 f_1\simeq m_1\rho_1,\ \ 
D_{E_{11}} \simeq m_1^2f_1 \simeq m_1\rho_1\nonumber \\
D_{EE_{12}} &\simeq& m_2^2 f_2\simeq m_2\rho_2,\ \ 
D_{E_{12}} \simeq m_1m_2f_2 \simeq m_1\rho_2\nonumber.
\end{eqnarray}
If $m_2\rho_2\gg m_1\rho_1$, $D_{EE_{12}} \gg D_{EE_{11}}$; in other
words, self-scattering is negligible compared with scattering off of BHs.
The first-order coefficients are smaller than $D_{EE_{12}}$ by factors
of $(m_1\rho_1)/(m_2\rho_2)$ and $m_1/m_2$ and can also be ignored.
The evolution equation for the light component becomes in this limit
\begin{equation}
\frac{\partial f_\mathrm{MS}}{\partial t}
\approx
\frac{1}{4\pi^2p}
\frac{\partial}{\partial E} 
\left(D_{EE}\frac{\partial f_\mathrm{MS}}{\partial E}\right)
\end{equation}
with $D_{EE}$ given by equation\,(\ref{Equation:FPc}) 
after setting $m_j=m_\mathrm{BH}$.
The steady-state solution is obtained by setting the term in
parentheses (the flux) to zero, yielding
\begin{equation}
\frac{\partial f_\mathrm{MS}}{\partial E} =0,\ \ \ \ 
f_\mathrm{MS}=\mathrm{const}
\end{equation}
independent of $f_\mathrm{BH}$.
A constant $f_\mathrm{MS}$ corresponds, in a point-mass potential,
to a density
\begin{equation}
\rho_\mathrm{MS} \propto r^{-3/2},
\end{equation}
which describes the MS cusp that appears, at early times and at small radii, 
in the Fokker-Planck models.

The manner in which this steady state is reached will depend on the initial
conditions.
In the models considered here, 
the initial MS density is $\rho_\mathrm{MS}\propto r^{-1/2}$,
which corresponds to an $f(E)$ that tends
to zero near the MBH -- a ``hole'' in phase space at low energies.
Scattering of the MS component by the BHs rapidly ``fills in'' this hole
as it drives $f$ toward a constant value.
The result is a sharp increase in the MS density at early times at the smallest radii.

The condition that the evolution of the light (MS) component be
dominated by scattering off the heavy (BH) component -- as opposed to
self-interactions, which tend to build a steeper, Bahcall-Wolf cusp -- is 
%$m_\mathrm{BH}^2f_\mathrm{BH}\gap m_\mathrm{MS}^2f_\mathrm{MS}$, i.e. 
$\rho_\mathrm{BH}\gap
(m_\mathrm{MS}/m_\mathrm{BH})\rho_\mathrm{MS} \approx
0.1\rho_\mathrm{MS}$.  In Figure\,\ref{fig:FP_g0.5_slopes}, the first
time at which this condition is satisfied is marked by open circles.
That figure shows evolution of the local slope,
$|d\log\rho_\mathrm{MS}/d\log r|,$ at two radii, $0.05 \ri$ and $0.002
\ri$, for several different values of ${\cal R}$, as computed via the
Fokker-Planck equation.  At the larger radius, the BHs remain a small
fraction of the total in most of these models, and the evolution of
the MS component is not strongly affected.  But because the
Bahcall-Wolf cusp builds ``from the outside in,'' its initial growth
is hardly reflected at much smaller radii.  Here, as the lower panel
shows, the dominant effect is scattering by BHs, and the effect of the
scattering on the MS density profile is strongly dependent on (roughly
proportional to) ${\cal R}$.

We summarize our findings in this section as follows.
\begin{itemize}
\item[1] One- and two-component Fokker-Planck models reproduce well 
the behavior seen in $N$-body integrations with the same initial
parameters.
\item[2] In two-component (MS + BH) models, addition of the heavy
(BH) component results in a slightly lower MS density 
at radii $r\gap 0.05\ri$, due to heating of the stars by the BHs.
\item[3] 
The same heating more promptly modifies the MS density profile at small radii,
$r\lap 0.05 \ri$, at least in models that start from a flat core.
The rate of growth of this ``mini-cusp'' is strongly dependent on the 
BH fraction.
\end{itemize}

The fact that the effects of scattering of MS
stars by BHs is restricted to  small radii, $r\lap 0.05\ri$, 
is consistent with the fact that we do not observe this phenomenon
in the $N$-body simulations: for instance, the density profiles of
Figure\,\ref{fig:profiles} extend down to only $\sim 0.05 \ri$.
Only a handful of particles were ever present at smaller radii.
At the radii resolvable by the $N$- body simulations, the Fokker-Planck 
models predict that the BHs should
{\it retard} the growth of the Bahcall-Wolf cusp in the MS component
(Figure\,\ref{fig:FP_g0.5}), 
not accelerate it, and this is what we see in the $N$-body simulations.

\subsection{Time dependence of the number of BHs near the MBH}

%%% figure 18 %%%
\begin{figure}
  \begin{center}
    \includegraphics[width=8cm]{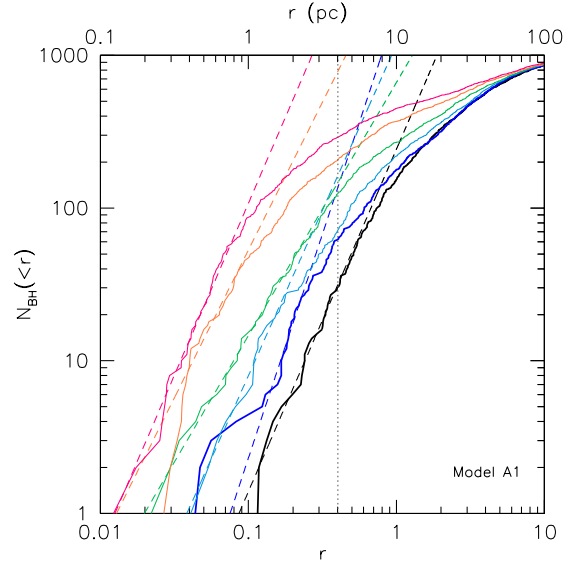}
  \end{center}
  \caption{Cumulative radial distribution of stellar-mass BHs (solid
    lines) in Model A1.  Different curves refer to different times:
    (0, 0.2, 0.5, 1, 2, 3) $T_r$ from right to left, as in
    Figure\,\ref{fig:profiles}. The dotted vertical line indicates
    $\ri$.  Dashed lines show fits to $N_{\rm BH} (<r)$ in the radial
    range [$r_1$, $r_2$] such that $N_{\rm BH} (r_1) = 5$ and $N_{\rm
      BH} (r_2) = 25$. }
  \label{fig:fitA1}
\end{figure}

%%% figure 19 %%%
\begin{figure}
  \begin{center}
    \includegraphics[width=8cm]{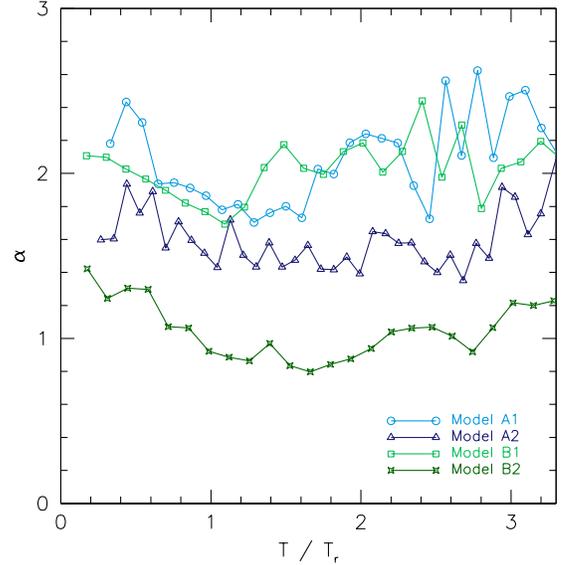}
  \end{center}
  \caption{Evolution of the slope, $\alpha \equiv -d\log
    N_\mathrm{BH}/d\log r$, derived from regression fits to the post-merger
    $N$-body data, as shown in Figure\,\ref{fig:fitA1} for Model A1.
    The corresponding density-profile slope is $\gamma=3-\alpha$.}
  \label{fig:fitslope}
\end{figure}

%%% figure 20 %%%
\begin{figure*}
  \begin{center}
    \includegraphics[width=8cm]{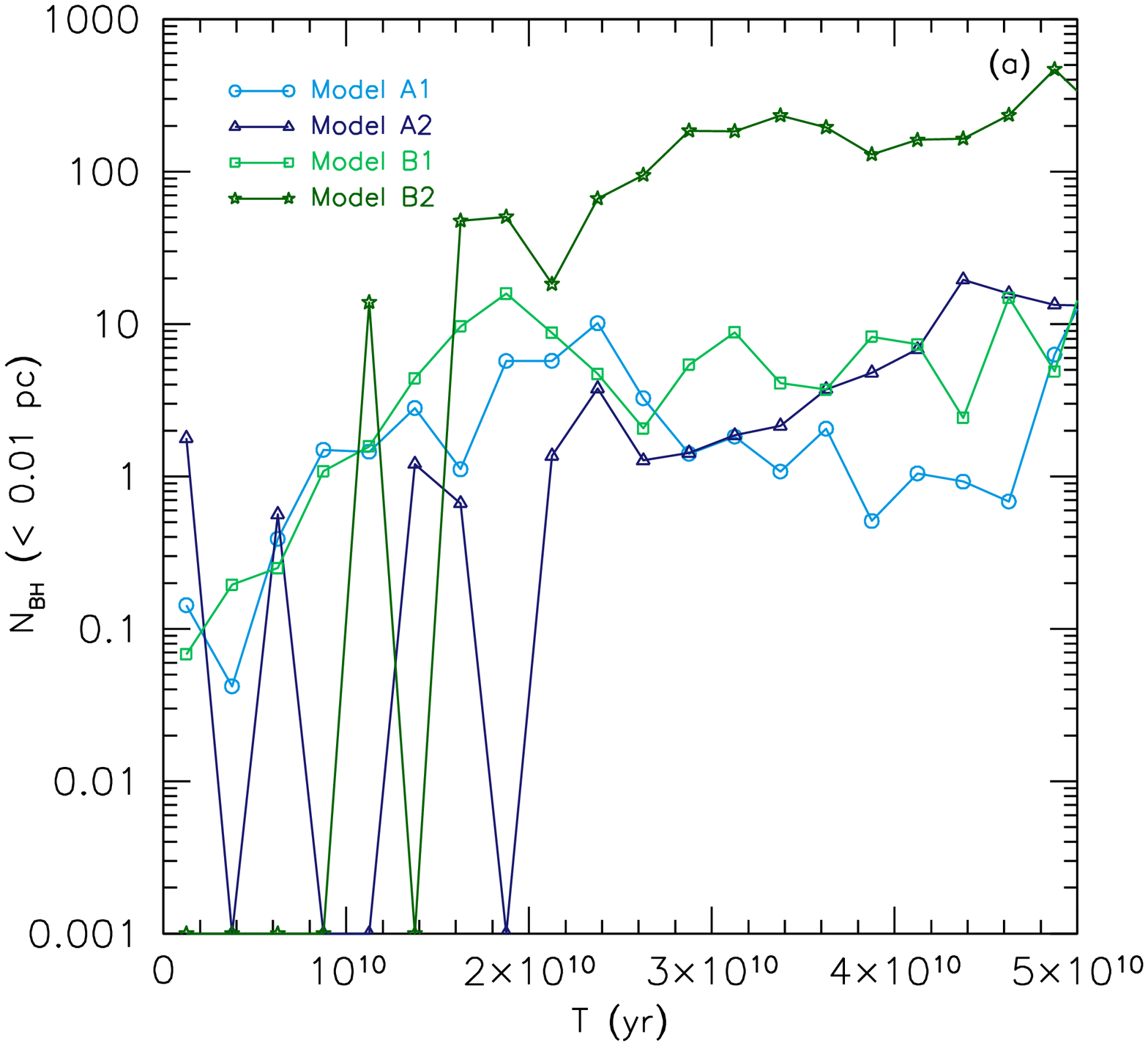}
    \includegraphics[width=8cm]{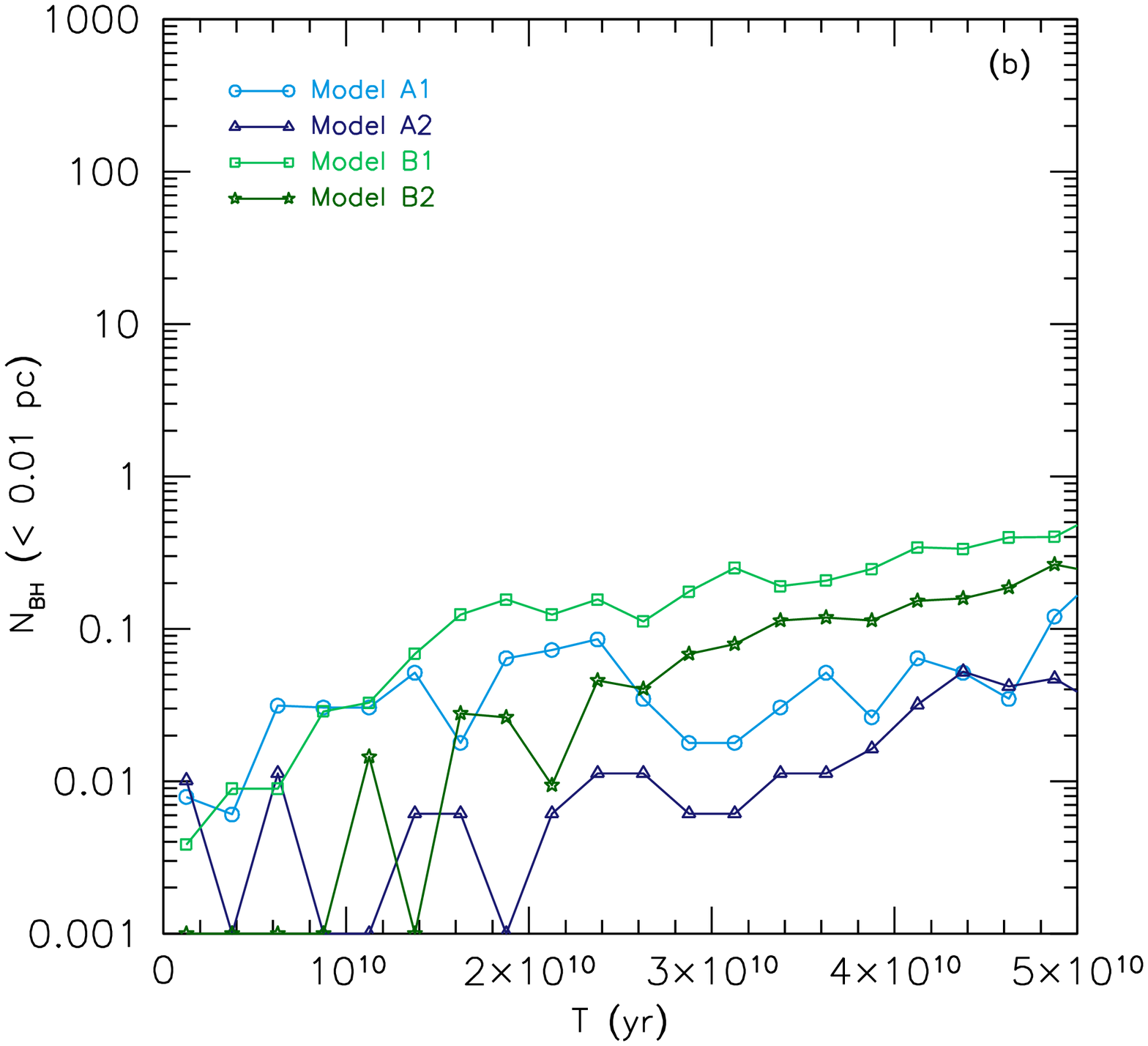}
  \end{center}
  \caption{Number of stellar-mass BHs within $0.01\pc$ versus time in
    our models, scaled to the Milky Way nucleus, assuming (a) a
    density profile of constant logarithmic slope at small radii,  and (b) a
    constant density inside $0.1\ri$.}
  \label{fig:BHs}
\end{figure*}

We now return to a discussion of the post-merger $N$-body
integrations.  Figure\,\ref{fig:massp} showed the evolution of the mass
enclosed within $\ri$ for each of the four species in each of the
$N$-body integrations.  In this section, we look more closely at how
the number of BH particles evolves with time on smaller scales.

Figure\,\ref{fig:fitA1} plots the cumulative radial distribution of BHs
in Model A1 at different times.
Time zero in this plot corresponds to the moment that the two,
MBH particles were combined into one.
The smallest radii at which there are any BH particles in the $N$-body models
decreases from
$\sim 0.1\ri$ at early times to $\sim 0.01\ri$ at late times. 
In order to estimate BH numbers at smaller radii, we carried out
regression fits of $\log N_\mathrm{ BH}$ to $\log r$. The
fits were performed in the radial interval [$r_1$, $r_2$] such that
$N_{\rm BH} (<r_1) = 5$ and $N_\mathrm{BH} (<r_2) = 25$.  
The resulting slopes, $\alpha\equiv |d\log N_\mathrm{BH}/d\log r|$,
are plotted in Figure\,\ref{fig:fitslope} for all models as a
function of time in units of the initial relaxation time.  We find
slopes $1 \lap \alpha \lap 3$ for $N_\mathrm{BH}(r)$, which imply
slopes in the mass density profile in the range $0\lap
-d\log\rho/d\log r \lap 2$.

Figure\,\ref{fig:BHs} shows the inferred number of BHs at $r<0.01\pc$
as a function of time.  In making this plot, we did a rough scaling of
our models to the nucleus of the Milky Way, assuming an influence
radius $\ri$ of $3\pc$.  The factor: \beq \frac{4\times
  10^6\msun}{10\msun} \times \frac{m_\mathrm{BH}}{\msbh} =
3.3\times10^3 \nonumber \eeq was used to convert the number of BH
particles in the simulations to the actual number; the first of these
factors contains masses in physical units, the second in the units of
the $N$-body code.  Since the number of BH particles at these radii is
small, we extrapolated inward under two assumptions: a density profile
of constant power-law index, and a constant density inside $r =
0.1\ri$.  In the former case, we used the slopes derived from the fits
to $N_\mathrm{BH}(r)$.  Since the fitted slopes are typically large,
the former assumption results in much larger, inferred numbers of BHs.

It is tempting to compare the numbers so obtained with estimates of
$N_\mathrm{BH}$ obtained from steady-state Fokker-Planck models of the
Milky Way nucleus \citep{HA06,Freitag06}.  Here we note one ambiguity
associated with such comparisons.  In the Milky Way, the two influence
radii $\rh$ and $\ri$ are similar.  The first is
\begin{equation}
\rh \equiv \frac{G\,M_\bullet}{\sigma^2}\approx 3.5\pc
\left(\frac{M_\bullet}{4\times 10^6\msun}\right)
\left(\frac{\sigma}{70\,\mathrm{km\ s}^{-1}}\right)^{-2}.
\end{equation}
The second is somewhat less certain, but dynamical estimates of the
mass distribution in the inner few parsecs
(e.g. \citet{SME2009,oh2009}) give $\ri \approx 2\pc$, consistent
within a factor of two with $\rh$.  The near-equality of $\rh$ and
$\ri$ in the Milky Way is due to the fact that the density profile is
similar to that of the singular isothermal sphere, $\rho\sim r^{-2}$;
the radius of the Milky Way's core is $\sim 0.5\pc$, substantially
smaller than both $\rh$ and $\ri$.  In our post-merger $N$-body
models, on the other hand, core radii and $\ri$ are both substantially
larger than $\rh$.  This fact precludes a unique scaling of our models
to the Milky Way -- at least in the Galaxy's current state.  (The
Milky Way's core may have been larger in the past
\citep{merritt2010}.)

With this caveat in mind, we assume that $\ri$ determines the scaling
of our models to the Milky Way.  Figure\,\ref{fig:BHs} is based on this
scaling.  In their collisionally-relaxed models, \citet{HA06} found
\beq N_\mathrm{BH}(r<0.01\pc) = 150.  \eeq Those authors assumed a
mass function with the same four species as in our models, and with
the same relative numbers as in our models A2 and B2
(Table\,\ref{tab:models1}).  In Figure\,\ref{fig:BHs}, the inferred
number of BHs inside $0.01\pc$ is very uncertain at the relevant
times, i.e. $t\approx 10^{10}$ yr, fluctuating between zero, and
maximum values of $\sim 1$ (Model A2) and $\sim 10$ (Model B2).  These
upper limits are factors of $\sim 10^2$ and $\sim 10^1$, respectively,
smaller than in the \citet{HA06} models.  There is an independent way
to reach a similar conclusion.  After several relaxation times,
$N_\mathrm{BH}(<0.01\pc)$ is $\sim 10$ (Model A2) and $\sim 100$
(Model B2).  These numbers, presumably representing the mass
distribution in a near steady-state, are $\sim 10$ times larger than
their values at $t=10^{10}$ yr.

It is interesting that the BH distribution takes so long to reach this
(nearly) steady state -- a time of at least twice the relaxation time
as defined by the dominant (MS) population
(Figure\,\ref{fig:profiles}).  This may be due in part to the fact
that the MS distribution is also continuously evolving.  Another
reason is the persistence of a core in the dominant component.
Chandrasekhar's (1943) dynamical friction coefficient, 
in its most widely-used form \citep{bt87},
predicts that the frictional force near a MBH drops essentially to
zero if $\rho(r)$ increases more slowly than $r^{-1/2}$ toward the
center.  The (isotropic) phase-space density $f(E)$ corresponding to
that density profile has no stars moving more slowly than the local
circular speed, and Chandrasekhar's formula identifies the frictional
force exclusively with field stars moving more slowly than the massive
body.  This property of the dynamical friction force was automatically
incorporated into the Fokker-Planck calculations presented above,
since that equation is based on Chandrasekhar's coefficients in their
standard forms \citep{RMJ57}.  In reality, some part of the frictional
force in the $N$-body simulations will come from stars moving faster
than the test star, and including the contribution from these stars
keeps the frictional force from falling identically to zero \citep{AntoniniMerritt2011}.  Nevertheless, one expects dynamical friction
to act more slowly in these models than expected based on the
application of standard formulae that neglect the special properties
of $f$.

\section{Shape and kinematics of the stellar spheroid}
\label{sec:shape}

%%% figure 21 %%%
\begin{figure*}
  \begin{center}
    \includegraphics[width=8cm]{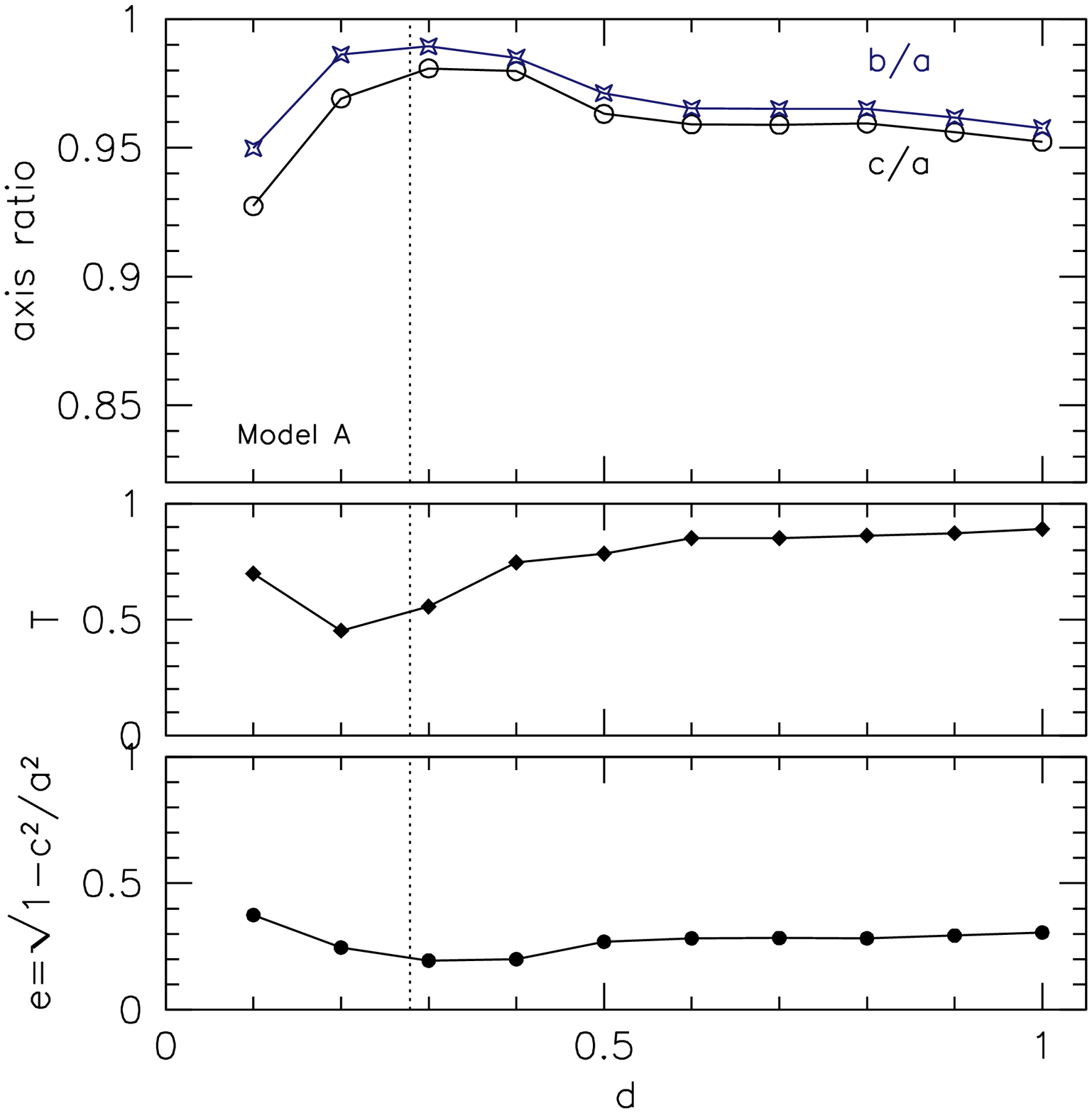}
    \includegraphics[width=8cm]{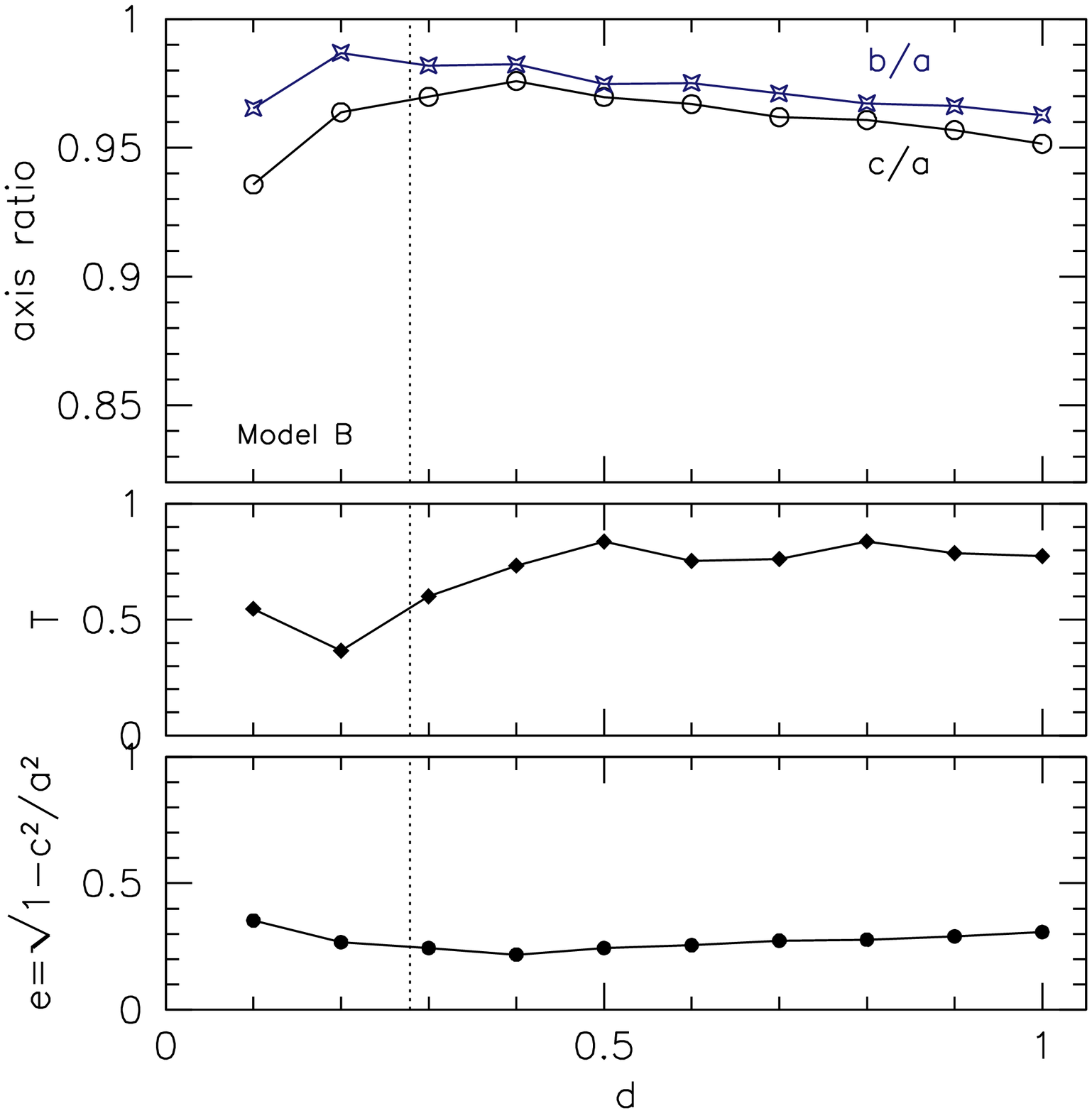}
  \end{center}
  \caption{Axis ratios, triaxiality parameter and ellipticity for
    models A (left) and B (right) as a function of distance from the
    center of the binary, at the time when $a \sim a_h$. The vertical
    lines indicate $\ri$.}
  \label{fig:axis}
\end{figure*}

\begin{figure}
  \begin{center}
    \includegraphics[angle = -90, width=8cm]{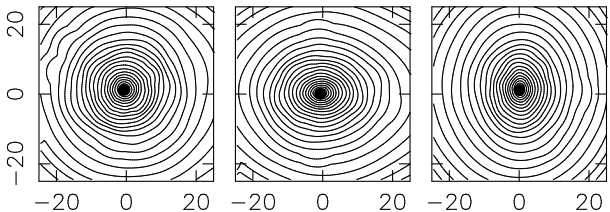}
  \end{center}
  \caption{Projected density contours for Model B1 at the beginning of
    the post-merger phase. Left: $xy$ plane. Middle: $xz$ plane. Right: $zy$
    plane. The position of the BH is indicated by the filled
    circle. The merger remnant is flattened in the direction
    perpendicular to the plane of the merger.}
  \label{fig:isophotes}
\end{figure}

We determined the shape of the merger remnants during the binary
hardening phase by computing the axis ratios at different distances
from the center. We followed the procedure described in
\citet{katz1991} and \citet{ant2009}.  We selected all particles
within a sphere of radius $d$ centered on the binary center of
mass. The axis ratios were then determined from the eigenvalues
$I_{ii}$ of the inertia tensor $I$ as
\begin{equation}
\xi = \sqrt{I_{11} / I_{\rm max}}, \quad \eta = \sqrt{I_{22} / I_{\rm
    max}}, \quad \theta = \sqrt{I_{33} / I_{\rm max}} 
\end{equation}
where $I_{\rm max} = \rm max \{I_{11}, I_{22}, I_{33}\}$.
New axis ratios were computed considering only particles
enclosed in the ellipsoidal volume having the previously
determined ratios, i.e. all particles satisfying the condition
$q_i < d$, where
\begin{equation}
q_i^2 = \left(\frac{x_i}{\xi}\right)^2
+\left(\frac{y_i}{\eta}\right)^2 +  \left(\frac{z_i}{\theta}\right)^2 \,.
\end{equation}
The last step was iterated until the axis ratios converged.  If we
define the axes $a,b,c$ such that $a > b > c$, we find that $c/a$ and
$b/a$ correspond, respectively, to the minimum and intermediate values
of $\xi, \eta, \theta$.  From the axis ratios it is possible to define
a triaxiality parameter
\begin{equation}
T = \frac{a^2-b^2}{a^2-c^2} \,,% = \frac{1-(b/a)^2}{1-(c/a)^2}\,,
\end{equation}
$0< T < 1$, such that $T=0$ for an oblate spheroid, $T=1$ for a prolate
spheroid and $T=0.5$ corresponds to maximum triaxiality.  The
ellipticity $e = \sqrt{1 - c^2/a^2}$ measures the degree of flattening
of the system.  The axis ratios, triaxiality parameter and ellipticity
for models A and B are shown in Figure\,\ref{fig:axis} as a function of
distance from the binary center of mass, at the time when the binary
becomes hard. A moderate triaxiality is present in both models at
distances of the order of $1-2~\ri$ from the binary center of mass.
At larger distances, the models appear axisymmetric with a small
flattening in the direction perpendicular to the binary orbital plane.
These features persist throughout the hardening phase.  The
flattening, due to rotation of the merger remnant, is also visible in
the isophotes shown in Figure\,\ref{fig:isophotes}, which are computed
at the beginning of the post-merger phase.

%%% figure 22 %%%

%%% figure 23 %%%
\begin{figure*}
  \begin{center}
    \includegraphics[width=5.5cm]{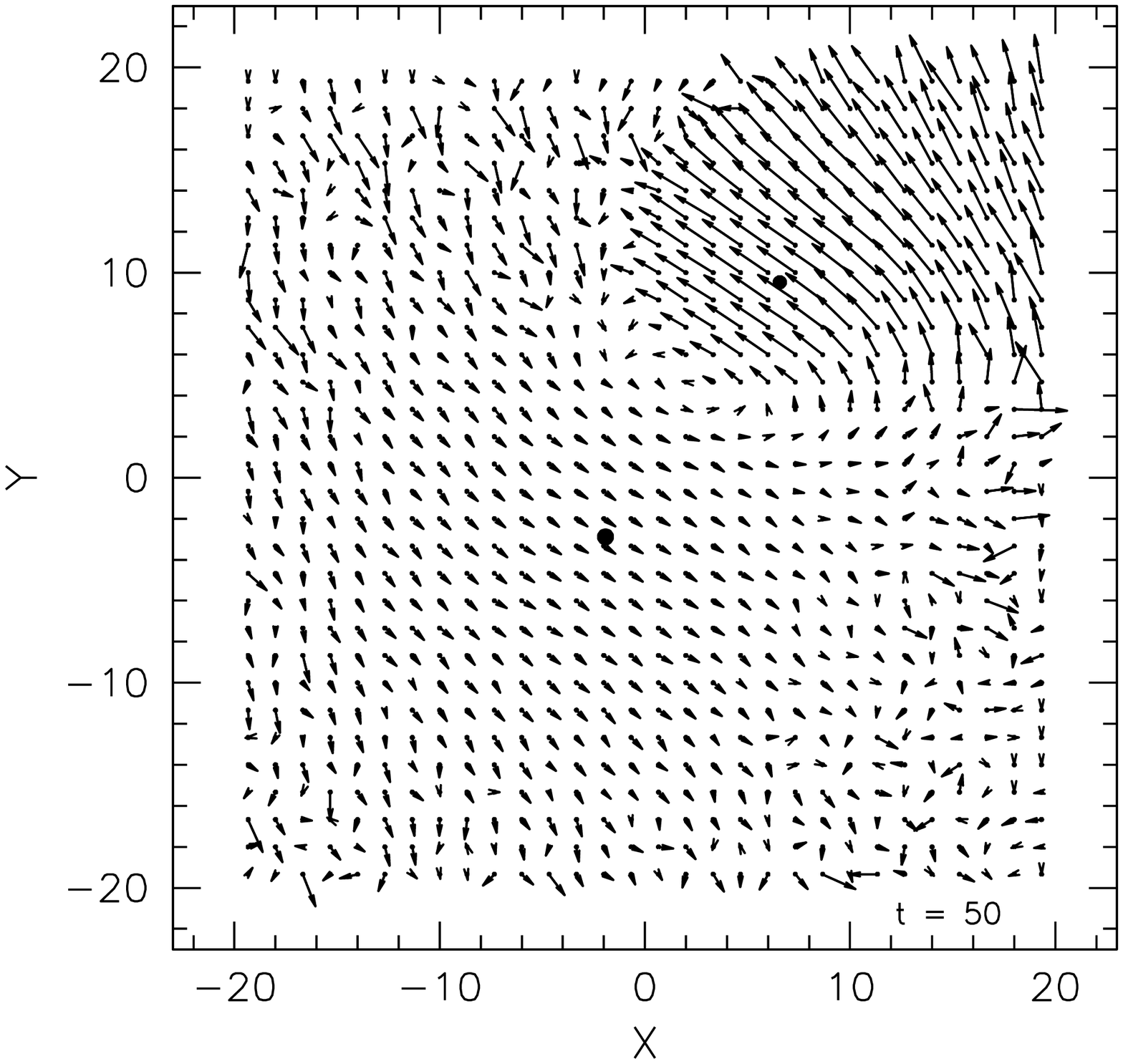}
    \includegraphics[width=5.5cm]{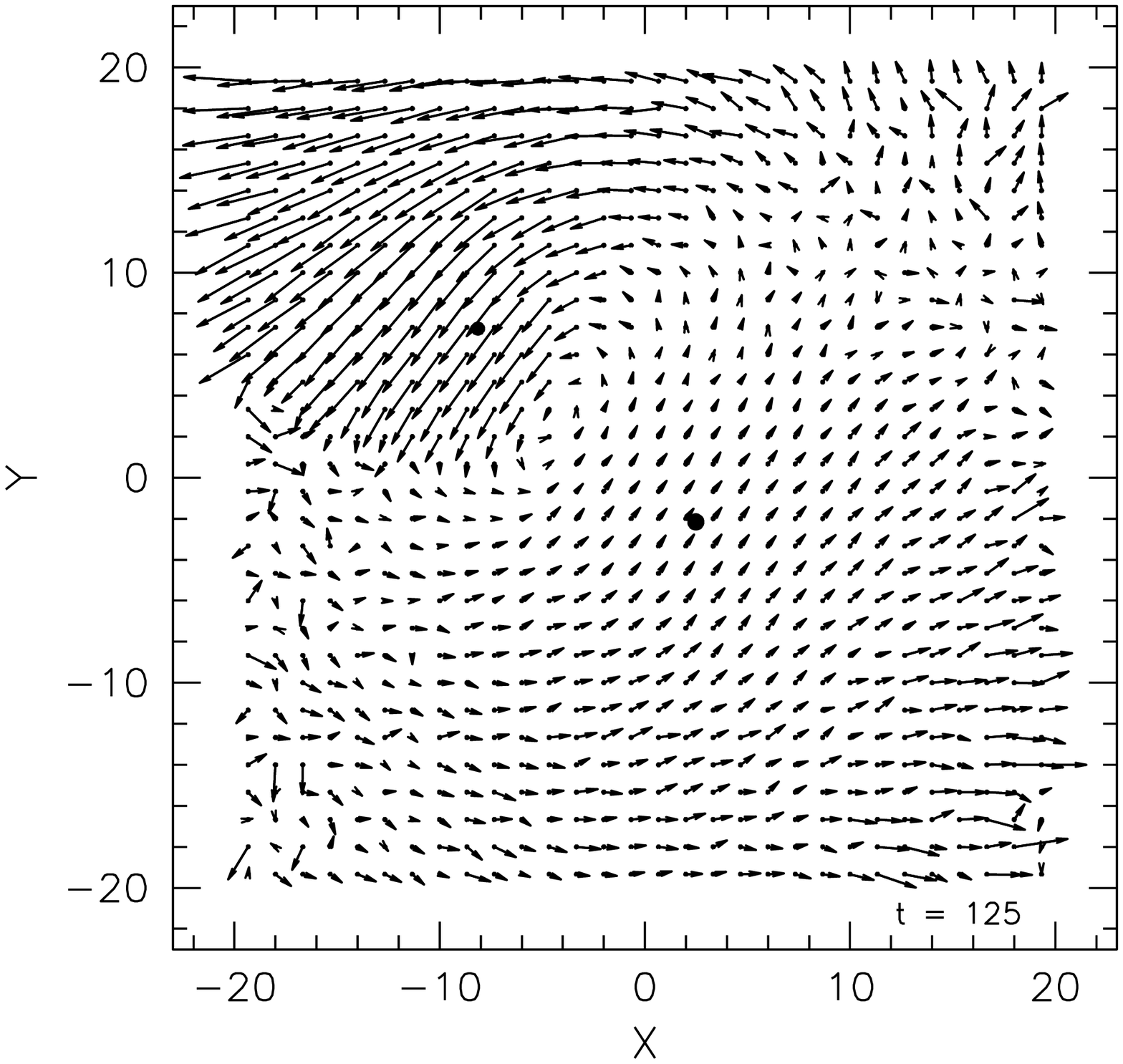}
    \includegraphics[width=5.5cm]{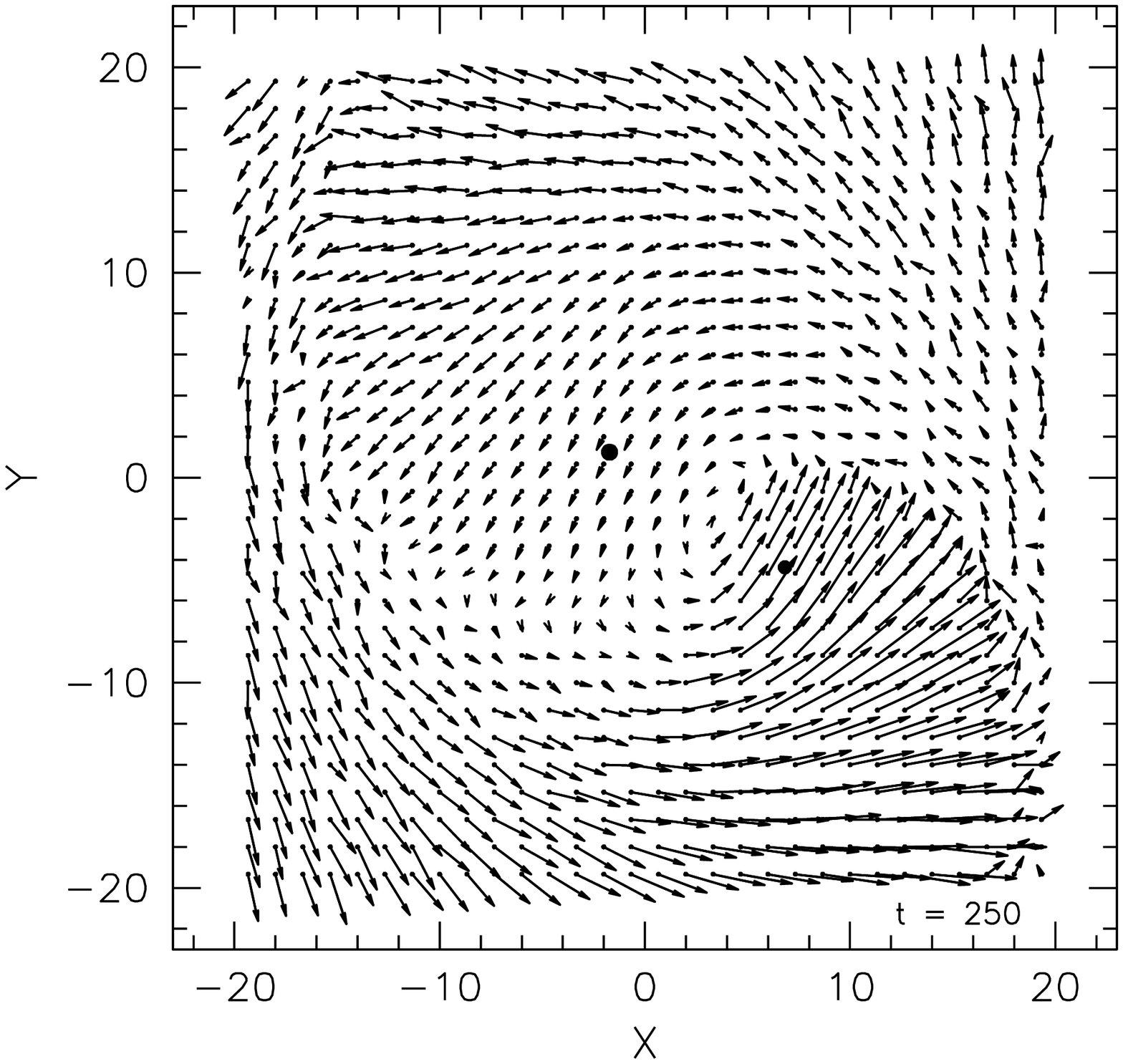}
    \includegraphics[width=5.5cm]{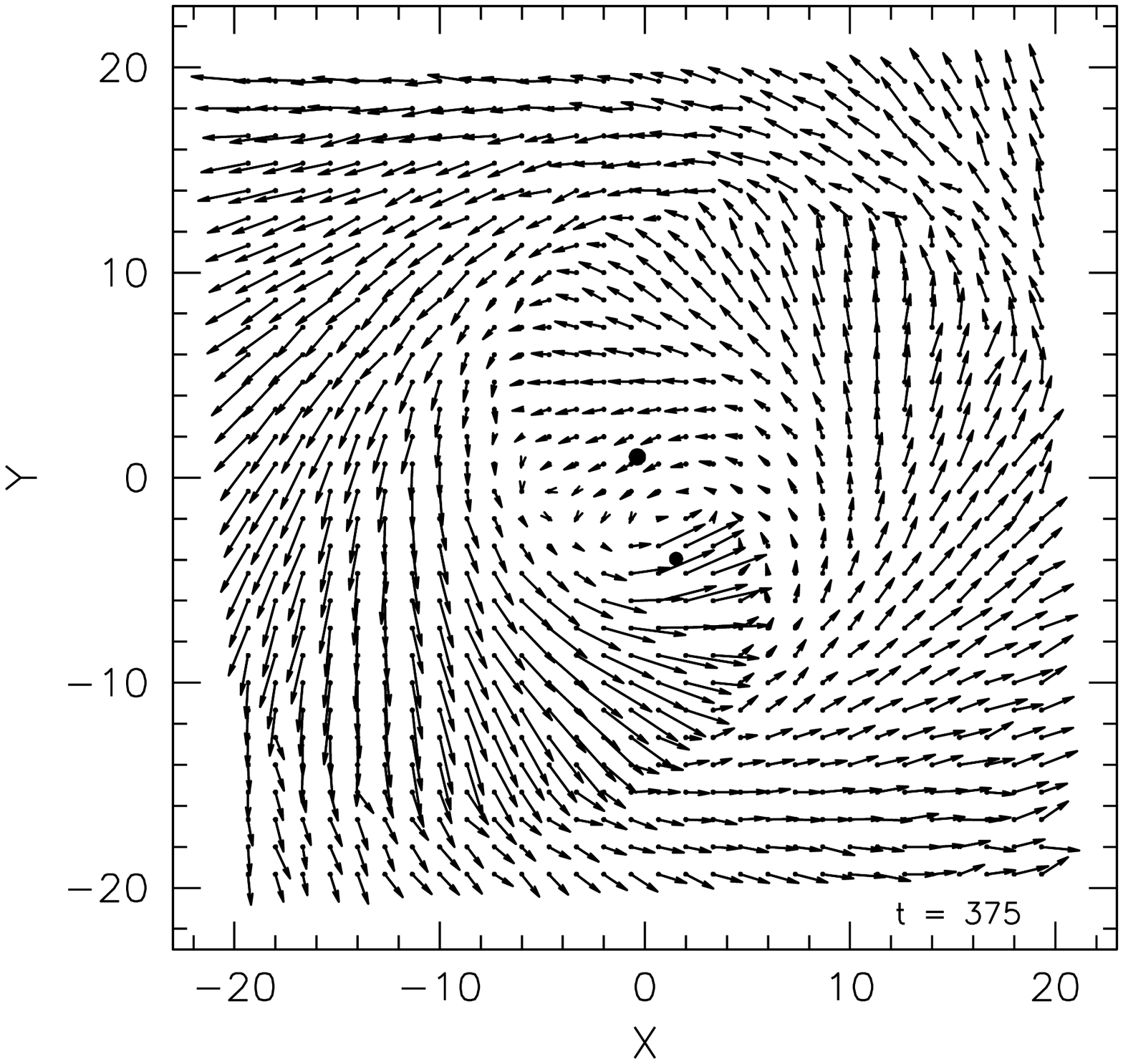}
    \includegraphics[width=5.5cm]{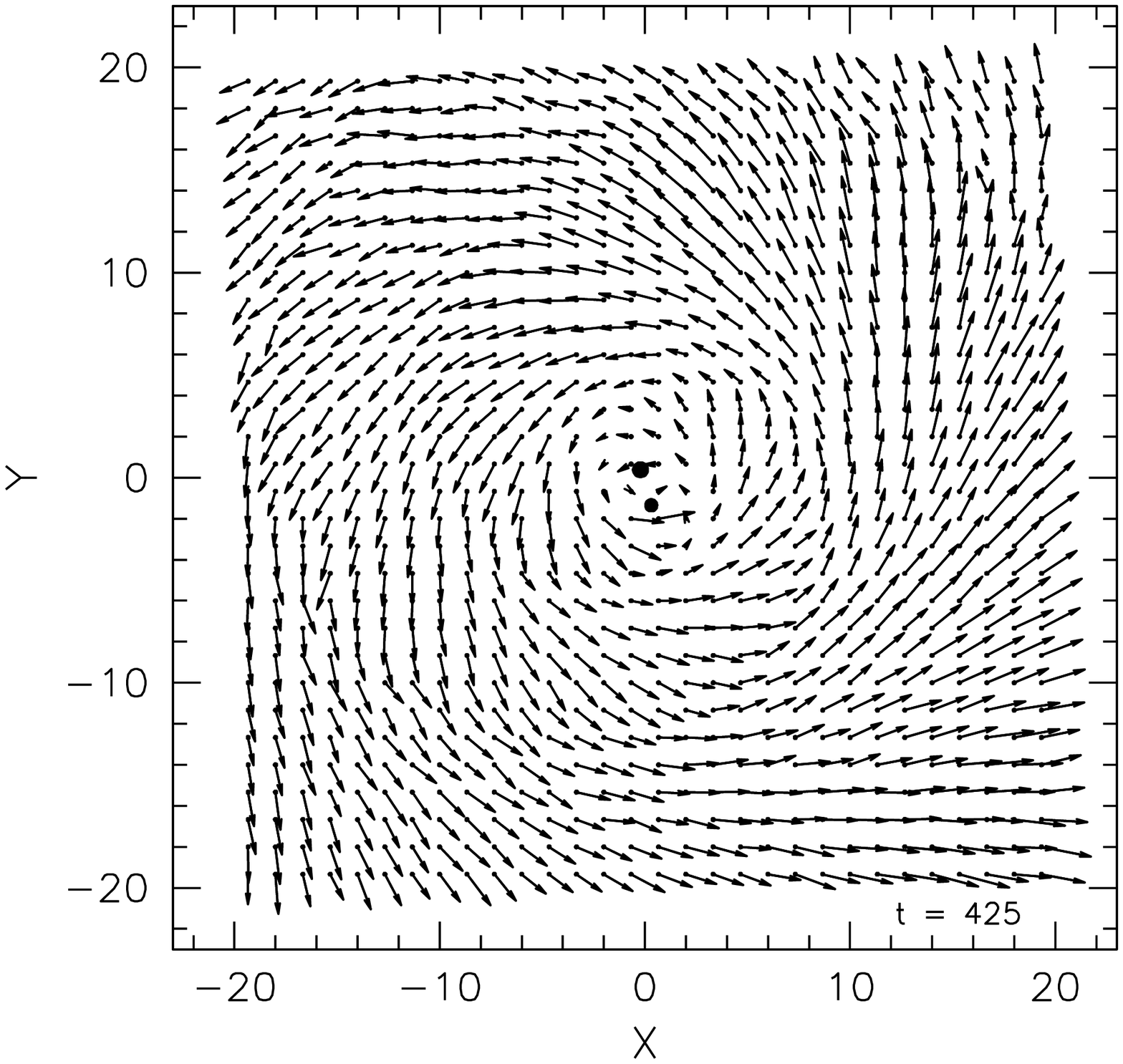}
    \includegraphics[width=5.5cm]{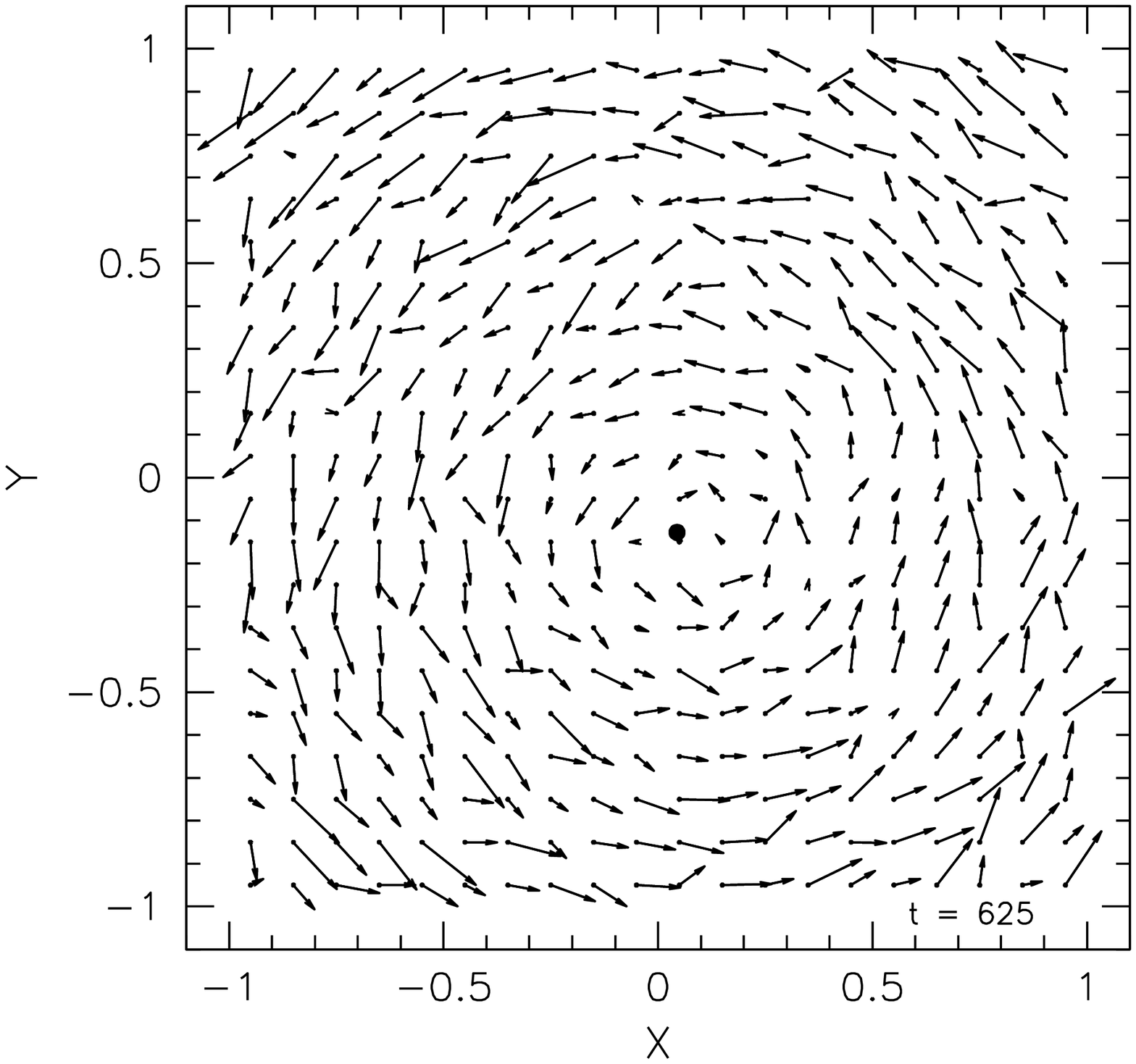}
  \end{center}
  \caption{Velocity vectors in the orbital plane ($x-y$ plane) of the
    binary at different times during the merger of Model A ($t=50,
    125, 250, 375, 425, 625$, from the top left to the bottom right).
    The origin corresponds to the initial position of the binary
    center of mass. The velocity vectors represent the in-plane
    velocities of all stars in the chosen area. The circles represent
    the locations of the massive black holes. The final frame is shown
    at higher resolution.}
  \label{fig:vel}
\end{figure*}

Rotation is introduced by the merger process, as illustrated by the
velocity map in Figure\,\ref{fig:vel} for Model A. The figure shows the
direction and magnitude of the velocities in the binary's orbital
plane. By the time the binary reaches the hard binary separation, a
well defined rotation pattern has been established.

The departures from spherical symmetry that we observe in the
merger models are probably responsible for the efficient hardening
of the massive binary \citep{merrittpoon2004,ber06}.

\section{Summary and Discussion}
\label{sec:disc}

\subsection{(Re-)growth of Bahcall-Wolf cusps in multi-component systems}
In galaxies containing a single stellar population, a $\rho\propto
r^{-7/4}$ \citet{BW76} cusp is expected to appear at radii $r\le
r_\mathrm{BW}\approx 0.2 \ri$ around the MBH.  While the growth time
of the cusp is dependent on the initial conditions, simulations
starting from a shallow cusp inside $\ri$ show that the stellar
density will have reached an approximate steady state after roughly
one relaxation time at $\ri$.
We presented an example of such evolution in Figure\,\ref{fig:C2}.

In nuclei with two mass groups, e.g. solar-mass stars (MS) and
$10\msun$ black holes (BHs), evolution toward a steady state near
the MBH depends on the relative numbers and masses in the two groups,
as well as on the initial conditions.  The results of our $N$-body and
Fokker-Planck integrations of two-component models were presented in
Section~\ref{sec:FP}.  Figure\,\ref{fig:FP_g0.5_slopes} showed that
addition of the BHs reduces slightly the rate of formation of a
Bahcall-Wolf cusp in the MS (light) component, and also affects the
final value of the density-profile slope, which varies from
$\rho_\mathrm{MS}\sim r^{-7/4}$ when
$\rho_\mathrm{BH}/\rho_\mathrm{MS}$ is small, to $\sim r^{-3/2}$ as
$\rho_\mathrm{BH}/\rho_\mathrm{MS}$ is increased \citep{BW77}.  In
addition, if the initial MS distribution is very flat near the MBH,
scattering by the heavier BHs can dominate the evolution of the MS
component at early times for $r\lap 0.05 \ri$, converting
$\rho_\mathrm{MS}\sim r^{-1/2}$ to $\rho_\mathrm{MS}\sim r^{-3/2}$ at
these small radii, even before the Bahcall-Wolf cusp has fully formed
at larger radii.

The full galaxy merger simulations presented here allowed us, for the
first time, to evaluate the evolution of multi-component nuclei
starting from initial conditions that were motivated by a well-defined
physical model.  Our pre-merger galaxies contained mass-segregated
nuclei with four mass components, representing an evolved stellar
population.  These initial distributions were modified both by the
galaxy merger, and by the formation of a MBH binary, which created a
large core in each of the components (Figures\,\ref{fig:lagr1},
\ref{fig:density}).  The core radius of the heaviest (BH) component
was somewhat smaller than the cores in the three lighter components, a
relic of the earlier mass segregation, and of the incomplete cusp
destruction process during the binary MBH phase
(Figure\,\ref{fig:density}).  Evolution during the merger phase set the
``initial conditions'' of the nucleus at the time when the two MBHs
were combined into one.  Unlike the rather ad hoc initial conditions
used in many earlier studies of nuclear evolution
\cite[e.g.][]{Freitag06,PAS2010}, our post-merger models have cores
that should be reasonable representations of the cores in real
galaxies that formed via dissipationless mergers, with mass densities
that decline toward the center and anisotropic kinematics that reflect
the action of the massive binary on the stellar orbits
(Figure\,\ref{fig:vel}).

By continuing the evolution of these multi-component models for a time
greater than $T_r(\ri)$ after coalescence of the two MBHs, we found that
the cores characterizing the distribution of the dominant, MS
component persisted; in fact the MS density at radii $\sim \ri$
decreased gradually with time (Figure~\ref{fig:massp}) -- 
a predictable consequence of the
initial extent of the cores, several times $\ri$, implying
$T_r(r_c)> T_r(\ri)$, and of continued ``heating'' by the heavier
BHs.  Nevertheless, a Bahcall-Wolf cusp gradually reformed in the
lighter components at radii $r\lap 0.2\ri\ll r_c$; growth times were
found to be  $\gap T_r(\ri)$, somewhat greater than in earlier
models with more idealized initial conditions
\cite[e.g.][]{Freitag06}.

Our pre-merger galaxies (models A1, B1) contained larger numbers of
remnants than would be expected based on a standard IMF; this was done
in order to better resolve the distributions of those components near
the MBH.  We tested the dependence of the {\it post}-merger evolution
on the assumed mass function by carrying out a second set of
integrations in which we decreased the relative numbers of remnants
(NS, WD, BH) by factors of a few, to values more consistent with
standard IMFs (models A2, B2).  Evolution of the dominant, MS
component in the latter models differed only modestly from its
evolution in the models with larger remnant fractions; the main
difference was a lower rate of core expansion reflecting a lower rate
of heating by the BHs (Figure\,\ref{fig:massp}).  The rate of growth of
the Bahcall-Wolf cusp in the MS component was essentially unchanged
(Figure\,\ref{fig:profiles}).

Our results on the regeneration of Bahcall-Wolf cusps following
dissipationless mergers are consistent with those obtained in
simulations of single-component galaxies \citep{MS06}.  We can
summarize these results by stating that regrowth of a cusp in the
dominant stellar component requires a time comparable with, or
somewhat longer than, the relaxation time of that component measured
at the MBH influence radius.  The new simulations presented here
suggest that time scales for cusp regrowth are only weakly dependent
on the number of heavy remnants (BHs), at least if the fraction of
mass in the heavier population does not exceed a few percent of the
total.

The Milky Way nucleus is near enough that a Bahcall-Wolf cusp could be
resolved if present, and the dominant stellar population is believed
to be old.  Since $\rh\approx \ri\approx 2-3\pc$ in the Milky Way, the
expected, outer radius of the Bahcall-Wolf cusp is
$r_\mathrm{BW}\approx 0.5\pc \approx 10''$.  As is well known, number
counts of the dominant, old stellar population show no evidence of a
rise in density at this radius; instead the number counts are flat, or
even falling, from $\sim 10''$ into at least $1''$ projected radius
\citep{buchholz09,Do09,Bartko10}.

Assuming solar-mass stars, the relaxation time at the influence radius
of Sgr A$^*$ is 20-30 Gyr \citep{merritt2010}, while the mean stellar
age in the nuclear star cluster is estimated to be $\sim 5$ Gyr
\citep{figer2004}.  The time available for formation of a cusp is
therefore $(5/25) T_r(\ri)\approx 0.2T_r(\ri)$.  Our simulations
(e.g. Figure\,\ref{fig:profiles}) suggest that a Bahcall-Wolf cusp in
the dominant component is unlikely to have formed in so short a time,
and this is consistent with the lack of a Bahcall-Wolf cusp at the
Galactic center.

The core in the Milky Way has a radius of $\sim 0.5\pc$, somewhat
smaller than $\rh$ or $\ri$, while the cores formed in our merger models are
somewhat larger than $\ri$ (Table 4).  It has been argued that the Milky Way
core is small enough that gravitational encounters would cause it to
shrink appreciably in 10 Gyr, as the stellar distribution evolves
toward a Bahcall-Wolf cusp \citep{merritt2010}.  The cores in our
$N$-body models are so large that they do not evolve appreciably after
the binary MBH has been replaced by a single MBH; the Bahcall-Wolf
cusp forms at radii smaller than $r_c$, leaving the core structure
essentially unchanged.  Without necessarily advocating a merger model
for the origin of the Milky Way core (it is unclear whether our galaxy
even contains a bulge; \citet{Valpuesta2011}), we note that the sizes
of cores formed by binary MBHs scale with the binary mass ratio.
Presumably, we could have produced cores more similar in size to the
Milky Way's if we had adjusted this ratio.

It has been suggested \citep{LBK2010} that heating by BHs could be
responsible for the lack of a Bahcall-Wolf cusp at the Galactic
center.  We do not find support for this hypothesis, either in our
four-component $N$-body models, nor in our two-component Fokker-Planck
models.  The latter showed that even a quite large BH population still
allowed a Bahcall-Wolf cusp to form in the MS stars on a time scale of
$\sim T_r(\ri)$; the main effect of the BHs is to decrease the
asymptotic slope of the cusp from $\sim r^{-7/4}$ to $\sim r^{-3/2}$.

\subsection{The distribution of massive remnants in galaxy nuclei}

A dense cluster of stellar-mass BHs has been invoked as a potential
solution to a number of problems of collisional dynamics at the
Galactic center.  Examples include randomization of the orbits of
young stars via gravitational scattering \citep[e.g.][]{perets09},
production of hyper-velocity stars through encounters with BHs
\citep[e.g.][]{OL08}, and warping of young stellar disks \citep{KT11}.
These treatments typically assume a collisionally-relaxed state for
the Galactic center.  In the relaxed models, the mass in BHs inside
$0.1\pc$ is $\sim 10^4\msun$, i.e. $N_\mathrm{BH} (r<0.1\pc) \approx
10^3$ and $N_\mathrm{BH}(r<0.01\pc)\approx 10^2$
\citep{HA06,Freitag06}, assuming ``standard'' IMFs.  A high density of
stellar BHs at the centers of galaxies like the Milky Way is also
commonly assumed in discussions of the EMRI (extreme-mass-ratio
inspiral) problem \citep{EMRIreview}.

Models like these are called into question by the lack of a
Bahcall-Wolf cusp in the late-type stars at the center of the Milky
Way \citep{buchholz09,Do09,Bartko10}.  If the Galactic center is not
collisionally relaxed, as these observations seem to suggest, then
computing the distribution of the heavy remnants becomes a more
difficult, time-dependent problem \citep{merritt2010}, and knowledge
of the initial conditions is essential.

Dissipationless mergers imply ``initial conditions'' characterized by
a core in the dominant stellar component.  The cores formed in our
(major) merger simulations are large enough that they do not evolve
appreciably (i.e. shrink) even after $\sim 3$ relaxation times at the
MBH influence radius.  As a result, evolution of the BH distribution
takes place against a stellar background with a very different density
profile than in the steady-state models.  In a core around a MBH,
dynamical friction is much weaker than one would estimate by plugging
the local density into standard formulae for orbital decay, due to the
absence of stars moving more slowly than the local circular velocity
\citep{AntoniniMerritt2011}.  Scaling our $N$-body models to the
Milky Way, we predicted numbers of BHs inside $\rh$ that were
substantially smaller than in the collisionally relaxed models; in the
inner $10^{-2}\pc$ -- the radii most relevant to the EMRI problem
\citep{MAMW2010} -- the number of BHs was, at most, 10-100 times
smaller than predicted by these models, even after 10 Gyr
(Figure\,\ref{fig:BHs}).

It is unclear how relevant merger models are to the center of the
Milky Way.  If the core observed at the Galactic center had some other
origin, the distribution of stellar BHs might have little connection
with the distribution of the giant stars.  However, if cores of size
$r_c\approx\rh$ are common features of galactic nuclei, and if at some
early time both the BHs and the stars had a common core radius, our
models imply that the distribution of BHs should be considered very
uncertain, even in galaxies for which nuclear half-mass relaxation
times are as short as the age of the universe.

\acknowledgements
DM acknowledges support from the National Science Foundation under
grants no. AST 08-07910, 08-21141 and by the National Aeronautics and
Space Administration under grant no. NNX-07AH15G.
We thank Tal Alexander and Eugene Vasiliev for useful discussions.

\bibliographystyle{apj}
%\bibliography{biblio}

\begin{thebibliography}{0}
\expandafter\ifx\csname natexlab\endcsname\relax\def\natexlab#1{#1}\fi

\end{thebibliography}


\begin{thebibliography}{85}
\expandafter\ifx\csname natexlab\endcsname\relax\def\natexlab#1{#1}\fi

\bibitem[{{Alexander}(2005)}]{Alexander2005}
{Alexander}, T. 2005, \physrep, 419, 65

\bibitem[{{Alexander} \& {Hopman}(2009)}]{AH2009}
{Alexander}, T., \& {Hopman}, C. 2009, \apj, 697, 1861

\bibitem[{{Amaro-Seoane} {et~al.}(2007){Amaro-Seoane}, {Gair}, {Freitag},
  {Miller}, {Mandel}, {Cutler}, \& {Babak}}]{EMRIreview}
{Amaro-Seoane}, P., {Gair}, J.~R., {Freitag}, M., {Miller}, M.~C., {Mandel},
  I., {Cutler}, C.~J., \& {Babak}, S. 2007, Classical and Quantum Gravity, 24,
  113

\bibitem[{{Antonini} {et~al.}(2009){Antonini}, {Capuzzo-Dolcetta}, \&
  {Merritt}}]{ant2009}
{Antonini}, F., {Capuzzo-Dolcetta}, R., \& {Merritt}, D. 2009, \mnras, 399, 671

\bibitem[{{Antonini} \& {Merritt}(2011)}]{AntoniniMerritt2011}
{Antonini}, F., \& {Merritt}, D. 2011, ArXiv e-prints 1108.1163

\bibitem[{{Bahcall} \& {Wolf}(1976)}]{BW76}
{Bahcall}, J.~N., \& {Wolf}, R.~A. 1976, \apj, 209, 214

\bibitem[{{Bahcall} \& {Wolf}(1977)}]{BW77}
---. 1977, \apj, 216, 883

\bibitem[{{Bartko} {et~al.}(2010){Bartko}, {Martins}, {Trippe}, {Fritz},
  {Genzel}, {Ott}, {Eisenhauer}, {Gillessen}, {Paumard}, {Alexander},
  {Dodds-Eden}, {Gerhard}, {Levin}, {Mascetti}, {Nayakshin}, {Perets},
  {Perrin}, {Pfuhl}, {Reid}, {Rouan}, {Zilka}, \& {Sternberg}}]{Bartko10}
{Bartko}, H., {et~al.} 2010, \apj, 708, 834

\bibitem[{{Baumgardt} {et~al.}(2004){Baumgardt}, {Makino}, \&
  {Ebisuzaki}}]{BME2004a}
{Baumgardt}, H., {Makino}, J., \& {Ebisuzaki}, T. 2004, \apj, 613, 1133

\bibitem[{{Begelman} {et~al.}(1980){Begelman}, {Blandford}, \& {Rees}}]{BBR80}
{Begelman}, M.~C., {Blandford}, R.~D., \& {Rees}, M.~J. 1980, \nat, 287, 307

\bibitem[{{Berczik} {et~al.}(2005){Berczik}, {Merritt}, \& {Spurzem}}]{ber05}
{Berczik}, P., {Merritt}, D., \& {Spurzem}, R. 2005, \apj, 633, 680

\bibitem[{{Berczik} {et~al.}(2006){Berczik}, {Merritt}, {Spurzem}, \&
  {Bischof}}]{ber06}
{Berczik}, P., {Merritt}, D., {Spurzem}, R., \& {Bischof}, H.-P. 2006, \apjl,
  642, L21

\bibitem[{{Binney} \& {Tremaine}(1987)}]{bt87}
{Binney}, J., \& {Tremaine}, S. 1987, {Galactic dynamics}, ed. J.~{Binney} \&
  S.~{Tremaine}

\bibitem[{{Boeker}(2010)}]{Boeker2010}
{Boeker}, T. 2010, {The Impact of HST on European Astronomy}, ed. {Macchetto,
  F.~D.}

\bibitem[{{Brown} {et~al.}(2005){Brown}, {Geller}, {Kenyon}, \&
  {Kurtz}}]{brown05}
{Brown}, W.~R., {Geller}, M.~J., {Kenyon}, S.~J., \& {Kurtz}, M.~J. 2005,
  \apjl, 622, L33

\bibitem[{{Brown} {et~al.}(2006){Brown}, {Geller}, {Kenyon}, \&
  {Kurtz}}]{brown06}
---. 2006, \apj, 647, 303

\bibitem[{{Buchholz} {et~al.}(2009){Buchholz}, {Sch{\"o}del}, \&
  {Eckart}}]{buchholz09}
{Buchholz}, R.~M., {Sch{\"o}del}, R., \& {Eckart}, A. 2009, \aap, 499, 483

\bibitem[{{C{\^o}t{\'e}} {et~al.}(2007){C{\^o}t{\'e}}, {Ferrarese},
  {Jord{\'a}n}, {Blakeslee}, {Chen}, {Infante}, {Merritt}, {Mei}, {Peng},
  {Tonry}, {West}, \& {West}}]{cote2007}
{C{\^o}t{\'e}}, P., {et~al.} 2007, \apj, 671, 1456

\bibitem[{{Cuadra} {et~al.}(2009){Cuadra}, {Armitage}, {Alexander}, \&
  {Begelman}}]{cuadra09}
{Cuadra}, J., {Armitage}, P.~J., {Alexander}, R.~D., \& {Begelman}, M.~C. 2009,
  \mnras, 393, 1423

\bibitem[{{Dehnen}(2005)}]{Dehnen2005}
{Dehnen}, W. 2005, \mnras, 360, 892

\bibitem[{{Do} {et~al.}(2009){Do}, {Ghez}, {Morris}, {Lu}, {Matthews}, {Yelda},
  \& {Larkin}}]{Do09}
{Do}, T., {Ghez}, A.~M., {Morris}, M.~R., {Lu}, J.~R., {Matthews}, K., {Yelda},
  S., \& {Larkin}, J. 2009, \apj, 703, 1323

\bibitem[{{Dotti} {et~al.}(2007){Dotti}, {Colpi}, {Haardt}, \&
  {Mayer}}]{dotti07}
{Dotti}, M., {Colpi}, M., {Haardt}, F., \& {Mayer}, L. 2007, \mnras, 379, 956

\bibitem[{{Escala} {et~al.}(2005){Escala}, {Larson}, {Coppi}, \&
  {Mardones}}]{escala05}
{Escala}, A., {Larson}, R.~B., {Coppi}, P.~S., \& {Mardones}, D. 2005, \apj,
  630, 152

\bibitem[{{Fakhouri} {et~al.}(2010){Fakhouri}, {Ma}, \&
  {Boylan-Kolchin}}]{Fakhouri2010}
{Fakhouri}, O., {Ma}, C.-P., \& {Boylan-Kolchin}, M. 2010, \mnras, 406, 2267

\bibitem[{{Ferrarese} {et~al.}(1994){Ferrarese}, {van den Bosch}, {Ford},
  {Jaffe}, \& {O'Connell}}]{ferr1994}
{Ferrarese}, L., {van den Bosch}, F.~C., {Ford}, H.~C., {Jaffe}, W., \&
  {O'Connell}, R.~W. 1994, \aj, 108, 1598

\bibitem[{{Ferrarese} {et~al.}(2006{\natexlab{a}}){Ferrarese}, {C{\^o}t{\'e}},
  {Dalla Bont{\`a}}, {Peng}, {Merritt}, {Jord{\'a}n}, {Blakeslee}, {Ha{\c
  s}egan}, {Mei}, {Piatek}, {Tonry}, \& {West}}]{Ferrarese2006}
{Ferrarese}, L., {et~al.} 2006{\natexlab{a}}, \apjl, 644, L21

\bibitem[{{Ferrarese} {et~al.}(2006{\natexlab{b}}){Ferrarese}, {C{\^o}t{\'e}},
  {Jord{\'a}n}, {Peng}, {Blakeslee}, {Piatek}, {Mei}, {Merritt},
  {Milosavljevi{\'c}}, {Tonry}, \& {West}}]{ferr2006}
---. 2006{\natexlab{b}}, \apjs, 164, 334

\bibitem[{{Figer} {et~al.}(2004){Figer}, {Rich}, {Kim}, {Morris}, \&
  {Serabyn}}]{figer2004}
{Figer}, D.~F., {Rich}, R.~M., {Kim}, S.~S., {Morris}, M., \& {Serabyn}, E.
  2004, \apj, 601, 319

\bibitem[{{Freitag} {et~al.}(2006){Freitag}, {Amaro-Seoane}, \&
  {Kalogera}}]{Freitag06}
{Freitag}, M., {Amaro-Seoane}, P., \& {Kalogera}, V. 2006, \apj, 649, 91

\bibitem[{{Gaburov} {et~al.}(2009){Gaburov}, {Harfst}, \& {Portegies
  Zwart}}]{sapporo2009}
{Gaburov}, E., {Harfst}, S., \& {Portegies Zwart}, S. 2009, \na, 14, 630

\bibitem[{{Graham}(2004)}]{graham2004}
{Graham}, A.~W. 2004, \apjl, 613, L33

\bibitem[{{Graham} {et~al.}(2003){Graham}, {Erwin}, {Trujillo}, \& {Asensio
  Ramos}}]{graham2003}
{Graham}, A.~W., {Erwin}, P., {Trujillo}, I., \& {Asensio Ramos}, A. 2003, \aj,
  125, 2951

\bibitem[{{Graham} \& {Guzm{\'a}n}(2003)}]{GG2003}
{Graham}, A.~W., \& {Guzm{\'a}n}, R. 2003, \aj, 125, 2936

\bibitem[{{Graham} \& {Spitler}(2009)}]{GS2009}
{Graham}, A.~W., \& {Spitler}, L.~R. 2009, \mnras, 397, 2148

\bibitem[{{Gualandris} \& {Portegies Zwart}(2007)}]{GPZ07}
{Gualandris}, A., \& {Portegies Zwart}, S. 2007, \mnras, 376, L29

\bibitem[{{Gualandris} {et~al.}(2005){Gualandris}, {Portegies Zwart}, \&
  {Sipior}}]{GPS05}
{Gualandris}, A., {Portegies Zwart}, S., \& {Sipior}, M.~S. 2005, \mnras, 363,
  223

\bibitem[{{Gvaramadze} {et~al.}(2009){Gvaramadze}, {Gualandris}, \& {Portegies
  Zwart}}]{gva2009}
{Gvaramadze}, V.~V., {Gualandris}, A., \& {Portegies Zwart}, S. 2009, \mnras,
  396, 570

\bibitem[{{Harfst} {et~al.}(2007){Harfst}, {Gualandris}, {Merritt}, {Spurzem},
  {Portegies Zwart}, \& {Berczik}}]{Harfst2007}
{Harfst}, S., {Gualandris}, A., {Merritt}, D., {Spurzem}, R., {Portegies
  Zwart}, S., \& {Berczik}, P. 2007, New Astronomy, 12, 357

\bibitem[{{Hopman}(2009)}]{Hopman2009}
{Hopman}, C. 2009, Classical and Quantum Gravity, 26, 094028

\bibitem[{{Hopman} \& {Alexander}(2006)}]{HA06}
{Hopman}, C., \& {Alexander}, T. 2006, \apjl, 645, L133

\bibitem[{{Katz}(1991)}]{katz1991}
{Katz}, N. 1991, \apj, 368, 325

\bibitem[{{Kauffmann} \& {Haehnelt}(2000)}]{KH00}
{Kauffmann}, G., \& {Haehnelt}, M. 2000, \mnras, 311, 576

\bibitem[{{Khan} {et~al.}(2011){Khan}, {Just}, \& {Merritt}}]{khan2011}
{Khan}, F.~M., {Just}, A., \& {Merritt}, D. 2011, \apj, 732, 89

\bibitem[{{King}(1962)}]{King1962}
{King}, I. 1962, \aj, 67, 471

\bibitem[{{Kocsis} \& {Tremaine}(2011)}]{KT11}
{Kocsis}, B., \& {Tremaine}, S. 2011, \mnras, 412, 187

\bibitem[{{Kormendy}(1985)}]{Kormendy1985}
{Kormendy}, J. 1985, \apj, 295, 73

\bibitem[{{Lauer} {et~al.}(1995){Lauer}, {Ajhar}, {Byun}, {Dressler}, {Faber},
  {Grillmair}, {Kormendy}, {Richstone}, \& {Tremaine}}]{Lauer95}
{Lauer}, T.~R., {et~al.} 1995, \aj, 110, 2622

\bibitem[{{L{\"o}ckmann} {et~al.}(2010){L{\"o}ckmann}, {Baumgardt}, \&
  {Kroupa}}]{LBK2010}
{L{\"o}ckmann}, U., {Baumgardt}, H., \& {Kroupa}, P. 2010, \mnras, 402, 519

\bibitem[{{Makino} \& {Funato}(2004)}]{makinofunato2004}
{Makino}, J., \& {Funato}, Y. 2004, \apj, 602, 93

\bibitem[{{Maness} {et~al.}(2007){Maness}, {Martins}, {Trippe}, {Genzel},
  {Graham}, {Sheehy}, {Salaris}, {Gillessen}, {Alexander}, {Paumard}, {Ott},
  {Abuter}, \& {Eisenhauer}}]{Maness2007}
{Maness}, H., {et~al.} 2007, \apj, 669, 1024

\bibitem[{{Martinez-Valpuesta} \& {Gerhard}(2011)}]{Valpuesta2011}
{Martinez-Valpuesta}, I., \& {Gerhard}, O. 2011, \apjl, 734, L20+

\bibitem[{{Merritt}(2006)}]{merritt06}
{Merritt}, D. 2006, \apj, 648, 976

\bibitem[{{Merritt}(2009)}]{merritt2009}
---. 2009, \apj, 694, 959

\bibitem[{{Merritt}(2010)}]{merritt2010}
---. 2010, \apj, 718, 739

\bibitem[{{Merritt}(2012)}]{DEGN}
---. 2012, {Dynamics and Evolution of Galactic Nuclei} (Princeton University
  Press)

\bibitem[{{Merritt} {et~al.}(2010){Merritt}, {Alexander}, {Mikkola}, \&
  {Will}}]{MAMW2010}
{Merritt}, D., {Alexander}, T., {Mikkola}, S., \& {Will}, C.~M. 2010, \prd, 81,
  062002

\bibitem[{{Merritt} \& {Cruz}(2001)}]{mc2001}
{Merritt}, D., \& {Cruz}, F. 2001, \apjl, 551, L41

\bibitem[{{Merritt} {et~al.}(2007{\natexlab{a}}){Merritt}, {Harfst}, \&
  {Bertone}}]{mhb07}
{Merritt}, D., {Harfst}, S., \& {Bertone}, G. 2007{\natexlab{a}}, \prd, 75,
  043517

\bibitem[{{Merritt} {et~al.}(2007{\natexlab{b}}){Merritt}, {Mikkola}, \&
  {Szell}}]{mms07}
{Merritt}, D., {Mikkola}, S., \& {Szell}, A. 2007{\natexlab{b}}, \apj, 671, 53

\bibitem[{{Merritt} \& {Poon}(2004)}]{merrittpoon2004}
{Merritt}, D., \& {Poon}, M.~Y. 2004, \apj, 606, 788

\bibitem[{{Merritt} \& {Szell}(2006)}]{MS06}
{Merritt}, D., \& {Szell}, A. 2006, \apj, 648, 890

\bibitem[{{Merritt} \& {Vasiliev}(2010)}]{merrittvasiliev2010}
{Merritt}, D., \& {Vasiliev}, E. 2010, \apj, 726, 61

\bibitem[{{Mikkola} \& {Valtonen}(1992)}]{MikkolaValtonen1992}
{Mikkola}, S., \& {Valtonen}, M.~J. 1992, \mnras, 259, 115

\bibitem[{{Milosavljevi{\'c}} \& {Merritt}(2001)}]{mm01}
{Milosavljevi{\'c}}, M., \& {Merritt}, D. 2001, \apj, 563, 34

\bibitem[{{Milosavljevi{\'c}} {et~al.}(2002){Milosavljevi{\'c}}, {Merritt},
  {Rest}, \& {van den Bosch}}]{mm02}
{Milosavljevi{\'c}}, M., {Merritt}, D., {Rest}, A., \& {van den Bosch}, F.~C.
  2002, \mnras, 331, L51

\bibitem[{{Oh} {et~al.}(2009){Oh}, {Kim}, \& {Figer}}]{oh2009}
{Oh}, S., {Kim}, S.~S., \& {Figer}, D.~F. 2009, Journal of Korean Astronomical
  Society, 42, 17

\bibitem[{{O'Leary} \& {Loeb}(2008)}]{OL08}
{O'Leary}, R.~M., \& {Loeb}, A. 2008, \mnras, 383, 86

\bibitem[{{Perets} {et~al.}(2009){Perets}, {Gualandris}, {Kupi}, {Merritt}, \&
  {Alexander}}]{perets09}
{Perets}, H.~B., {Gualandris}, A., {Kupi}, G., {Merritt}, D., \& {Alexander},
  T. 2009, \apj, 702, 884

\bibitem[{Peters(1964)}]{peters64}
Peters, P.~C. 1964, Physical Review, 136, 1224

\bibitem[{{Preto} \& {Amaro-Seoane}(2010)}]{PAS2010}
{Preto}, M., \& {Amaro-Seoane}, P. 2010, \apjl, 708, L42

\bibitem[{{Preto} {et~al.}(2011){Preto}, {Berentzen}, {Berczik}, \&
  {Spurzem}}]{preto2011}
{Preto}, M., {Berentzen}, I., {Berczik}, P., \& {Spurzem}, R. 2011, \apjl, 732,
  L26+

\bibitem[{{Preto} {et~al.}(2004){Preto}, {Merritt}, \& {Spurzem}}]{preto2004}
{Preto}, M., {Merritt}, D., \& {Spurzem}, R. 2004, \apjl, 613, L109

\bibitem[{{Quinlan}(1996)}]{quinlan1996}
{Quinlan}, G.~D. 1996, New A, 1, 35

\bibitem[{{Rosenbluth} {et~al.}(1957){Rosenbluth}, {MacDonald}, \&
  {Judd}}]{RMJ57}
{Rosenbluth}, M.~N., {MacDonald}, W.~M., \& {Judd}, D.~L. 1957, Physical
  Review, 107, 1

\bibitem[{{Sch{\"o}del}(2011)}]{Schoedel2011}
{Sch{\"o}del}, R. 2011, in Astronomical Society of the Pacific Conference
  Series, Vol. 439, Astronomical Society of the Pacific Conference Series, ed.
  {M.~R.~Morris, Q.~D.~Wang, \& F.~Yuan}, 222--+

\bibitem[{{Sch{\"o}del} {et~al.}(2009){Sch{\"o}del}, {Merritt}, \&
  {Eckart}}]{SME2009}
{Sch{\"o}del}, R., {Merritt}, D., \& {Eckart}, A. 2009, Astron. Astrophys.,
  502, 91

\bibitem[{{Sesana} {et~al.}(2011){Sesana}, {Gualandris}, \& {Dotti}}]{SGD2011}
{Sesana}, A., {Gualandris}, A., \& {Dotti}, M. 2011, \mnras, 415, L35

\bibitem[{{Sesana} {et~al.}(2006){Sesana}, {Haardt}, \& {Madau}}]{sesana2006}
{Sesana}, A., {Haardt}, F., \& {Madau}, P. 2006, \apj, 651, 392

\bibitem[{{Seth} {et~al.}(2008){Seth}, {Ag{\"u}eros}, {Lee}, \&
  {Basu-Zych}}]{Seth2008}
{Seth}, A., {Ag{\"u}eros}, M., {Lee}, D., \& {Basu-Zych}, A. 2008, \apj, 678,
  116

\bibitem[{{Spitzer}(1987)}]{spitzer1987}
{Spitzer}, L. 1987, {Dynamical evolution of globular clusters}, ed. {Spitzer,
  L.}

\bibitem[{{Sridhar} \& {Touma}(1999)}]{sridhartouma1999}
{Sridhar}, S., \& {Touma}, J. 1999, \mnras, 303, 483

\bibitem[{{Szell} {et~al.}(2005){Szell}, {Merritt}, \&
  {Kevrekidis}}]{Szell2005}
{Szell}, A., {Merritt}, D., \& {Kevrekidis}, I.~G. 2005, Physical Review
  Letters, 95, 081102

\bibitem[{{Terzi{\'c}} \& {Graham}(2005)}]{tg05}
{Terzi{\'c}}, B., \& {Graham}, A.~W. 2005, \mnras, 362, 197

\bibitem[{{Volonteri} {et~al.}(2003){Volonteri}, {Haardt}, \& {Madau}}]{VHM03}
{Volonteri}, M., {Haardt}, F., \& {Madau}, P. 2003, \apj, 582, 559

\bibitem[{{Walcher} {et~al.}(2006){Walcher}, {B{\"o}ker}, {Charlot}, {Ho},
  {Rix}, {Rossa}, {Shields}, \& {van der Marel}}]{Walcher2006}
{Walcher}, C.~J., {B{\"o}ker}, T., {Charlot}, S., {Ho}, L.~C., {Rix}, H.-W.,
  {Rossa}, J., {Shields}, J.~C., \& {van der Marel}, R.~P. 2006, \apj, 649, 692

\end{thebibliography}

\end{document}